*Review*

# The Pnictogen Bond Formation Ability of Bonded Bismuth Atoms in Molecular Entities in the Crystalline Phase: A Perspective

Pradeep R. Varadwaj [1,2,*], Arpita Varadwaj [1,*] Helder M. Marques [2] and Koichi Yamashita[1]

[1] Department of Chemical System Engineering, School of Engineering, The University of Tokyo 7-3-1, Tokyo 113-8656, Japan; varadwaj.arpita@gmail.com (A.V.); yamasita@chemsys.t.u-tokyo.ac.jp (K.Y.)
[2] Molecular Sciences Institute, School of Chemistry, University of the Witwatersrand, Johannesburg 2050, South Africa; helder.marques@wits.ac.za
* Correspondence: varadwaj.arpita@gmail.com; pradeep@t.okayama-u.ac.jp

**Abstract:** A bismuth bond, a type of pnictogen bonding interaction, occurs in chemical systems when there is evidence of a net attractive interaction between the electrophilic region of a covalently or coordinately bonded bismuth atom and the nucleophilic region in another, or the same, molecular entity. In this review, we report on the signatory details of bismuth bonding in several crystalline systems, along with other non-covalent interactions such as hydrogen and halogen bonds, which are important driving forces in the rational design of various types of materials. Illustrative crystal structures were retrieved through careful inspection of the Inorganic Crystal Structure Database (ICSD) and Cambridge Structural Database (CSD). Although thousands of crystal structures containing bismuth have been deposited in these databases, we selected only a number in which the covalently or coordinately bonded bismuth atoms have clearly conceived positive regions on their electrostatic surfaces. We show that these positive regions on bismuth electrostatically attract various Lewis bases, including, for example, O, N, F, P, Cl, Br, I, S, Se, Te, and Bi atoms, as well as those with regions of π-density in $C_\pi$ (arene) and $(C≡C)_\pi$ moieties, present in the same or different molecular entities, resulting in the formation of bismuth bonds. The characteristics of bismuth bonds were evaluated using several current state-of-the-art techniques, including geometric features such as inter- and intramolecular distances, which were also used to verify the "less than the sum of van der Waals radii concept," and the use of interaction angles, which indicate the presence of directionality. The results of molecular electrostatic potential model calculations were used to draw conclusions about the electrophilic and nucleophilic properties of specific regions on the electrostatic surface of certain isolated monomeric entities and to establish their link with bismuth bonds. The independent gradient model was utilized to calculate the pro- or actual-molecular charge density based iso-surface topology between interacting atomic basins and to demonstrate that the inter- and/or intramolecular interactions that occurred between them were genuine. Statistical analysis of various Bi-centered pnictogen bonding interactions formed by different electron density donors was performed to show the nature of the population of each interaction type in crystals and to provide insight into the nature of the intermolecular bond distance and bond angle ranges involved for each interaction type.

## 1. Introduction

Pnictogens refer to elements in Group 15 of the periodic table. They, especially the heavier pnictogen derivatives in molecular entities, have played a very important role in the development of high-density, clean energy, and alloy-based anode battery materials. The origin of the terms *pnictogen* (sometimes referred to as *pnigogen* or *pnicogen*) and *pnictide* may be traced back to an opinion provided in the early 1950s [1]. However, the term *pnictogen bond* (PnB) appeared relatively recently in the non-covalent chemistry and crystallography literature (for example [2-10]). It has similar connotations to, for instance, hydrogen bond, halogen bond (formed by Group 17 elements), tetrel bond (formed by Group 14 elements) and chalcogen bond (formed by Group 16 elements), among others. As we shall introduce below, the nature and characteristics of a pnictogen bond are similar to those of the hydrogen bond (HB) that was formally redefined in 2011 [11]. The halogen bond (XB) was formally defined by a IUPAC working group in 2013 [12], as was the chalcogen bond (ChB) in 2019 [13]; each definition was accompanied by a list of characteristics and features. However, since relatively little is known about pnictogen bonding, IUPAC has yet to propose a definition and list its characteristics; it is expected to be formally published in the future.

Some have suggested [14] that pnictogen bonds have been known since 2011 (with one experimental [15] and two theoretical [16,17] reports). However, we have found that numerous experimental studies dealing with PnBs have been available in the literature since the middle of the last century, and hence experimental evidence in support of the interaction in crystal systems are not rare [7-10]. In fact, there are thousands of crystals reported over many decades that are either the result of pnictogen bonding, or a joint involvement of pnictogen bonding and other non-covalent interactions (hydrogen bond, halogen bond, tetrel bond, or chalcogen bond, among others).

The similarity between a hydrogen bond, a halogen bond, and a chalcogen bond, all of which may be represented by the geometric motif R–A···D, is that atom A (a hydrogen, a halogen or a chalcogen) covalently (or coordinately) bound to R, has a positive site or sites on its electrostatic surface that is able to attract the negative sites of a donor atom D (such as a Lewis base like N in $NH_3$) to form a non-covalent interaction along the extension of the R–A bond, where R is the remaining part of the molecular entity R–A. If the interaction lies precisely along, or perhaps slightly off, the R–A bond extension it said to be a linear, or quasi-linear, interaction. If the positive site on A is not along the R–A bond extension, then the interaction between A and D is non-linear (curved, or bent). The latter nature of the bond also originates when secondary/tertiary interactions are involved, even though the positive site on A appears along (or slightly off) the R–A bond extension. These concepts apply to the covalently or coordinately bonded pnictogen (Group 15) elements provided they feature positive sites on their electrostatic surfaces that are able to act as electrophilic regions for nucleophilic regions in near neighbors to sustain an attractive engagement among one another, causing the local development of packing in crystals. The concept underlying the formation of non-covalent interactions is transferable to all elements of the periodic table, *albeit* with different names. Because of this similarity, hydrogen bonds, halogen bonds, chalcogen bonds, and pnictogen bonds are sometimes referred to as sister non-covalent interactions [18-21].

The main difference between hydrogen, halogen, chalcogen, and pnictogen bonds is really one of terminology; in the motif R–A···D, atom A involved in these bonds could belong to a different group of the periodic table. For this reason, these bonds have been given different names; so, for example, the name H-bond (HB) is given to attractive interactions formed by a covalently bound hydrogen atom, and an XB is formed by a covalently bound halogen [22,23], and so on.

Known non-covalent interactions have different bonding characteristics; some are called σ-hole interactions [24,25], others π-hole interactions [24,26-29], and van der Waals interactions [30-34]. The latter are generally dispersive, with a binding energy ≤ −1.0 kcal mol$^{-1}$, and can be of the σ-hole or the π-hole type. A σ-hole on atom A in a molecular entity R–A is a region of electron density deficiency along the R–A covalent/coordinate σ-bond extension when compared to the lateral regions of A [35,36]. Thus, a σ-hole on A can be positive, negative, or neutral, depending on the extent of electron density deficiency associated with it [37-39]. For instance, the σ-hole on F in HF [40], $CH_3F$ [41], $CH_3F$, $PF_3$ [8], and $C_6F_6$ [38] is negative, whereas that on F in $F_2$ [42], $NF_3$ [7], and $CF_4$ [43] is positive. There are rare examples in which a σ-hole is electrically neutral (such as on F in $AsF_3$) [9]. Similarly, a π-hole on the surface of a molecular entity is a charge density deficient region. It can be either positive or negative, and often appears perpendicular to a portion of a molecular framework [24,26-29]. Examples of simple chemical systems containing π-holes include arene moieties, such as the centroid region of $C_6X_6$ (X = H, a halogen [44,45]), multiple bonds such as P≡N, C≡C, C=O, N≡N, and P≡P [7,8], and delocalized bonds (as in $NO_3^-$ [46]). There have been many chemical systems theoretically reported that feature π-holes. σ- and/or π-holes have already shown great potential as chemically oriented supramolecular synthons for the rational design of various molecular and supramolecular complexes, crystals, and nanoscale materials.

Drawing on these concepts above, a pnictogen bond formed by a molecular entity can be σ-hole centered, or π-hole centered. When a positive σ-hole on the surface of a Pn atom in R–Pn forms a non-covalent interaction with a donor D, this would be referred to as a σ-hole interaction; the same would apply when a negative σ-hole plays an analogous role (counter-intuitive). When a π-hole involved with Pn makes a non-covalent interaction, it is called a π-hole interaction. The σ- and π-hole interactions formed by a covalently bound pnictogen atom featuring an electron deficient region, often but not always positive, would be referred to as σ- and π-hole centered pnictogen bonded interactions, respectively. Although pnictogen bonding is one of the least theoretically and computationally studied non-covalent chemical interactions [2,4,5,14,47-49], its implication is widespread in catalysis [50-53], crystal packing [54,55], coordination chemistry [4,56-60], photovoltaics [61-63] and supramolecular chemistry [14,49].

This review discusses the ability of bismuth in molecular entities to participate in non-covalent bonding interactions, and the features and properties of these interactions in crystal lattices and nanoscale materials. We have done so because of the frequency of its occurrence, which is often overlooked; synthons featuring these non-covalent interactions may well turn out to be important in the design of novel materials. Single crystals with no errors, with an *R*-factor ≤ 0.1, retrieved from the Cambridge Structural Database (CSD) [64,65] and the Inorganic Crystal Structure Databases (ICSD) [66,67], were explored. These chemical systems were either unimolecular or largely formed between two ions of opposite charge polarity to satisfy the charge-neutral condition.

We propose that *a bismuth bond, or a covalently/coordinately bound bismuth bond, in chemical systems occurs when there is evidence of a net attractive interaction between the electrophilic region associated with a covalently or coordinately bound bismuth atom in a molecular entity and a nucleophilic region in another, or the same, molecular entity*.

Accordingly, a bismuth bond can be regarded as an intermolecular or intramolecular non-covalent bond.

We have considered the inter- or intra-molecular distance [68-72] and directionality (angle of interaction that electrophilic Bi makes with donor site(s)) [70,73] as a major geometrical feature to characterize bismuth bonds in the illustrative crystal systems examined. Statistical analyses of the geometric data obtained from the CSD-based crystallographic survey were performed to demonstrate the frequency of occurrence of various types of bismuth bonds (Bi···O, Bi···N, Bi···F, Bi···S, and Bi···π, etc.), their distance ranges and angular features in the crystals. This was done in order to propose the involvement of various electron density donors that were responsible in the formation of Bi-centered pnictogen bonds.

The pro- or actual-molecular charge density isosurfaces that emerged from the recently proposed Independent Gradient Model (IGM) [74,75] were examined to validate the presence or absence of bismuth bonds. Moreover, the surface extrema of potential evaluated using the Molecular Electrostatic Surface Potential (MESP) model (for details see for example [26,27,35,68,76-79]) were used as descriptors to provide insight on the possibility of bonding modes of bismuth in and/or between molecular entities in the chemical systems examined.

## 2. The inter- and intra-molecular interactions and the sum of the van der Waals radii criterion

Historically [80], and also more recently [72,81-83], the *less than the sum of the van der Waals radii* concept has been widely utilized to identify and characterize non-covalent interactions in molecules, molecular complexes, and crystals [7-10,84]. Based on this concept, when the inter- or intra-molecular interaction distance in a molecular or supermolecular entity is less than the sum of the van der Waals (vdW) radii of interacting atomic basins, they should be regarded as being non-covalently bonded to each other. This is a necessary condition for a non-covalent interaction, but might not be sufficient.

While some have found the criterion is actually an impediment in the search for non-covalent interactions [69], others have argued that it is very useful for rationalizing such interactions in chemical systems, including molecules, molecular complexes, and crystals. Alvarez observed that the vdW radii and associated vdW surfaces are extensively used for crystal packing and supramolecular interaction analysis [72]. Politzer and Murray [68], and others [72,85], have argued that where the Pn···D inter- or intramolecular distance exceeds the sum of the vdW radii by several tenths of an Ångstrom in a crystal system, this can still be recognized as a non-bonded interaction. This is understandable since values proposed for the vdW radii of atoms have a typical uncertainty of ± 0.2 Å; hence "less than the sum of the vdW radii" concept will necessarily miss a significant number of non-covalent interactions if treated as a strict criterion to identify a non-covalent interaction. This uncertainty is not surprising since a hard sphere model with spherical symmetry of electron density was assumed in proposing the vdW radii of atoms. In reality, the charge density profile of atoms in molecules is anisotropic, and hence the vdW "radius" of an atom in a molecular entity is likely to vary between molecular entities. Chernyshov and co-workers [86] have recently attempted to describe an efficient and universal approach for the analysis of non-covalent interactions and determination of vdW radii using the line-of-sight (LoS) concept. The authors argued that this approach is able to unambiguously identify and classify the "direct" interatomic contacts in complex molecular systems, and hence is an improved theoretical base to molecular "sizes" but also enables the quantitative analysis of specificity, anisotropy, and steric effects of intermolecular interactions.

With the limitations in the accuracy of the vdW radii of atoms firmly in mind, we have used the "less than the sum of the vdW radii" concept to examine the possibility of Bi-centered non-covalent interactions in the crystals explored. Concomitantly, we examined the directional feature that is often invoked, the angle of interaction [73], in rationalizing whether a covalently or coordinately bound Bi atom in a molecular entity is involved attractively with a negative site, or a positive site, and whether such an attraction exhibiting the directional feature contributes to the overall stability of the crystal lattice.

## 3. Directionality

As mentioned before [9,10], and reproduced here for the reader's convenience, to maximize the integrity of our identification and subsequent characterization of pnictogen bonding in the crystal systems examined, we took into account the angle of interaction.

The putative site of interaction on donor D was inspected to determine whether it was indeed a nucleophilic or an electrophilic domain. Type-I interactions (Scheme 1a), which can be subdivided into Type-Ia and Type-Ib, appear when the regions of the interacting atomic domains have either both a positive or negative local polarity. They are non-linear interactions, and non-directional. There are many chemical systems deposited in the CSD where the angles $\theta_1$ and $\theta_2$ (Scheme 1a, left) are virtually equal (and range between 110° and 150°), and should therefore be regarded as Type-Ia (*trans*) depending on the charge polarity of both Pn and D in the R–Pn···D motif. For a Type-Ib topology of bonding to appear between interacting atomic domains in chemical system, the angles $\theta_1$ and $\theta_2$ should be reasonably different from one another, thus following a configuration similar to that depicted in Scheme 1a (right).

The angle $\theta$ of approach, $\theta = \angle R–Pn(Bi)\cdots D$, of the electrophilic region on the electrostatic surface of Bi towards the nucleophilic region of the donor, D, was determined and classified as a linear, quasilinear, or bent (non-linear) interaction. There could be three σ-holes, for instance, along the extensions of three covalently/coordinately bound R–Bi bonds in a molecular entity containing a trivalent bismuth, so there would be three angles between the electrophiles on bismuth and D. The angle, $\theta$, corresponding to each of the three R–Bi$\cdots$D σ-hole interactions may follow a common Type-IIa pattern of bonding (Scheme 1b, left).

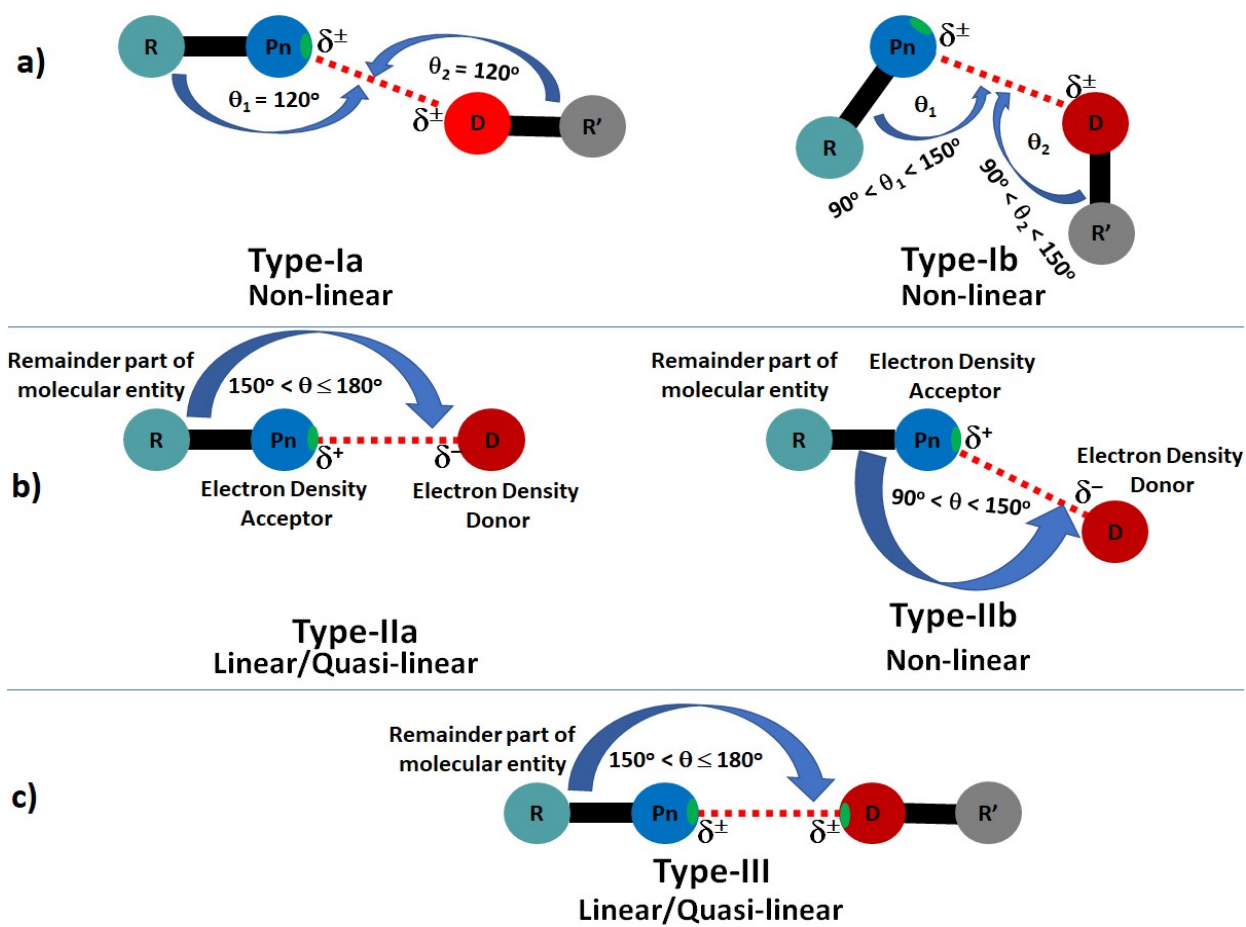

**Scheme 1.** A schematic view of a) Type-I, b) Type-II and c) Type-III topologies (geometrical) of non-covalent bonding interactions (Pn = Bi in this case). The $\theta$s represent the angle of interaction between the interacting atomic domains, and Pn (i.e., Bi in this case), D, R and R′ refer to the covalently bound pnictogen atom, the interacting atomic domain (generally nucleophilic), and the remaining part of the molecular entities associated with Pn and D atomic basins, respectively. $\delta^\pm$ signifies the local polarity (positive or negative), and the small region on atom Pn along R–Pn bond extension colored in green indicates a σ-hole. This classification scheme has been discussed elsewhere [40,70].

The directional interactions that are linear or quasi-linear generally follow a Type-IIa topology of bonding (Scheme 1b, left). This is a topology of bonding that typifies the name "pnictogen bonding"; the topology also resembles hydrogen bonding, halogen bonding, chalcogen bonding and so on, in which a positive site on a covalently bound atom (in this case hydrogen, halogen, and chalcogen and so on) is linearly/quasi-linearly in attractive engagement with a negative site on another site on the same or on a different molecular entity. As such, the interaction is linear when $\theta = \angle$R–Pn···D = 180°, and quasi-linear when 150° < $\theta$ < 180°. In either case, the electrostatic surface of covalently bound Bi must feature an electrophilic region along the extension of the R–Bi covalent or coordinate bond, and D is a Lewis base (such as N in $NH_3$, O in $OH_2$, and F in HF). There could be exceptions, for example, when the angle $\theta$ < 150°, but the covalently bound Pn atom (Bi this case) in a molecular entity that has an electrophile on it is non-linearly attracting a negative site on an interacting molecular entity. Thus, when 90° < $\theta$ < 150°, and the electrostatic surface of Bi still features a positive region, we recognize the interaction as being of Type-IIb. Both Type-IIa and Type-IIb interactions are of coulombic origin.

Type-III interactions occur when the angle of interaction follows a Type-IIa topology of bonding, but the interacting regions on Pn and D are either both positive or both negative. This type of interaction occurs not only between anions, or between cations, in chemical adducts in the crystalline phase, but also in chemical systems, in which two covalently bound positive Bi atoms in molecular entities attract each other due to differences in the charge density on their electrostatic surfaces (see succeeding sections for examples).

This study highlights chemical systems that feature all three types of non-covalent interaction delineated above.

**4. Computational approaches**

The following section contains several illustrative crystal systems retrieved from CSD or ICSD to highlight instances where bismuth bonding can be anticipated. In order to provide insight into evidence of Bi-centered pnictogen bonding formed by Bi containing molecular entities and the negative site in interacting species, a few systems were considered for the calculation of Molecular Electrostatic Surface Potential (MESP). The simplest chemical systems chosen include the bismuth trihalides, $BiX_3$ (X = F, Cl, Br, I), trimethyl bismuth, $Bi(CH_3)_3$, and molecular Bi, $Bi_2$.

Electronic structure calculations at the MP2 level theory [87,88], in conjunction with def2-TZVPPD pseudopotential basis set [89], were performed to obtain their equilibrium geometries. Normal mode vibration analysis was performed for each of the six systems mentioned above and positive eigenvalues were found. We then calculated the resulting wavefunctions at the same level of theory to compute the electrostatic surface potential of the molecular entities. The Gaussian 16 suite of programs was used [90]. The 0.001 a.u. (electrons Bohr$^{-3}$) isoelectron density envelope that arbitrarily defines the van der Waals surface of a molecular entity was used on which to calculate the potential. As has been shown many times [7-10,26,35,76,77], the application of the MESP model to a molecular entity results in two types of extrema called the local most minimum and the local most maximum of potential ($V_{S,min}$ and $V_{S,min}$, respectively) that appear on the molecular surface. The sign of both $V_{S,min}$ and $V_{S,min}$ could either be positive, or negative, or sometimes even neutral. When it positive, it is generally assumed that the region on the surface that accompanies this is electrophilic, and hence may be suitable for accepting electron density from an interacting electron donor in close proximity. When it is the negative, the reactive feature is opposite, meaning that the region on the surface of the molecular entity that accompanies a negative $V_{S,min}$ or $V_{S,min}$ is nucleophilic, and hence may be capable of donating electron density to an interacting electrophile in its vicinity. However, it should be kept in mind that all negative or positive sites on the surface of the molecular entity may or may not always be capable of engaging in attractive interaction with a region that features the opposite reactive profile. The usefulness of the MESP model to understand the surface reactivity has been demonstrated on many occasions (for example [27,38,78,91]).

The theoretical details of the promolecular charge density-based IGM model called IGM-$\delta g$ has been discussed elsewhere [74,75] and its usefulness in understanding intra- and inter-molecular interactions in chemical systems has been demonstrated many times (for example [7-10,70]). Since this model can separate intramolecular and inter-fragment interactions in a molecular entity, one can plot this in two (spikes) or three dimensions

(isosurface volumes) to reveal the presence and nature of intramolecular or intramolecular interactions between bonded atomic basins in molecular entities. From the shape of the isosurface volume, we can infer the localized or delocalized nature of the interactions involved between interacting domains, and the charge density responsible for the shape of the volume is a measure of strength of the interaction. The colors of these volumes, blue and green, generally represent strong and weak attractions, respectively, and red represents a repulsive interaction.

As a special case, in order to clarify whether the Bi–I and B···O close contacts in host-guest complexes such as that in [BiI$_3$][15-crown-5], and [BiI$_3$][Benzo-15-crown-5], are coordinate or pnictogen bonds, we have energy minimized the geometry of the two complexes in the gas phase at the [$\omega$B97XD/def2-TZVPPD] level of theory. Quantum Theory of Atoms in Molecules (QTAIM) [92] calculations were performed to explore the nature of the bond path and bond critical point topologies of the charge density. Examination of this, together with that of the other three descriptors of theory, the Laplacian of the charge density, the potential energy density, and the total energy density, has enabled us to suggest that Bi–I are typical coordinate bonds and the Bi···O close contacts are pnictogen bonds.

Analysis and drawing of geometries of various molecular entities and crystals were performed using the Mercury 4.0 [93] and VMD [94] suite of programs. AIMAll [95] and MultiWfn [96] codes were used for calculation and analysis of MESP and QTAIM graphs, and VMD [94] was used for drawing of IGM-$\delta g$ based isosurfaces.

## 5. Bismuth in Crystallography: Materials Design and Discovery

Bismuth compounds have played a very significant role in the development of many functional materials [97,98], including photocatalysts and photovoltaics. Our constrained search (*R*-factor ≤ 0.1) of the R–Bi geometrical motif in the CSD resulted in 4924 (5026) hits that feature Bi in its variable oxidation states; the parenthesis value was obtained from an unconstrained CSD search; R represents any atom of the periodic table. Fig. 1 illustrates the frequency of appearance of such crystals in terms of the number of publications per year. It shows a systematic increase in the number of publications with respect to time, with very significant growth in the last five years.

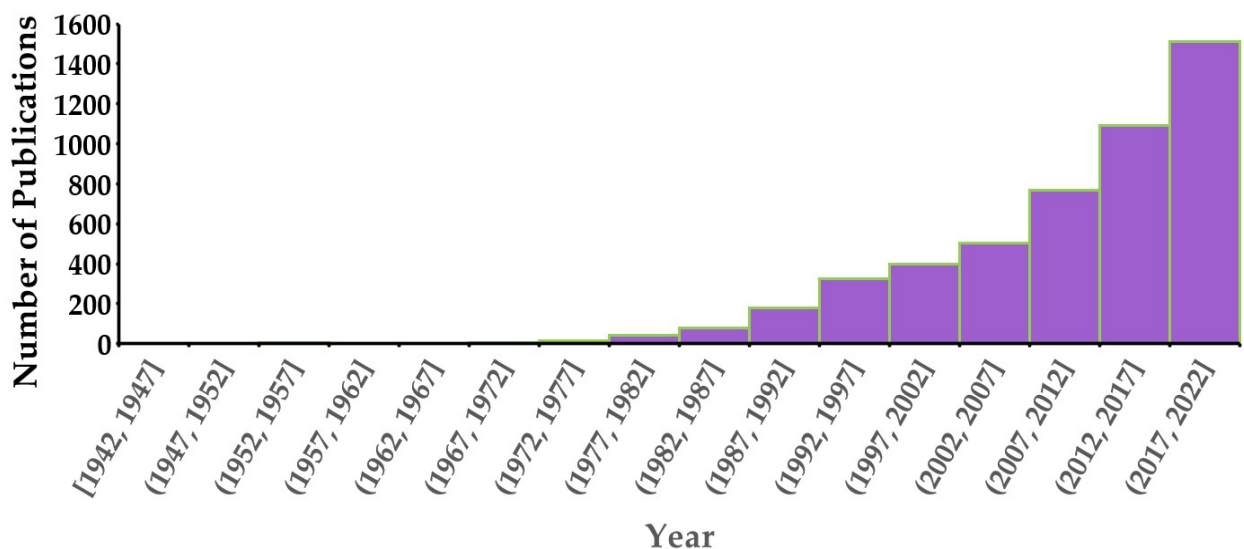

**Figure 1.** A histogram showing a systematic growth in the number of publications of crystalline solids containing the element Bi. CSD (version 5.43 updates (July 2022)) search gave 4924 single crystals that contain Bi.

Although the solid state structure of many bismuth compounds have been reported, the basic chemical reactivity of Bi when forming complexes in the solid state is yet to be fully understood [99]. In many of these compounds, there is clear evidence of the ability of Bi to engage in non-covalent interactions when in suitable proximity to a nucleophilic site.

Illustrated in Fig. 2 are a set of crystals containing covalently or coordinately bonded Bi; in some Bi is involved in forming bismuth bonds, but not in others, thus featuring the variability in the nature of coordination/interaction modes of Bi in molecular entities. In the first two geometries shown in Figs. 2a and 2b, Bi in one electrically neutral molecular entity (viz. as in $C_{22}H_{21}BiBrN$) interacts favorably with an equivalent Bi in an identical neighboring entity to form a Bi-centered pnictogen bond [100], including the involvement of a Bi⋯π(arene) interaction. In these systems, bismuth bonds occur between the coordinately bonded Bi atoms; they are long-ranged, and quasi-linear. The quasi-linear nature of the Bi⋯Bi interaction may arise from repulsion between the Bi atoms causing a mutual shift to positions where they can maximize their mutual attraction. This suggests that the positive region on the electrostatic surface of Bi in one molecular entity is in an attractive engagement with the positive region on Bi in another similar molecule that has a different charge density. The Bi⋯Bi interaction is therefore a Type-III interaction (Scheme 1c).

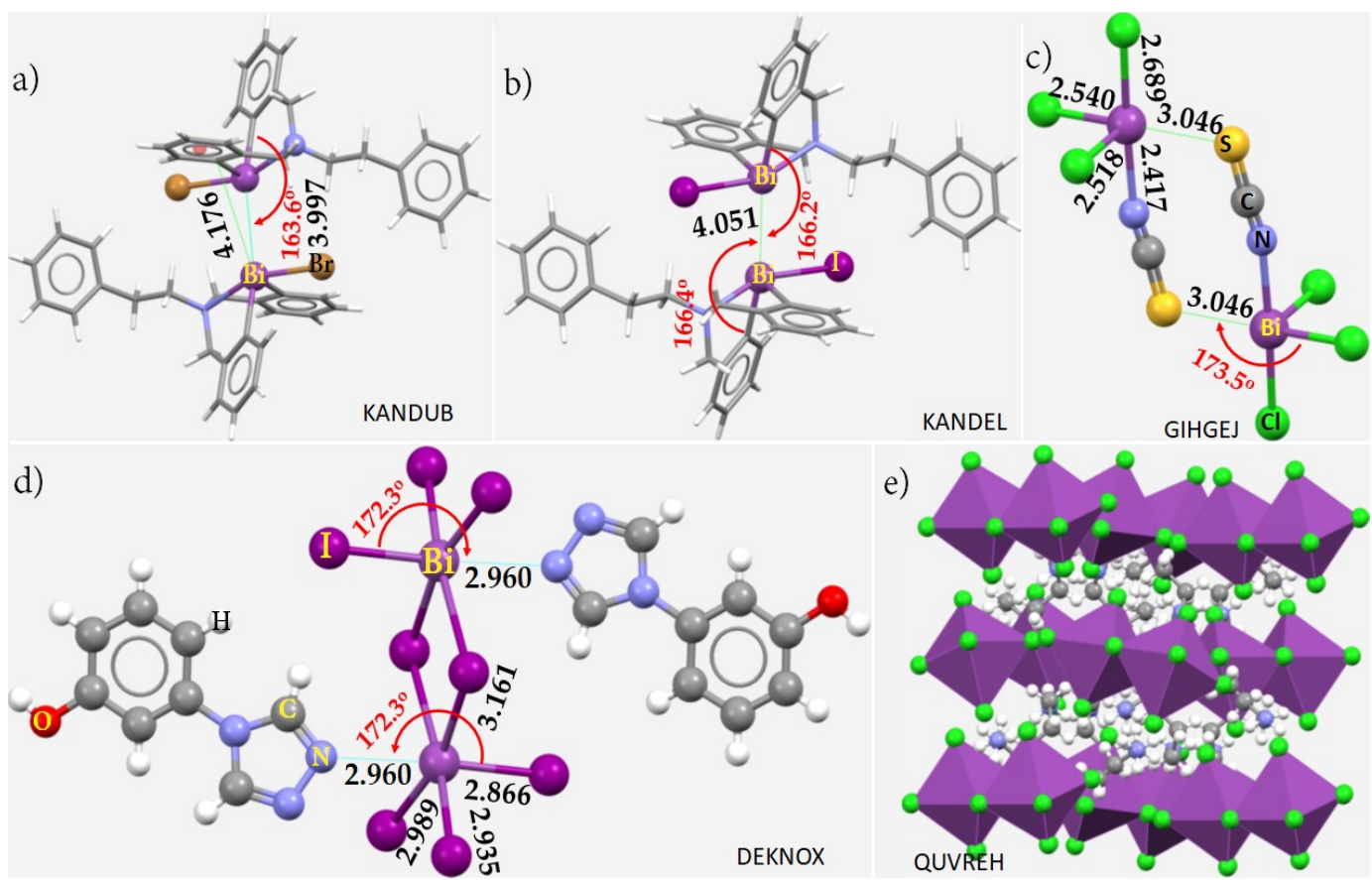

**Figure 2.** Illustration of Bi-centered non-covalent interactions in some randomly selected chemical systems. a) (2,2'-[[(2-phenylethyl)azanediyl]bis(methylene)]di(benzen-1-yl))-bromo-bismuth(III) ($C_{22}H_{21}BiBrN$) [100]; b) (2,2'-[[(2-phenylethyl)azanediyl]bis(methylene)]di(benzen-1-yl))-iodo-bismuth(III) ($C_{22}H_{21}BiIN$) [100]; c) (18-crown-6)-potassium trichloro-isothiocyanato-bismuth(ii) ($C_{12}H_{24}KO_6^+,CBiCl_3NS^-$) [101],; d) bis((*m*-phenol)-1,2,4-triazolium) bis(*m*-phenol)-1,2,4-triazole bis($\mu$-iodo)-hexaiodo-di-bismuth ($2(C_8H_8N_3O^+),Bi_2I_8^{2-},2(C_8H_7N_3O)$) [102]; e) bis(propylammonium) bis($\mu$-chloro)-octachloro-bismuth(iii) ($4(C_3H_{10}N^+),Bi_2Cl_{10}^{4-}$) [103]. Selected bond distances and bond angles are in Å and degrees, respectively. CSD ref. is shown for each case in upper-case letters. 18-crown-6-potassium in c) is removed for clarity. Selected atom types are marked.

In Fig. 2c, which illustrates part of the crystal of $[C_{12}H_{24}KO_6]^+[CBiCl_3NS]^-$ [101], the Bi···S close contact is not a Type-IIa pnictogen bond since the charges on Bi and S are both negative, even though the directional feature is satisfied. The intermolecular distance associated with this interaction ($r$(Bi···S) = 3.046 Å) is significantly longer than the formal Bi–N and Bi–Cl coordinate bonds ($r$(Bi–Cl) = 2.540 Å; $r$(Bi–Cl) = 2.689 Å; $r$(Bi–Cl) = 2.518 Å; $r$(Bi–N) = 2.417 Å) that occur in the $[CBiCl_3NS]^-$ anion. Because the Bi···S close contact appears along the Cl–Bi bond extension, and is substantially longer than the other four coordinate bonds, we recognize this $Bi^{\delta-}$···$S^{\delta-}$ close contact to be a Type-III pnictogen bond. By the same logic, the $Bi^{\delta-}$···$S^{\delta-}$ close contact in Fig. 2c is also a Type-III interaction. In case of the organic-inorganic hybrid structure, $(2(C_8H_8N_3O^+),Bi_2I_8^{2-},2(C_8H_7N_3O))$, shown in Fig. 2d [102], Bi is hexacoordinate and one of the Bi(III) ions is involved in a Bi···N close contact; while ∠I–Bi···N is quasi-linear, it may be a nitrogen-centered Type-IIb N···Bi pnictogen bond since ∠C–N···Bi is non-linear (∠C–N···Bi = 127.6°), and since Bi in $Bi_2I_8^{2-}$ acts as an electron density donor while N in the interacting organic cation is positive. Although the Bi–I bond distances in Fig. 2d are comparable to the Bi···N bond distance, the latter is clearly a nitrogen-centered pnictogen bond, whereas the former is a coordinate bond. This is understandable given that the vdW radius of I is substantially larger than that of N ($r_{vdW}$(N) = 1.66 Å; $r_{vdW}$(I) = 2.04 Å).

The bonding topology shown in Figs. 2a-d is not the same as in found in the crystal of $4(C_3H_{10}N^+)\cdot Bi_2Cl_{10}^{4-}$, Fig. 2e [103], where each Bi is coordinated by six chloride ions, forming a corner-shared octahedron, leading to the formation of a 2D inorganic layered framework; the organic cation connects the inorganic layers via a network of H···I hydrogen bonds and other non-covalent interactions, giving stability to the overall geometry of the 2D system. Given the coordination geometry of Bi(III), we found no evidence for a bismuth bond in this crystal. It is worth noting that the polynuclear $[Bi_2I_8]^{2-}$ anion in the organic-inorganic hybrid system in Fig. 2d consists of two BiI₅ square pyramids as inorganic layers. The compound is a semiconducting material since the experimentally determined optical absorption spectra features a sharp optical bandgap of 2.07 eV. Comparable properties were observed for similar compounds, viz. $[HL_1]_4[Bi_6I_{22}]\cdot[L_1]_4\cdot 4H_2O$ ($L_1$=3-(1,2,4-triazole-4-yl)-1H-1,2,4-triazole); $[HL_2]_4[Bi_6I_{22}]\cdot 6H_2O$ ($L_2$=($m$-phenol)-1,2,4-triazole). These comprise polynuclear $[Bi_6I_{22}]^{4-}$ anions to build up the inorganic layers and substituted 1,2,4-triazoles as the organic layers. There exist several hydrogen bonding and I···I halogen-halogen bonded interactions in all these three structures, and that they feature optical gaps of 1.77, 1.77, and 2.07 eV, respectively. These demonstrate that a basic understanding of pnictogen bonds in chemical systems is necessarily required since their presence enables compounds to behave as semiconductors; they therefore should be taken into account in the *de novo* design of functional materials.

A search of the ICSD and CSD databases produced thousands of compounds containing Bi, in which, $Bi^{3+}$ has a flexible coordination number that is usually anywhere between 3 and 10 [104]. Coordination to π systems is not unusual. For instance, the compound $[BiCl_3(1,2,3-Me_3C_6H_3)]$ contains quasi-dimeric units of arene-coordinated $BiCl_3$ fragments that are further associated *via* additional Bi–Cl contacts to form coordination-polymeric layers. In this system, $Bi^{3+}$ has three primary coordinate bonds with $Cl^-$, three secondary contacts, and is associated with an arene moiety. The Bi-arene bonding in this crystal is characterized by Bi–C distances in the range 3.168 (7)-3.751 (8) Å [105], suggesting $\eta^6$ coordination of the $Bi^{3+}$ ion with the arene. Another example is its π bonding to a pyrrole ring in the compound 5,10,15,20-tetraphenylporphyrindium dibromo-trichloro-bismuth (CSD refs. EPAQOE and EPARUL, Fig. 3a) [106]). An example of 4, 6 and 10 coordinated $Bi^{3+}$ is in the crystal of the 2-amino-4-methylpyridinium tetrachloro-bismuth (Fig. 3b), octakis(pyrrolidin-1-ium) hexachloro-bismuth (μ-chloro)-decachloro-di-bismuth dihydrate (CSD ref. AKEMUB; Fig. 3c) and polyoxopalladate, $Na_7[Bi(III)Pd_{15}O_{40}(PPh)_{10}]\cdot 39H_2O$ (CSD ref. NADDAB) [107]) respectively.

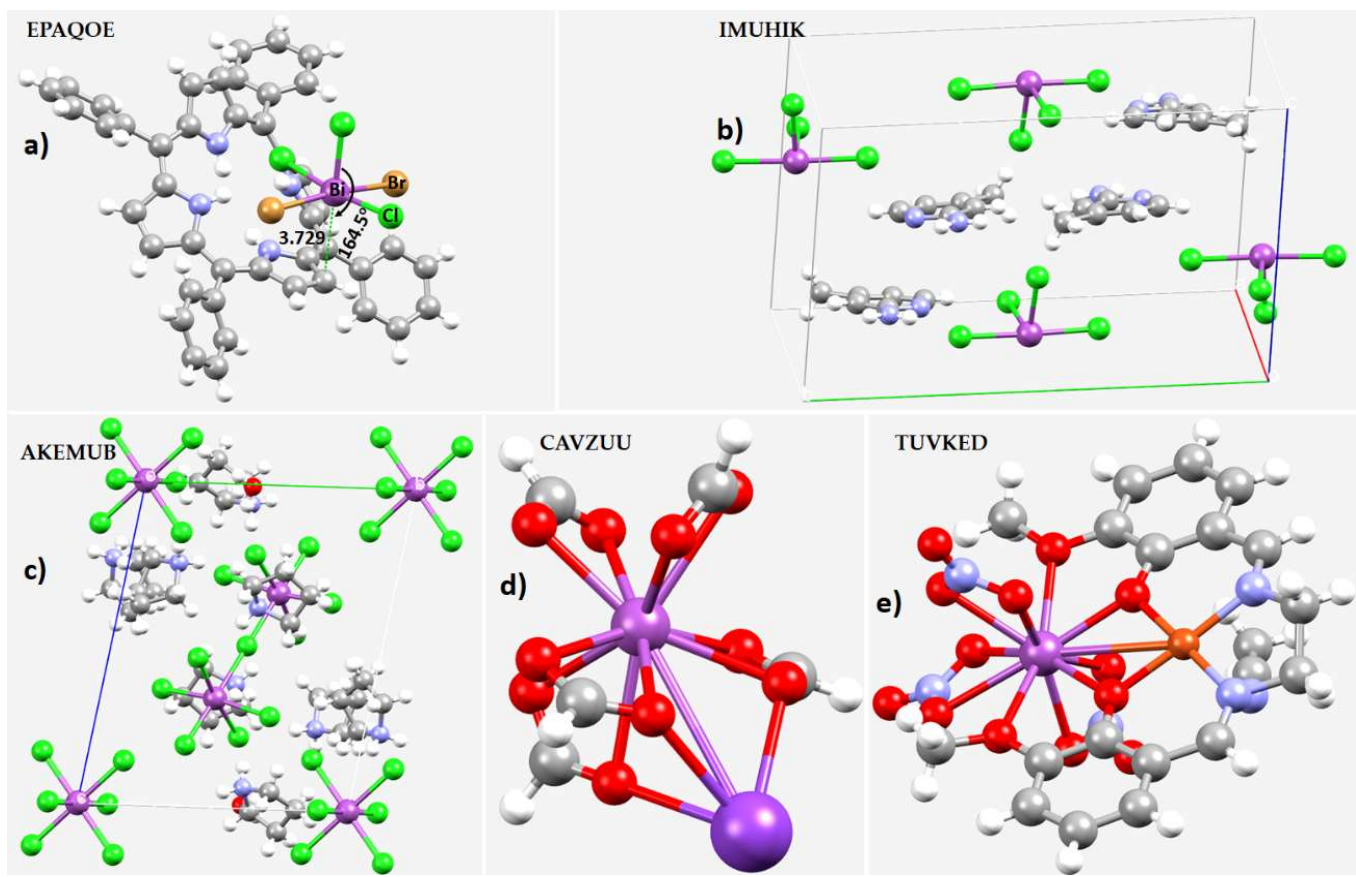

**Figure 3.** Examples illustrating the diverse coordination modes of Bi in crystals: a) 5,10,15,20-tetraphenylporphyrindium dibromo-trichloro-bismuth (CSD refs: EPAQOE and EPARUL [106]); b) 2-amino-4-methylpyridinium tetrachloro-bismuth (CSD ref: IMUHIK [108]); c) octakis(pyrrolidin-1-ium) hexachloro-bismuth (μ-chloro)-decachloro-di-bismuth dihydrate (CSD ref. AKEMUB [97]); d) dipotassium (pentaformato-O,O')-bismuth (CSD ref: CAVZUU [109]); e) (μ-2,2'-[ethane-1,2-diylbis(azanylylidenemethanylylidene)]bis(6-methoxyphenolato))-tris(nitrato)-copper(II)-bismuth(III) acetonitrile solvate (CSD ref: TUVKED [110]). The Bi…π(C=C) bond length and ∠Cl–Bi…C bond angles are indicated only for the structure shown in a) to clarify that there is probably tetrel bonding between Bi and π(C-C)-moieties, and there is no pnictogen bond in this system since Bi is entirely negative in [BiBr$_2$Cl$_3$]$^{2-}$.

Higher coordination numbers of Bi are certainly known; crystals such as the potassium salt of the pentaformate complex of Bi$^{3+}$ (Fig. 3d) [109] and a bimetallic Cu$^{2+}$, Bi$^{3+}$ salen complex, possibly featuring bonding between the two metal ions (Fig. 3e) [110], are examples. Three types of Bi–M bonding are known in transition metal complexes displaying a bismuth species in the coordination sphere of the transition metal M: dative Bi→M interactions (with Bi acting as a donor), dative Bi←M interactions (with Bi acting as an acceptor), and covalent Bi–M interactions [111]. The nature of Bi–M bonding, trends in the geometric parameters, and in the coordination chemistry of the Bi-containing compounds have been reviewed, focusing on the reactivity of bismuth species in the coordination sphere of transition metal complexes in stoichiometric and catalytic reactions [111].

Our study does not consider the transition metal coordination chemistry of Bi, per se. We have just highlighted the coordination ability of Bi in some chemical systems in Fig. 3 to demonstrate below that a coordinately bound Bi site in molecular entities has an exceptional potential to engage with a versatile number of electron density donors to form pnictogen bonds.

## 6. Statistical analysis of bismuth bonds in crystal lattices

We considered the electronegative elements D = O, N, F, Cl, Br, I, S, Se, and Te, as well as $C_6(\pi)$ in arene moieties, as potential electron density donor sites in molecular entities for covalently bonded Bi to gauge the extent of occurrence of Bi⋯D close contacts in the crystals deposited in the CSD. Our searches involved intermolecular interactions in crystals, comprising the geometric motifs such as R–Bi⋯D, where R is the remainder part of the molecular entity. The geometric data (the bond distance, $r$(Bi⋯D), and the bond angle, ∠R–Bi⋯D) obtained from the CSD searches were statistically analyzed to infer the range of intermolecular distances and directional features for a variety of donor sites for a bismuth bond. Our search was limited to intermolecular distances between 2.6 and 4.5 Å for most cases, and bond angles between 140° and 180°. Only single crystals were selected that were free of errors and distortions, and that had an $R$-factor ≤ 0.1. The geometric fragments R–Bi⋯D–R' and R–Bi⋯D were chosen for the searches (R' = any element; R, D = selected elements of Groups 15, 16, and 17 and include carbon in aromatic rings; the bond between D and R' was of any type). The upper limit of the intermolecular distance was chosen depending on the sum of the vdW radii of Bi and D, with a flexibility of ±0.2 Å.

The charge on Bi in the crystals resulted from our searches was either (formally) positive, neutral, or negative. For instance, the Bi⋯O interactions in the crystal, $(C_8H_{20}N^+)_2$ $(C_{13}Bi_4Fe_4O_{13})^{2-}$ [112], was found to be existing between negative potentials on the surfaces of Bi and O atoms since both are responsible for the $(C_{13}Bi_4Fe_4O_{13})^{2-}$ anion and is probably unavoidable during the search of the motif in crystals in CSD. If one imposes charge as a search criterion, a large number of real interactions will be missed.

Many crystals in the search of R–Bi⋯O resulted in close contacts that were formed between Bi and O in the building blocks that were entirely negative, and were driven by a counter ion of the crystal concerned. One such crystal, for example, is tetramethylammonium (μ2-carbonyl)-decacarbonyl-di-bismuth-tetra-cobalt $(C_4H_{12}N^+,C_{11}Bi_2Co_4O_{11}^-)$ (CSD ref. FOGGEN) [113], in which, Bi⋯O distances vary between 3.5 and 3.8 Å, and ∠Co/Bi–Bi⋯D are in the range 145-172°. Similarly, in the tetrakis(tetraphenyl-bismuth) bis(μ3-iodo)-tetrakis(μ2-iodo)-decaiodo-tetra-bismuth acetone solvate crystal, $[4(C_{24}H_{20}Bi^+),Bi_4I_{16}^{4-},2(C_3H_6O)]$ (CSD ref: HUJCIZ [114]), the Bi⋯O intermolecular interaction is formed between negative Bi in $Bi_4I_{16}^{4-}$, and negative O in acetone $(C_3H_6O)$ ($r$(Bi⋯O) = 3.094 Å and ∠C–Bi⋯O = 177.1°), this is probably an example of a Type-III bonding feature. Nevertheless, our statistical analysis was largely limited to the category of Type-II topology of bismuth bonds in crystals, and we have eliminated crystals with close contacts where Bi could be intrinsically negative (as in an anion).

*A. Bi⋯O close contacts, with oxygen as donor*

Our CSD search for the R–Bi⋯O geometric motif with $r$(Bi⋯O) in the range 2.6 – 3.8 Å and $A$ (∠R–Bi⋯O) in the range 140 – 180.0° resulted in 376 single crystals with 871 close-contacts. A close inspection of these crystals revealed that Bi in 39 crystals contain 81 close contacts that occurred between negative sites and are directional. We removed them from the list since they cannot be Type-II bismuth bonds. The normal distribution of the remaining close contacts in 337 crystals is shown in Fig. 4. The results suggest that the great majority of Bi⋯O close contacts follow a Type-IIb topology of bonding rather than Type-IIa (Fig. 4a). The non-linearity in the Bi⋯O close contacts is because the fragment R' covalently attached with the donor atom O is prone to form other interactions (such as hydrogen bonds) with its nearest neighbor; the donor atom therefore shifts to a position to maximize its non-covalent interaction with Bi, resulting in a non-linear close contact. In any case, the peaks of the normal distributions occur around 155° and 3.2 Å in bond angle and bond distance, respectively. When a non-constrained search was performed with $r$(Bi⋯O) in the range 2.6 – 4.1 Å that has an upper limit of the intermolecular distance slightly longer than the sum of the vdW radii of Bi and O, 4.04 Å, and $A$ in the range 140 – 180.0°, the number of hits increased to 463, with 1184 close contacts; these are not shown in Fig. 4. The peaks of the normal distribution occurred at 153.5° and 3.4 Å for bond angle and bond distance, respectively. The largest population of these features occurred in the range

142 – 153° and 2.9 – 3.3 Å, respectively; this is very similar to the constrained results shown in Fig. 4.

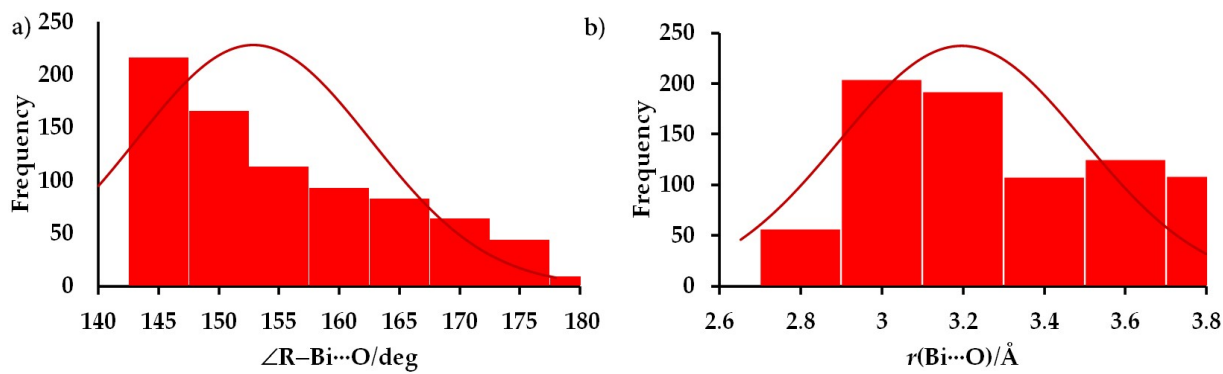

**Figure 4.** Histograms showing a) the angular distribution and b) intermolecular distance of 788 Bi⋯O close contacts in 337 crystals that emerged from a CSD search. The Bi⋯O intermolecular distance $r$ (and intermolecular angle ∠R–Bi⋯O) in the range of 2.6–3.8 Å (140–180°) was used as a geometric criterion during the CSD search, where R represents any atom. Bond lengths and angles are in Å and degrees, respectively. The normal distribution curve is shown in dark-red in each plot.

*B. Bi⋯S (Bi⋯Se, and Bi⋯Te) close contacts, with heavy chalcogen derivatives as donors*

Our search of the CSD for the R–Bi⋯S geometric motif with $r$(Bi⋯S) in the range 2.6 – 4.3 Å and ∠R–Bi⋯S in the range 140 – 180.0° resulted in 119 single crystals with 188 close-contacts. When Se and Te were included as donors, the search resulted in 128 crystals that contain 204 close contacts, showing that the frequency of Se and Te as electron donors is quite low in forming bismuth bonds. Some of the Bi⋯Se and Bi⋯Te close contacts found in seven of these 128 crystals were not bismuth bonds since they appear between entirely negative Bi and Se (or Te) sites. For instance, the Bi⋯Se/Bi⋯Ch close contact in [$C_8H_{20}N^+,C_9BiFe_3O_9Se^-$] (CSD ref: COFGAI) [115]/[$C_8H_{20}N^+,C_6BiFe_2O_6Ch^-$] (Ch = Se, Te) (CSD ref. COFGEM and COFGIQ [115]) occurs between two identical anions ($C_9BiFe_3O_9Se^-$ or $C_6BiFe_2O_6Se^-$) and is driven by the tetraethylammonium ($C_8H_{20}N^+$) cation. This is indeed different from the Type-IIa bonding topology involving a positive site on Bi and negative S as in the crystal of diphenyl-phenylselenyl-bismuthine, $Ph_2NiSePh$ (CSD ref: GIPREC [116]); $r$(Bi⋯Se) = 3.897 Å and ∠Se–Bi⋯Se = 176.8°).

In some crystal structures several close contacts occur between Bi in one unit and S in a thiacetate. We regard them as false contacts; in reality, these are Bi⋯O close contacts [as in a dioxodibenzothiabismuth-phenyl acetate (CSD ref: IGESAR [117]); [(2,6-($CH_2NH_2$)Me)Ph)Bi(Me)$^+$][$CF_3SO_3^-$] (CSD ref: WUXHUV [118]); [$SO_2Ph_2Bi(OPh$-*p*--OMe)][*p*-OMe-phenol] (CSD ref. XADQUO [119]); and [2-(*N*-phenylcarbonylamino)phenyl-Bi(py)$_2^+$][$CFSO_3^-$] (CSD ref: YODWAS [120])]. Based on these observations, 32 of the 128 crystals were rejected as possessing false contacts. The geometric motifs associated with the Bi⋯Ch (Ch = S, Se, Te) close contacts in the remaining 96 crystals are plotted in Fig. 5, based on 156 close contacts. As can be seen, no close contacts were found below 3.0 Å. The normal distribution shows that the largest occurrence of $r$(Bi⋯S) and ∠R–Bi⋯S and is in range 3.4 – 4.0 Å and 150 – 170° and, respectively, with the peak of the normal distributions of corresponding geometries at 3.6 Å and 157°, respectively.

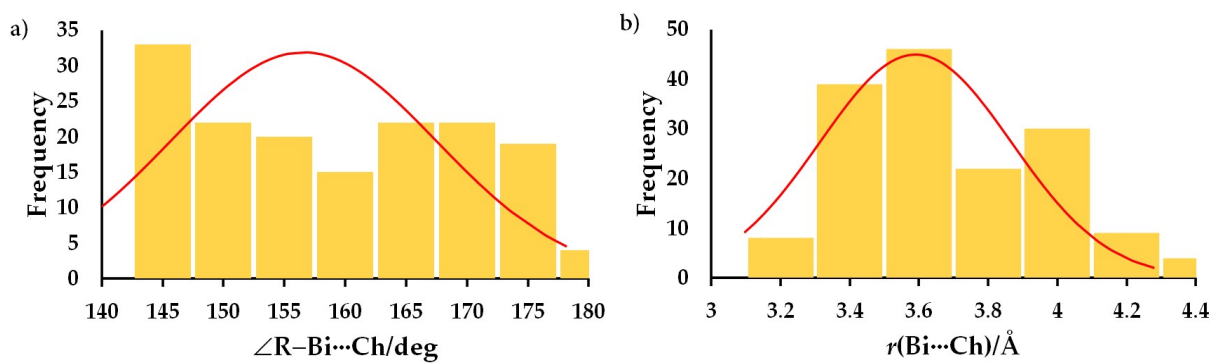

**Figure 5.** a) and b) Histograms showing, respectively, the angular and intermolecular distance distributions of 157 Bi···Ch (Ch = S, Se, Te) close contacts in 96 crystals that emerged from a CSD search. The Bi···Ch intermolecular distance $r$ (and intermolecular angle ∠R–Bi···Ch) in the range of 2.6–4.3 Å (140–180°) was used as a geometric criterion during the CSD search, where R represents any atom. Bond lengths and angles are shown in Å and degrees, respectively. The normal distribution Bell curve is shown in dark-red in each plot.

*C. Bi···N close contacts, with pnictogen nitrogen as donor*

A search for Bi···N close contacts in the CSD gave 64 crystals. The occurrence of close contacts was examined based on the search criteria set at $r$(Bi···N) = 2.6 – 3.8 Å and ∠R–Bi···N = 140 – 180.0°. To give an example, our search gave a Bi···N close contact between the negative Bi site in pentacyano-bismuth and the negative N in CH$_3$CN in the crystal of (N(PPh$_3$)$_2$$^+$)$_2$(Bi(CN)$_5$$^{2-}$)·CH$_3$CN (CSD ref: UBIQUU [121]), in which $r$(Bi···N) = 3.136 Å and ∠C–Bi···N = 157.5°. Similarly, we found Bi···N close contacts between Bi in Bi$_2$I$_8$$^{2-}$ and N in the $\mu$-phenol-1,2,4-triazolium cation (C$_8$H$_8$N$_3$O$^+$), in the crystal structure of [C$_8$H$_8$N$_3$O$^+$)$_2$][Bi$_2$I$_8$$^{2-}$]·2(C$_8$H$_7$N$_3$O) (CSD ref. DEKNOX [102]), in which, $r$(Bi···N) = 2.960 Å and ∠I–Bi···N = 172.3°); this is a charge-assisted N-centered pnictogen bond. We rejected this and other such close contacts from the list of close contacts in 64 single crystals. In addition, we also rejected several Bi···N false contacts found between N in nitrate and Bi in the partner cation that showed up because of the angular flexibility; in fact it is O of the nitrate anions that is linked non-covalently with Bi, as in, for example, [Bi(Ph)$_4$(OH$_2$)$^+$][Bi(Ph)$_4$$^+$][Bi(Ph)$_4$(ONO$_2$)][NO$_3$$^-$]$_2$ (CSD ref: VUZFEE [122]). Consequently, 24 crystals with 45 false contacts were not included in the histograms shown in Fig. 6. The population of ∠R–Bi···N is predominantly the range 150 – 175°, and there are only a few crystals that feature ∠R–Bi···N between 175 and 180° (Fig. 6a). Similarly, the histogram plot in Fig. 6b indicates that Bi···N close contacts span the range between 2.8 and 3.8 Å, with most contacts between 3.1 and 3.7 Å.

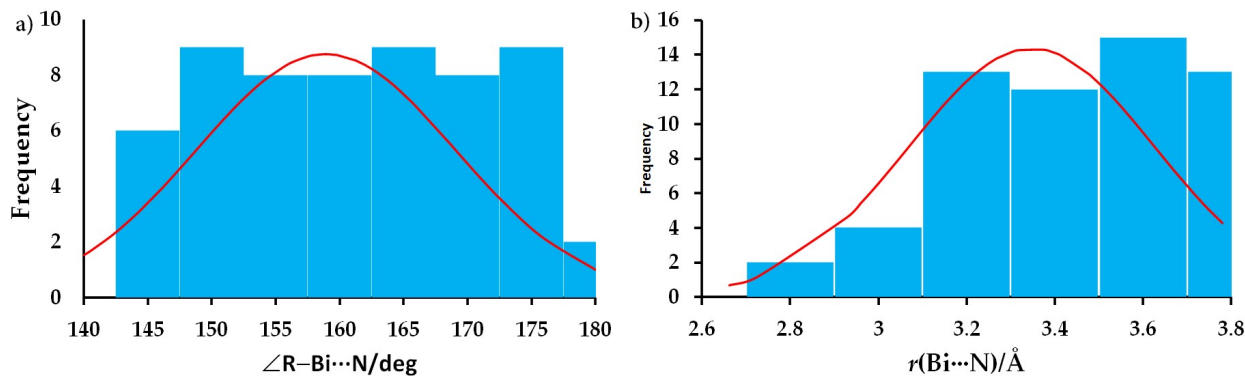

**Figure 6.** a) and b) Histograms showing, respectively, the angular and intermolecular distance distributions of 59 Bi···N close contacts in 40 crystals that emerged

from a CSD search. The Bi⋯N intermolecular distance $r$ (and intermolecular angle ∠R–Bi⋯N) in the range of 2.6–3.8 Å (140–180°) was used as a geometric criterion during the CSD search, where R represents any atom. Bond lengths and angles are shown in Å and degrees, respectively. The normal distribution curve is shown in dark-red in each plot.

The searches discussed above involved intermolecular distance less than the sum of the vdW radii of Bi and N, 4.20 Å. We performed a similar search with $r$(Bi⋯N) = 2.6 – 4.3 Å and ∠R–Bi⋯N = 140 – 180.0°. This gave 118 hits, with 214 close contacts (including false contacts). When the search did not involve any constraint, the number of hits and close contacts became 178 and 322, respectively. In these two cases, the peaks of the normal distributions, including false contacts, appeared around 3.73 Å and 155.5° for $r$(Bi⋯N) and ∠R–Bi⋯N, respectively.

*D. Bi⋯X (X = F, Cl, Br, I) close contacts, with halogen derivatives as donors*

Searches of the CSD for Bi⋯X (X = F, Cl, Br, I) were conducted with intermolecular distance $r$ limited to ranges between 2.7 Å and an upper limit of (depending on the identity of X), 3.8, 4.3, 4.5 and 4.6 Å, respectively, and in all cases with the intermolecular angle ∠R–Bi⋯X limited to 140–180°. No close contacts were found below 2.7, 2.9, 3.2 and 3.4 Å for X = F, Cl, Br and I, respectively, which is unsurprising given the vdW radii of these elements increase with the increasing size of the halogen derivative (vdW radii values 1.46, 1.82, 1.86 and 2.04 Å, respectively [72]). The results are shown in Fig. 7. In all the four cases, the Bi⋯X intermolecular distances were found to range from either smaller to markedly greater than the sum of the vdW radii of Bi and X, emphasizing that a strict adherence to the use of the less than the sum of the vdW criterion to identify non-covalent bonds would miss many genuine Bi⋯X close contacts in crystals, "a pitfall in the search for bonding" [69].

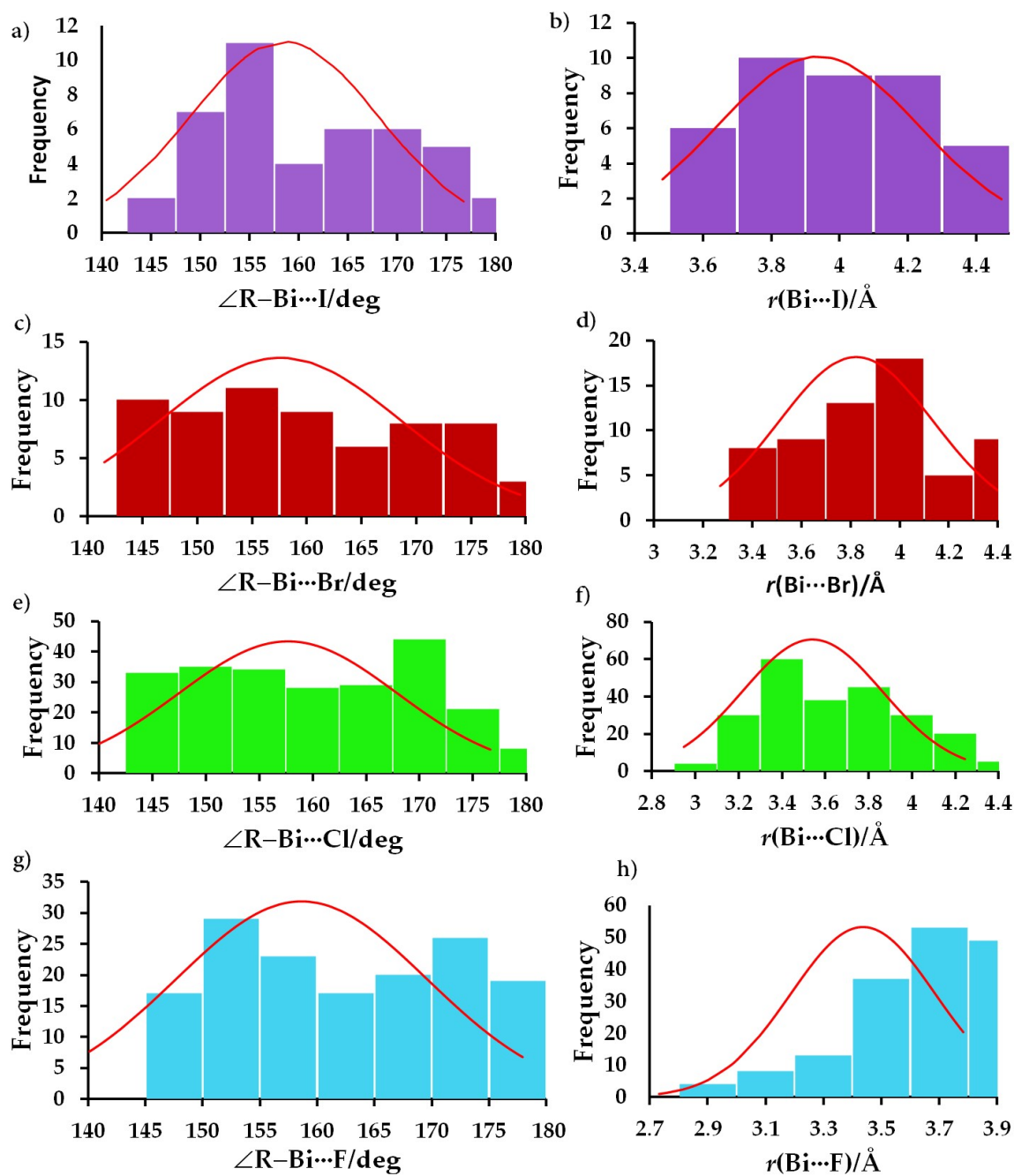

**Figure 7.** a) and b) Histograms showing, respectively, the angular and intermolecular distance distributions of 43 Bi⋯I close contacts in 30 crystals that emerged from a CSD search. c)-d), e)-f) and g)-h) are the corresponding plots for 64 Bi⋯Br close contacts in 41 crystals, 236 Bi⋯Cl close contacts in 136 crystals, and 168 Bi⋯Br close contacts in 89 crystals, respectively. The Bi⋯X (X = F, Cl, Br, I) intermolecular distance $r$ (and intermolecular angle ∠R–Bi⋯X) in the ranges of 3.0–4.5 Å (140–180°), 2.8–4.5 Å (140–180°), 2.6–4.3 Å (140–180°), and 2.6–3.8 Å (140–180°) was used as a geometric criterion during the CSD search for Bi⋯I, Bi⋯Br, Bi⋯Cl and Bi⋯F, respectively, where R represents any atom. Bond lengths and angles are shown in Å and degrees, respectively. The normal distribution curve is shown in dark-red in each plot.

From the histograms and normal distribution curves shown in Fig. 7a-h, it is clear that Bi⋯I, Bi⋯Br, Bi⋯Cl and Bi⋯F close contact distances occur approximately in the ranges 3.7 – 4.2, 3.6 – 4.1, 3.3 – 3.9 and 3.4 – 3.9 Å, respectively; each range is less than sum of the vdW radii of the respective atomic basins ($r_{vdW}$(Bi) + $r_{vdW}$(I) = 4.58 Å; $r_{vdW}$(Bi) + $r_{vdW}$(Br)

= 4.40 Å; $r_{vdW}$(Bi) + $r_{vdW}$(Cl) = 4.36 Å; $r_{vdW}$(Bi) + $r_{vdW}$(F) = 4.0 Å). The upper limit of $r$ used for each search was very close to the sum of the vdW radii sum of the appropriate atoms. We are cognizant that the vdW radii of atoms of the elements proposed separately by Batsanov [123] and Alvarez [72] are different, with $r_{vdW}$(Bi) = 2.3 and 2.54 Å, respectively. Note that an increase in the upper limit of the Bi⋯I contact distance from 4.5 Å to 4.8 (and 4.7) Å in the range 3.0–4.5 Å used in in the histogram in Fig. 7a, which is slightly longer than ($r_{vdW}$(Bi) + $r_{vdW}$(I) = 4.58 Å, did not give any additional crystals in our search of the CSD in the first case, and only increased the total number of Bi⋯I close contacts from 43 to 46. An un-constrained search with default $R$-factor gave 40 hits with 62 Bi⋯I close contacts. In the case of our search for Bi⋯Br with the change of criteria from ($r$(Bi⋯Br) [∠R–Bi⋯Br)] = 2.8–4.5 Å [140–180°]) to ($r$(Bi⋯Br) [∠R–Bi⋯Br)] = 2.8–4.7 Å [140–180°]), we found 64 Bi⋯Br close contacts in 41 crystals increased to 69 close contacts in 43 crystals.

It is clear from the plots in Fig. 7c and 7d that the number of Bi⋯Cl and Bi⋯F close contacts is greater than the number of Bi⋯I and Bi⋯Br close contacts in crystals. The peak of ∠R–Bi⋯X occurs between 150° and 165°, and there are a relatively very small number of such interactions that are linear. This is likely to be an effect of the involvement of other primary and/or secondary interactions associated with the electron density donors responsible for the formation of the bismuth bonds, as well as the ligating framework of covalently bound Bi.

*E. Bi⋯π(arene) close contacts, with arene as donor*

Given there is such an abundance of known π-systems, we limited our search of the Bi⋯π motif to cases where the π-density belongs to the centroid of a $C_6$ aromatic ring. Accordingly, and in our search, the Bi⋯$C_6$(centroid) distance and intermolecular angle ∠R–Bi⋯$C_6$(centroid) were constrained to the ranges of 2.8–4.5 Å and 140–180°, respectively. The search produced 216 crystals with 315 close contacts. Of these, eight crystals had nine false contacts that were rejected. We recognized them as false contacts since the positive site on covalently bound Bi in one molecular entity forms a bismuth bond with π-density on a single C atom (but not the centroid region) of the $C_6$ aromatic ring in another same or different molecular entity. For instance, this was observed in the crystal structures of tris(5-chloro-2-methoxyphenyl)-bismuth(III) (CSD ref. RAGBIO [124]), ($\mu_2$-oxydiethan-2-yl)-tetraphenyl-di-bismuth (CSD ref: AVANOB [125]), and bis($\mu$-chloro)-di-chloro-bis(6-(diphenylphosphinoyl)-1,2-dihydroacenaphthylen-5-yl)-di-bismuth dichloromethane solvate (CSD ref. COJDIR [126]). The $r$(Bi⋯$C_\pi$($C_6$)) [∠R–Bi⋯$C_\pi$($C_6$)] in the corresponding systems were 3.619 Å [161.6°], 3.591 Å [173.8°], and 3.947 Å [160.1°], respectively. In other cases a genuine Bi⋯N contact was falsely detected as Bi⋯π($C_6$), such as in tris(quinoline-8-thiolato)-bismuth (CSD ref: FUDMAS [127]). The false contacts show up in angular range 140–160°.

The normal distribution of the remaining close contacts in 208 crystals is shown in Fig. 8. The histogram in Fig. 8a shows that the population of ∠R–Bi⋯$C_6$(centroid) is relatively low in the range 140–145°, and high in the range between 155–180°, with a peak of the bell curve occurring around 160°. Similarly, as may be readily seen from Fig. 8b, the population of the Bi⋯$C_6$(centroid) close contact distance is relatively less dense in the range 2.8–3.4 Å than in the range 3.5–4.5 Å, even though the peak of the bell curve appears at 3.86 Å.

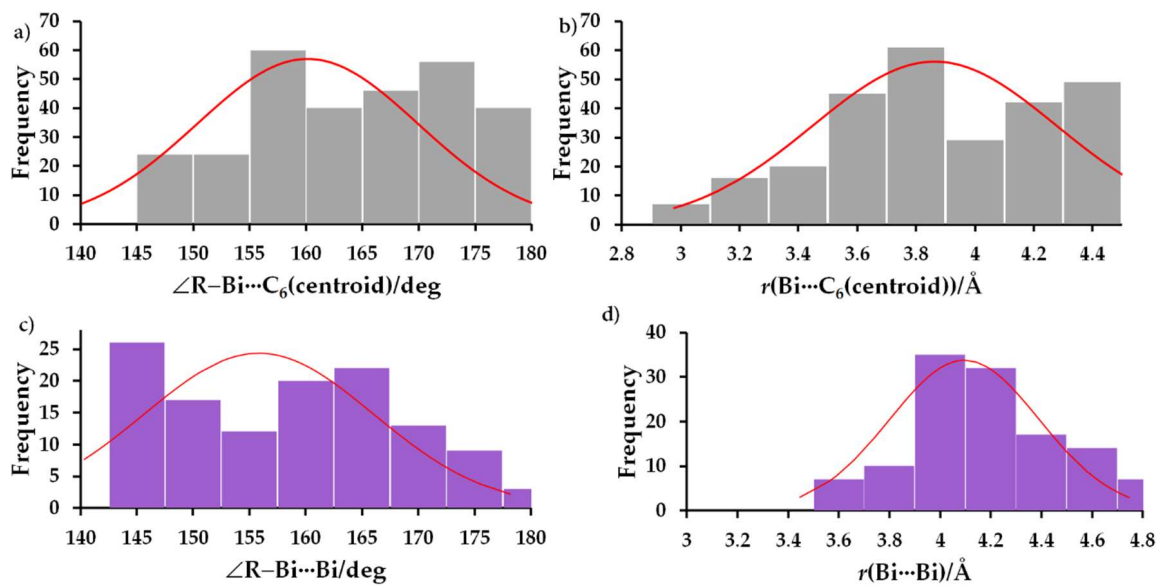

**Figure 8.** a) and b) Histograms showing, respectively, the angular and intermolecular distance distributions of 306 Bi···C$_6$(centroid) close contacts in 208 crystals originated from a CSD search. The Bi···C$_6$(centroid) intermolecular distance $r$ (and intermolecular angle ∠R–Bi··· C$_6$(centroid) in the range of 2.6–4.5 Å (140–180°) was used as a geometric criterion during the CSD search, where R is any atom. c) and d) Histograms showing, respectively, the angular distribution and intermolecular distance of 12 Bi···Bi close contacts in several crystals originated from a CSD search (see text for discussion). Bond lengths and angles are shown in Å and degrees, respectively. The normal distribution curve is shown in dark-red in each plot.

*F. Bi···Bi, and Bi···P close contacts, with pnictogen derivative as donor*

A further search of Bi···Pn (X = P, As, Sb, Bi) intermolecular distance (and intermolecular angle ∠R–Bi···Pn) was individually carried out. These geometrical constraints, limited to the range of 2.6–5.1 Å (140–180°), gave 383 crystals with 634 close contacts, where the upper limit of the distance range was very close to the sum of the vdW radius of Bi proposed by Alvarez, 5.1 Å ($r_{vdW}$(Bi) = 2.54 Å) [72]. A scrutiny of individual structures gave numerous false contacts probably because our search included both inter- and intra-molecular close contacts, and the angular flexibility was reasonably high. When the upper limit of the distance range was set equal to (or slightly larger than) the twice the vdW radius of Bi as proposed by Batsanov, 4.6 (4.8) Å ($r_{vdW}$(Bi) = 2.3 Å) [123], our search of the CSD resulted in 224 (293) hits that comprised 373 (485) close contacts. When intramolecular close contacts were not included, our search resulted in 192 (258) crystal structures with 275 (373) close contacts. This suggests that the occurrence of Bi···Bi intramolecular close contacts in crystals is not very rare. Individual analysis of 293 crystals hits that comprised 485 close contacts led us to reject 363 close contacts out of a total of 485. The remaining 122 close contacts are shown in the histogram plot in Fig. 8c-d; there is no close contact in the range between 2.6 – 3.45 Å, except a single intramolecular interaction that exists between the Bi atoms in the crystal bis(μ-naphthalene-1,8-diyl)-diphenyl-di-bismuth (C$_{32}$H$_{22}$Bi$_2$, CSD ref. KARZEM [128] ) that was reported this year; this was the shortest Type-III Bi···Bi close contact ($r$(Bi···Bi) = 3.227 Å and ∠R–Bi···Bi = 173.95°) among all close contacts identified. When the upper limit of (Bi···Bi) used in our search was set to 4.8 Å, a large number of close contacts subsequently identified to be false contacts arose. The criteria unequivocally detected Bi···Bi close contacts, even though the actual contacts in those crystals were Bi···O, Bi···S, Bi···N, Bi···C, Bi···C$_\pi$ and Bi···X types, among others. An illustrative example is the crystal structure of the thioacyl bismuth complex (4-MeC$_6$H$_4$)Bi(4-MeOC$_4$H$_4$COS)$_2$ (CSD ref: ACUPEU [129]). This shows a purported Bi···Bi close contact of 4.370 Å, with a S–Bi···Bi contact angle of 177.3°. Yet visual inspection shows two other, probably more relevant, close contacts with Bi: Bi···O at 3.013 Å to one of the acyl O atoms of a neighboring entity, and of 3.429 Å to the O atom of a methoxy group of another. Clearly the purported Bi···Bi close contact is a false contact.

Nevertheless, Fig. 8c and d suggests that the population of bond distances and bond angles responsible for peaks in the normal distribution is around 155.8° and 4.1 Å, respectively. The strong population of bond angle around 145° suggests the Bi···Bi close contacts featuring this were secondary interactions, and are a consequence of primary interactions such as Bi···O, Bi···S, Bi···N, Bi···C, Bi···C$_\pi$, or Bi···X. The majority of the remaining 122 Bi···Bi close contacts were Type-III. $r$(Bi···Pn) and ∠R–Bi···Pn were in the ranges 3.5 – 4.7 Å and 145 – 175°, respectively. An example of the occurrence of a Type-III Bi···Bi close contact between the Bi(CH$_3$)$_3$ molecular units in crystalline trimethylbismuthine (CSD ref: HUVQOG [130]) was readily apparent. In case of the crystals of Bi and Rh carbonyl anionic clusters (CSD ref: SOKVUM [131]) and Bi and Co anionic clusters (SOKWAT [131]), the Bi···Bi close contact occurs between the negative Bi sites of the interacting anions in the crystal.

A search of Bi···P inter- and intra-molecular distances (and inter- and intra-molecular angles ∠R–Bi···P) in the range of 2.6–4.5 Å (140–180°) gave 11 crystals with 23 close contacts. Of these, six crystals contained six false close contacts. These were found either between Bi in a cation and PF$_6^-$ (as in [C$_{20}$H$_{21}$BiCuN$_4$][PF$_6$] CSD ref: VEZCEL [132]; [C$_{48}$H$_{40}$BiP$_2$][PF$_6$] CSD ref: VISVUP [133]; and [C$_{20}$H$_{21}$AgBiN$_4$][PF$_6$] CSD ref: VUGVOL [134], or between Bi and P in a tert-butylphosphonato fragment (as in C$_{20}$H$_{39}$BiN$_2$O$_6$P$_2$, CSD ref. AQOLEY [135]).

The remaining 15 Bi···P close contacts were found in the crystals including tris(6-(diphenylphosphanyl)-1,2-dihydroacenaphthylen-5-yl)-bismuth(III) (CSD ref: GAJZOK [136]), (6-(di-isopropylphosphino)acenaphthylen-5-yl)-diphenyl-bismuth (CSD ref: OJAJIV [137]) , bis(6-(di-isopropylphosphino)acenaphthylen-5-yl)-phenyl-bismuth (CSD ref. OJAJOB [137]), (bismuthinetriyltris(1,2-dihydroacenaphthylene-6,5-diyl))tris(di-isopropylphosphine) dichloromethane solvate (CSD ref: OTINIQ [138]) and tris(2-(diphenylphosphanyl)benzenethiolato)-bismuth (CSD ref: QEFVOO [139]). The intramolecular distances and angles in all these chemical systems were in the ranges 3.19 – 3.37 Å, and 140.6 – 173.4°, respectively, and none of the crystals featured intermolecular Bi···P close contacts. They all appear along the extension of the C–Bi covalent coordinate bonds, however. Whether or not the electrostatic potential on the surface of coordinate P in these systems is positive is not clear, and hence requires further theoretical exploration.

Searches of Bi···Sb and Bi···As in crystals in CSD with the same geometric criteria as above gave a few hits, but with false close contacts.

## 7. Exemplary crystal systems

*a. Bismuth Trihalides and Bi···X Pnictogen Bonds*

Our interrogation of the CSD showed that bismuth trihalides BiX$_3$ (X = Cl, Br, I) are often responsible for shaping a variety of structures in the crystalline phase. The local coordinate environment in these systems is different from those shown in Figs. 3a-3c. The latter ones show how a Bi-containing counter ion that stabilizes a variety of organic systems. Although in all these systems Bi has a formal oxidation state of +3, it has (as discussed above) the ability to exhibit a variety of coordination numbers, and 4-, 5- and 6-fold coordination complexes are common. For instance, in the solid state structure of [2-amino-4-methylpyridinium]BiCl$_4$ (CSD ref. IMUHIK; Fig. 3b) [108], Bi$^{3+}$ is four-coordinate (approximately C$_{2v}$ symmetry), while [BiBr$_2$Cl$_3$]$^{2-}$ serves as the counter ion to a tetraphenylporphin dication (CSD ref: EPARUL) [106].

In order to understand the behavior of these BiX$_3$ building blocks, we carried out an MESP analysis of isolated BiX$_3$ (X = F, Cl, Br, I) molecules. These were energy minimized with MP2(full) and with the DFT-$\omega$B97XD functional (a range-separated version of Becke's 97 functional [140] with additional dispersion correction comprising 22% HF exchange at short range and 100% HF at long range), in conjunction with the def2-TZVPPD basis set. The results of the local most potential maxima and minima ($V_{s,max}$ and $V_{s,min}$, respectively) on the electrostatic surfaces of BiX$_3$ molecules are summarized in Table 1. Both methods give very similar values for $V_{s,max}$ and $V_{s,min}$, although they are predicted to be slightly larger in most cases with MP2(full).

As was previously noted for AsX$_3$ [9] and SbX$_3$ [10], we observed no potential maxima on F along the Bi–F bond extensions in BiF$_3$; F in this molecule is entirely negative (Fig. 9a). This is not unexpected since F is significantly less polarizable than Bi, even though it does have an electron-withdrawing power comparable to entities such as N, O and –C≡N in molecules.

The maxima of potential are negative on the surface of the Cl atoms along the extension of the three Bi–Cl bonds in BiCl$_3$ and equivalent ($V_{s,max}$ = –2.8 kcal mol$^{-1}$ with MP2(full) and –2.4 kcal mol$^{-1}$ with $\omega$B97XD), indicating the presence of three negative σ-holes on Cl (cf. Fig. 9b and Table 1). Both Br and I, when bonded to Bi, feature positive and negative regions on their surfaces along and around the Bi–X (X = Br, I) bond extensions, respectively (Fig. 9c and 9d, respectively). In all three cases, the axial portions are significantly more positive than the lateral sites of X (X = Cl, Br, I). The $V_{S,min}$ values around the lateral portions of X follow the trend BiF$_3$ > BiCl$_3$ > BiBr$_3$ > BiI$_3$, in agreement with the electronegativity trend of the halogen derivative.

**Table 1.** Comparison of MP2(full)/def2-TZVPPD level 0.001 *a.u.* isodensity envelope mapped potentials with those of $\omega$B97XD/def2-TZVPPD potentials computed on the electrostatic surfaces of the BiX$_3$ molecules (X = F, Cl, Br, I).

| Local most extrema on the surface of specific atom/bond | BiF$_3$ | | BiCl$_3$ | | BiBr$_3$ | | BiI$_3$ | |
|---|---|---|---|---|---|---|---|---|
| | MP2(full) | $\omega$B97XD | MP2(full) | $\omega$B97XD | MP2(full) | $\omega$B97XD | MP2(full) | $\omega$B97XD |
| $V_{s,min}$ On X (lateral portions) | -27.6 | -27.4 | -13.9 | -13.8 | -11.8 | -11.6 | -9.1 | -8.9 |
| $V_{s,min}$ on Bi (opposite to the triangular face formed by three X atoms) | 53.6 | 50.6 | 45.8 | 44.2 | 42.4 | 41.1 | 37.5 | 35.8 |
| $V_{S,max}$ (on Bi–X bond extensions) | | | -2.8 | -2.4 | 2.6 | 3.4 | 9.1 | 11.1 |
| $V_{S,max}$ (on X–Bi bond extensions) | 58.9 | 58.0 | 50.4 | 50.7 | 46.3 | 47.0 | 40.3 | 41.1 |
| $V_{S,max}$ (on the centroid of the triangular face formed by three X atoms) | 6.2 | 8.8 | 0.9 | 3.7 | 0.5 | 3.8 | 0.3 | 3.5 |

Bismuth presents three σ-holes, each along the extension of the X–Bi bond in all four monomers BiX$_3$ (Fig. 9). The largest and most stable σ-hole is found on Bi along the F–Bi bond extensions in BiF$_3$ ($V_{S,max}$ = 58.9 kcal mol$^{-1}$ with MP2(full) and 58.0 kcal mol$^{-1}$ with $\omega$B97XD) and the smallest and least stable of this is along the I–Bi bond extension in BiI$_3$ ($V_{S,max}$ = 40.3 kcal mol$^{-1}$ with MP2(full) and 41.1 kcal mol$^{-1}$ with $\omega$B97XD). The stability of the σ-hole on Bi is in the order BiF$_3$ > BiCl$_3$ > BiBr$_3$ > BiI$_3$. These σ-holes are significantly more positive than those found on the surfaces of Sb, As, P, and N along the X–Pn (Pn = Sb, As, P, and N) bond extensions [7-10]. For a given X, the stability of the σ-holes on the halogen bound pnictogen Pn along the pnictogen series is in the order: BiX$_3$ > SbX$_3$ > AsX$_3$ > PX$_3$ > NX$_3$.

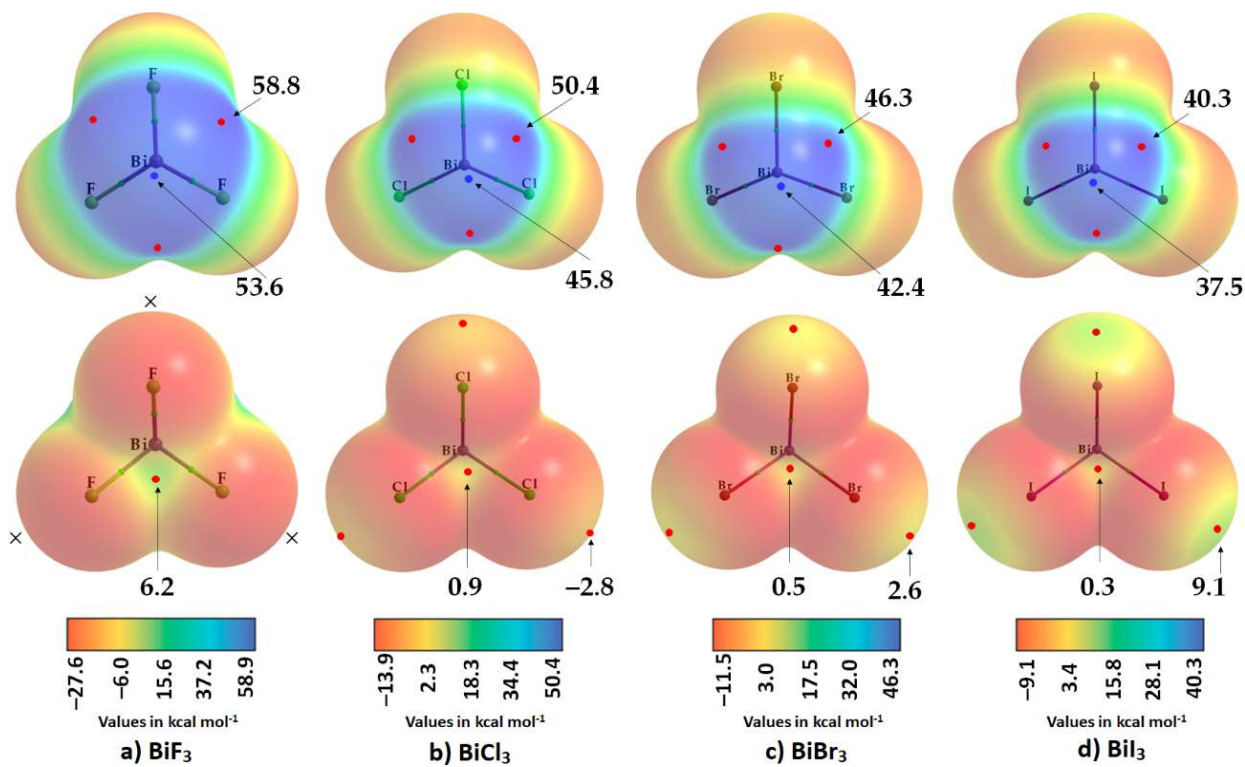

**Figure 9.** Comparison of MP2(full)/def2-TZVPPD calculated 0.001 *a.u.* isodensity envelope mapped potential on the electrostatic surfaces of BiX$_3$ (X = F, Cl, Br, I) molecules. Tiny circles in blue and red represent the local most minimum and maximum of potential, respectively. Selected $V_{S,min}$ and $V_{S,max}$ values corresponding to these extrema are in kcal mol$^{-1}$. Two views of each MESP graph for each molecule are displayed for each case: (Top) Coordinated Bi faces the reader; (Bottom) The three X-atoms forming a triangular architecture in BiX$_3$ faces the reader.

The crystal structure of BiF$_3$ has been known for a long time [141,142] and its structure in different space groups has been reported subsequently [143,144]. As shown in Fig. 10, the Bi$^{3+}$ ion in BiF$_3$ is eight-coordinate in the *Pnma* and *Fm3m* space groups (Figs. 10a-b), 7-coordinate in the *P-43m* structure (Fig. 10c) and 9-coordinate in the *P3-c1* structure (Fig. 10d). Similarly, the structure of BiF$_5$ has been known since 1971 [145]. It crystallizes tetragonally in the space group *I4/m* (Fig. 10e), with Bi–F bond distances varying between 1.90 and 2.11 Å.

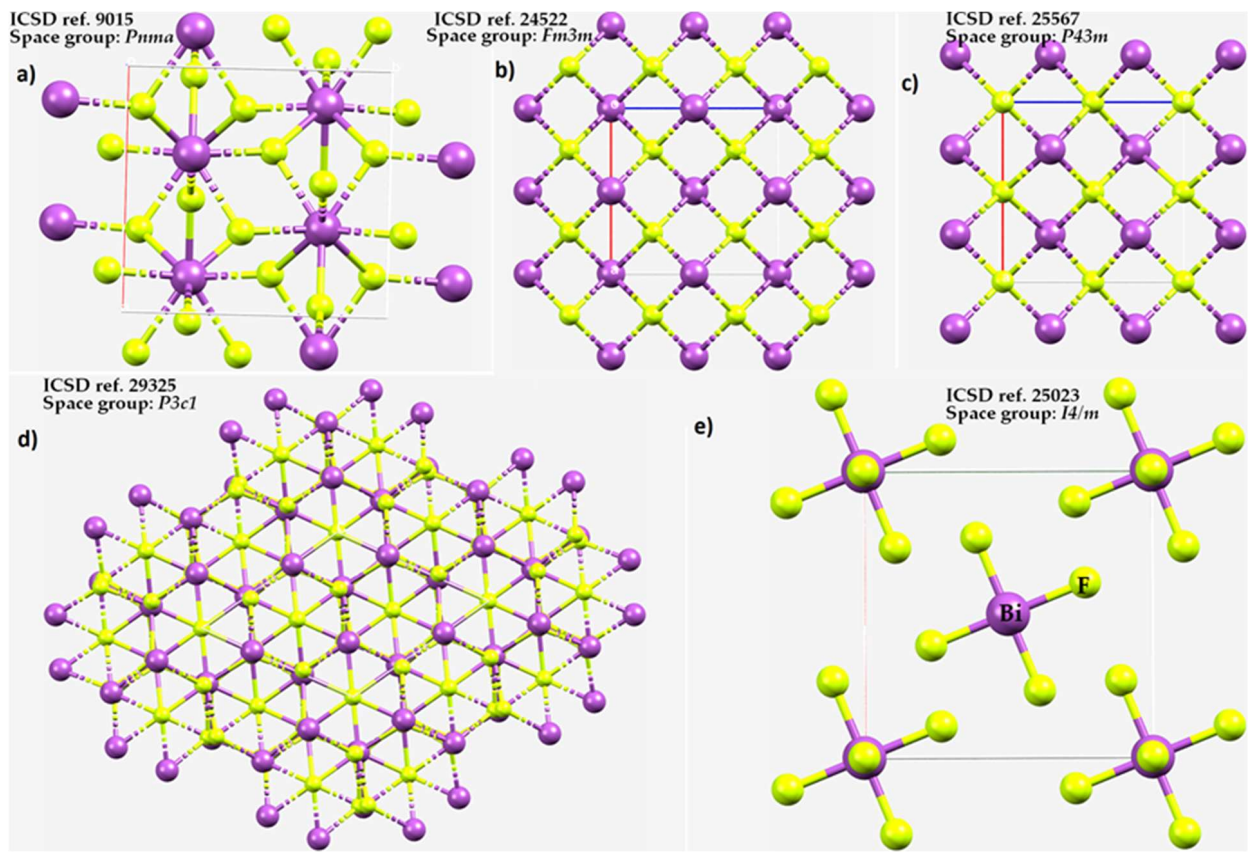

**Figure 10.** a)-d) Ball-and-stick models of the crystal structure of BiF$_3$ in different space groups; e) Ball-and-stick model of the crystal structure of BiF$_5$. The ICSD ref codes and space groups are shown for each case. Bonds are shown as sticks in atom color: Bi – purple; F – dark-yellow.

It has been suggested that the distortion of the coordinate sphere of Bi is a result of its stereochemically active lone pair [143]. However, the results of the MESP analysis shown in Fig. 9a suggests that the portion of Bi opposite to the triangular face formed by the three F atoms in BiF$_3$, as well as those along the F–Bi bond extensions, is fully positive. This is actually a region on Bi in BiF$_3$ that allows the entirely negative fluorine on another BF$_3$ moiety to engage with it attractively, and thereby to form an infinite crystalline network. While the Bi–F bond lengths in the first coordination sphere of Bi are not all equivalent, they are representatives of covalent coordinate bonds, and BiF$_3$ is a covalent species. It is probably for this reason that BiF$_3$ has a band gap. For instance, Feng *et al.* [146] have recently demonstrated that BiF$_3$ is a photocatalyst based on DFT calculations; the system has a highly positive valence band and a wide direct bandgap of 3.94 eV. Clearly, none of the crystals shown in Fig. 10 show any long-range pnictogen bonded interactions. From our knowledge of the strong electrostatic potential on Bi in BiF$_3$, it seems obvious that the formation of coordinate bonds, rather than pnictogen bonds, along the extension of the F–Bi is a likely consequence.

There were two entries for the *Pn2$_1$a* orthorhombic crystalline state of BiCl$_3$ in the ICSD (refs. 2866 and 41179), reported in 1972 [147] and 1982 [148], respectively. The arrangement of BiCl$_3$ molecules in the crystal is shown in Fig. 11a, and the local connectivity between them is depicted in Fig. 11b.

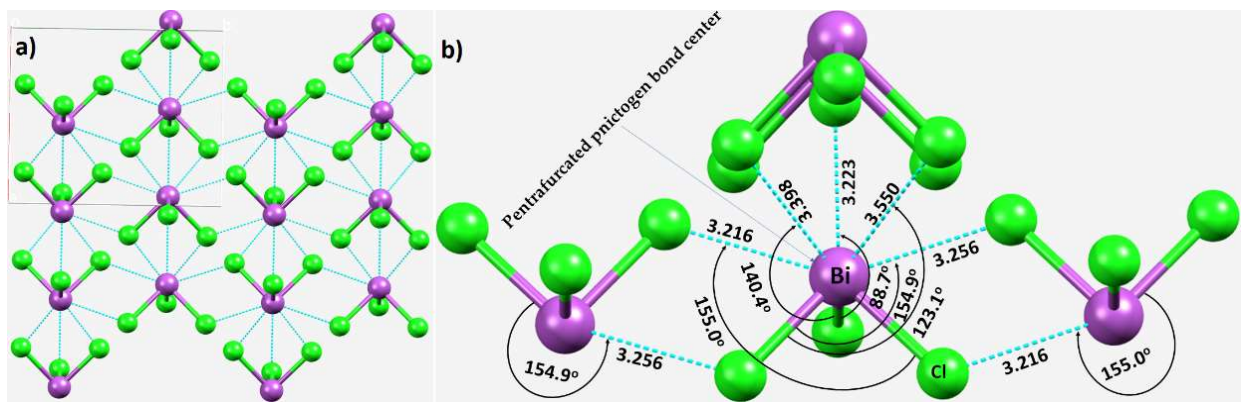

**Figure 11.** a) Ball-and-stick model of the geometric topology of Bi⋯Cl non-covalent bonding between interacting BiCl$_3$ monomers in crystalline BiCl$_3$ (ICSD ref. code: 2866). b) Illustration of the local pentafurcated topology of bismuth in the BiCl$_3$ monomer. Bond lengths and bond angles are in Å and degrees, respectively. Bonds are shown as sticks in atom color: Bi – purple; Cl – green. Non-covalent interactions are shown as dotted lines in cyan.

Bi in BiCl$_3$ acts as a pentafurcated pnictogen bond donor. The acceptors of these bonds are the lone-pair regions on Cl in the surrounding BiCl$_3$ molecules that are negative (see Table 1). Because the positive region on Bi in BiCl$_3$ is of the diffuse type (as, for example, is hydrogen atoms in molecules such as H$_2$O), the directionality of pnictogen bonds formed by covalently/coordinately bonded Bi is weaker (when compared to N and P). This is evident from the ∠Cl–Bi⋯Cl angles (Fig. 11b), which vary between 88.7° and 155.0°, indicating a significant flexibility on the part of Bi to enter into pnictogen bonding with negative sites. The Bi⋯Cl bond lengths between the monomers in the crystal vary between 3.216 and 3.450 Å. This is much smaller than the sum of the vdW radii of Cl and Bi, 4.43 Å ($r_{vdW}$(Bi) = 2.54 and $r_{vdW}$(Cl) = 1.89 Å), suggesting the possibility of significant covalency in the Bi⋯Cl bonds. This is in contrast with the three Bi–Cl coordinate bonds of each molecular BiCl$_3$ that are not only inequivalent (with $r$(Bi–Cl) values of 2.518, 2.513 and 2.468 Å), but also markedly shorter than the five Bi⋯Cl close contacts identified. The marked difference between two types of bond distances within and between BiCl$_3$ molecular unit(s) enables us to unequivocally demonstrate that the close contacts are certainly not coordinate bonds, but no other than pnictogen bonds. This also becomes evident by looking at the IGM-$\delta g^{inter}$ domains (isosurfaces in bluish and green) between bonded atomic basins, Fig. 12. These domains indicate that the charge density in the bonding regions between interacting Bi and Cl atomic basins is not negligible and are evident even at an isovalues > 0.02 a.u. The bluish nature of the isosurfaces indicate the presence of partial covalency in the intermolecular interactions involved. The three pnictogen bonds appearing along the Cl–Bi bond extensions are relatively stronger than the remaining two non-linear pnictogen bonds which are longer; this is attested to by the respective sizes of the isosurface volumes (cf. Fig. 12a).

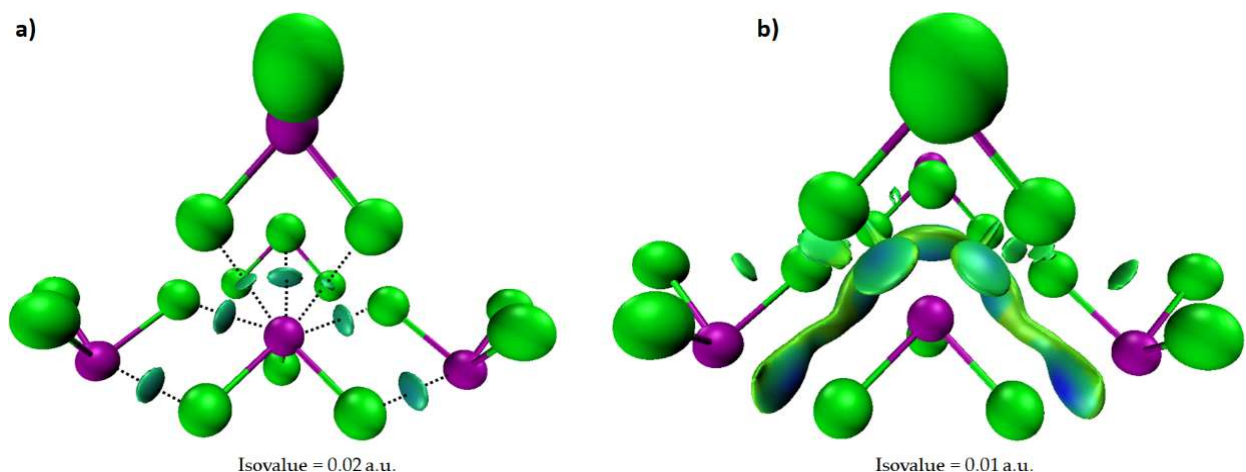

**Figure 12.** a) and b) Promolecular IGM-$\delta g^{inter}$ based isosurface topologies (colored bluish and green) between the BiCl₃ monomers feasible in the BiCl₃ crystal (ICSD ref. code: 2866), obtained using a part of geometry extracted from the crystal; two different isovalues were used. These show the prominence of intermolecular interactions in the low-density region. See Fig. 11b for the details of the Bi···Cl local bonding modes and bond distances involved.

Crystals of the other two members of the BiX₃ family are known in 2D and 3D in different space groups because of their sizes, including *C2/m*, *R-3*, and *Fm3m*. Fig. 13a and 13b illustrate two such examples, featuring BiBr₃ and BiI₆, respectively. It is clear that Bi³⁺ in these crystal systems is six-coordinate. The three Bi···X links between BiX₃ molecular entities are very similar. As discussed above, Fig. 9c-d, the MESP analysis shows that Bi in each BiX₃ unit has three potentially stable positive σ-holes along the Bi–X bond extensions, and the lateral portions of the Bi bonded X sites are negative. This suggests that the Bi···X links between the BiX₃ units in these crystals, Fig. 13, could be the result of pnictogen bonding. These interactions are either linear or quasi-linear (∠I–Bi···I = 179.1° in BiI₆ and ∠Br–Bi···Br = 167.1°–176.9° in BiBr₆) and that the Bi···X links between the BiX₃ molecular entities are probably a result of an attractive engagement between two sites of opposite electrostatic potential. Given that Bi is a brittle metal, and that the Bi-to-X link distances are too short, the Bi···X interactions may have significant covalent character, and hence can be regarded as coordinate bonds, but not pnictogen bonds.

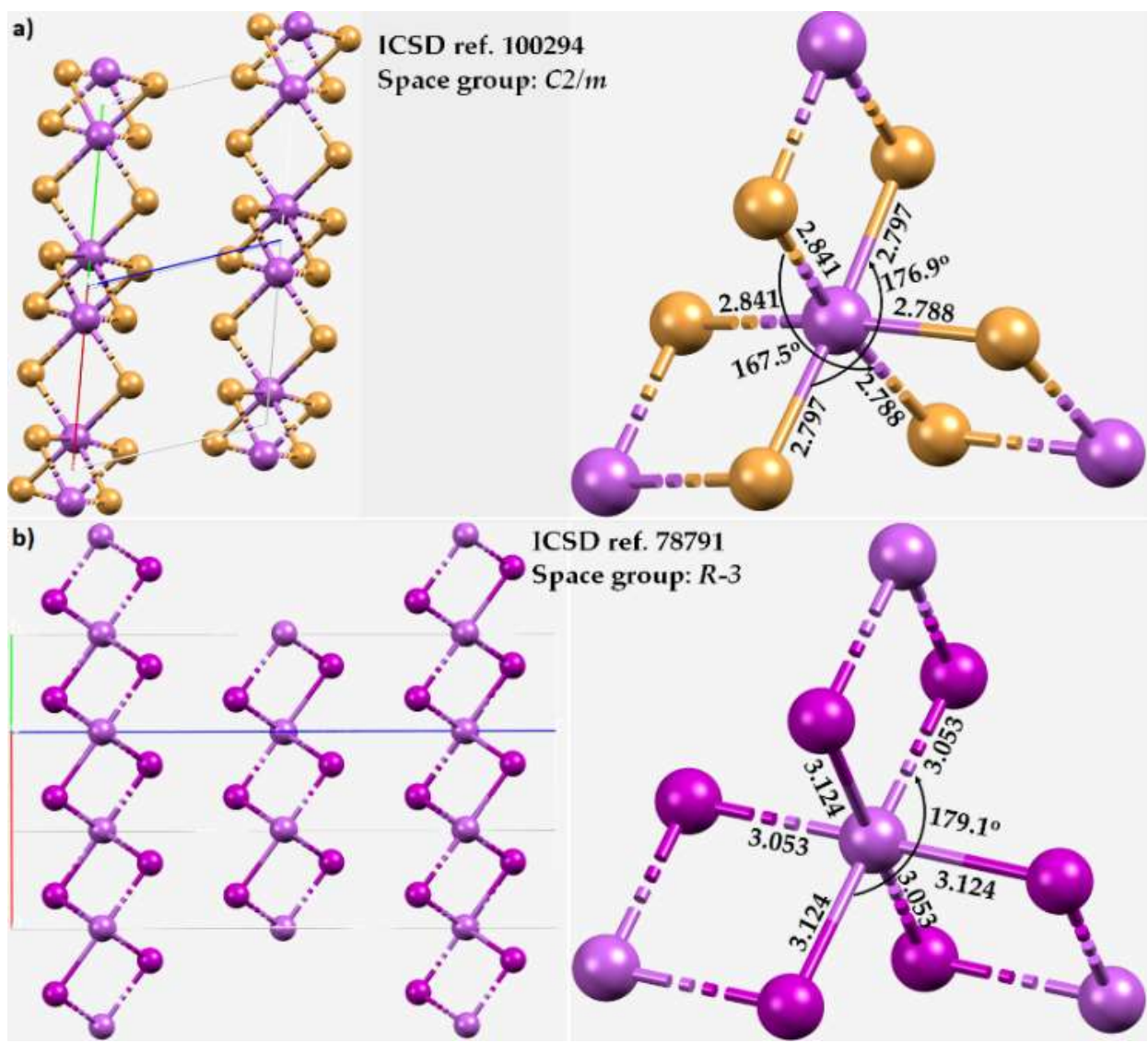

**Figure 13.** Ball-and-stick models of the crystal structures of a) BiBr₃ and b) BiI₃, showing layer-like structures in 2D. The ICSD ref codes and space groups are shown for each case. Selected bond distances and angles are in Å and degrees, respectively. Bonds are shown as sticks in atom color: Bi – faint-purple; I – purple; Br – faint-brown. Non-covalent interactions are shown as dotted lines in cyan.

*b. Bi⋯O Pnictogen Bonds in Host-Guest Complexes*

Like the crown ether complexes of AsX₃ [9] and SbX₃ (X = Cl, Br) [149] discussed elsewhere [9,10], BiCl₃ and BrBr₃ have also played an important role in Bi-centered host-guest chemistry [149-151]. The Bi³⁺ cation does not adopt a pseudo-octahedral coordination geometry in either of four complex geometries shown in Fig. 14a-d for BiCl₃, but its coordination number increases with an increase in the size of the cavity of the crown ether. The largest number of links of Bi with the surrounding negative sites is found in its complex with μ₂-dibenzo-24-crown-8 [149]. Except for the crystal structure of its complex with 18-crown-6 [151], where the coordination of the Bi with the surrounding negative sites is strongest, the Bi⋯O links in all other complexes are longer than the three genuine Bi–Cl bonds in BiCl₃ (as shown in Fig. 14a-b and d). The longest Bi⋯O non-covalent interaction is 3.221 Å in the complex with μ₂-dibenzo-24-crown-8 [149] and most of ∠Cl–Bi⋯O contacts are found between 135° and 173°; they follow Type-IIa or Type-IIb topologies of bonding. We suggest that these contacts between Bi and O in the complexes shown in Fig.

14a,b,d have the characteristics of pnictogen bonding, in agreement with the views of Prokudina and coworkers reported for the SbBr$_3$-Py complexes in the solid state [152].

The boundary between coordinate and pnictogen bonds may be drawn with knowledge of the Bi⋯D bond distances. For instance, from Fig. 14a, it is apparent that the three Bi–Cl bond distances in BiCl$_3$ are nearly equivalent, with $r$(Bi–Cl) = 2.505, 2.503 and 2.539 Å. These are all less than 2.55 Å, which is true for all systems shown in Fig. 14 (except for Fig. 14c). On the other hand, the five Bi⋯O close contacts formed by the Bi center in BiCl$_3$ with the O-site of the crown ether are substantially longer, with $r$(Bi⋯O) = 2.871, 2.733, 3.008, 2.934, 3.221 and 2.871 Å. Given that the van der Waals radius of Cl ($r_{vdW}$(Cl) = 1.89 Å) is substantially longer than that of O ($r_{vdW}$(O) = 1.50 Å [72]), and that O is more electronegative, the (Bi⋯O) bond distances would have had to be much shorter to recognize these as coordinate bonds (as found for a Bi–O coordinate bond in the system in Fig. 14c). We therefore characterize the Bi⋯O close contacts in the $\mu_2$-dibenzo-24-crown-8 system as strong pnictogen bonds rather than coordinate bonds. They are strong since the Bi⋯O bond distances are appreciably shorter than the sum of the van der Waals radii of Bi and O, 4.05 Å. This is also valid for the Bi⋯O close contacts in host-guest complexes shown in Fig. 14b and 14d, but not that in Fig. 14c. In the latter, there is a single Bi⋯O close contact ($r$(Bi⋯O) = 2.490 Å) in the cation (BiCl$_2$·18-crown-6)$^+$ of the ionic adduct [2(BiCl$_2$·18-crown-6)$^+$·(Bi$_2$Cl$_8$)$^{2-}$], which is not only shorter than the two Bi–Cl coordinate bonds, but the remaining five Bi⋯O contacts are also close to the coordinate bond formation limit. The exception arises because of the BiCl$_2$$^+$ cation attracts the neutral ligand 18-crown-6 leading to the formation of a cationic complex locally. In this sense, the bismuth cation is eight-coordinate involving all six oxygen atoms of the crown and two chloride ions in a bicapped trigonal prismatic geometry. These bonding features are markedly different from those observed for the crown ether complexes of BCl$_3$, BiCl$_3$·12-crown-4 (Fig. 14d) and BiCl$_3$·15-crown-5 (Fig. 14b) that are a neutral adducts with the pyramidal BiCl$_3$ linked to all four (five) oxygen atoms of the crown in a half-sandwich structure.

Another argument in favor of our view on the formation of pnictogen bonds in the complexes shown in Fig. 14a,b,d is that all Bi⋯O close contacts are present along and off the extension of the C–Bi bonds in BiCl$_3$. This is because the electrostatic potential on Bi opposite the triangular face formed by the three chlorine atoms is completely positive (see Figure 9b). These positive potential regions are adequate in accepting electron density from the surrounding oxygen atoms of the crown ether. This scenario is required to form a Type-II topology of the pnictogen bonding interaction (Scheme 1b).

We surveyed the Bi-O distances of compounds deposited in CSD. In the range 1.8 to 4.5 Å, we 2559 contacts in 597 crystals. The peak of the normal distribution curve was at approximately 2.45 Å. There were no contacts below 2.013 Å or above 3.15 Å, indicating that all covalent, coordination, and strong pnictogen bonds are within this bond distance range. Considering that covalent and coordination bonds are generally stronger than pnictogen bonds, the best Bi–O coordination bonds in the crystal are expected to be around 2.6 Å, the boundary that tentatively separates Bi⋯O pnictogen bonds from the Bi–O coordinate bonds. When both Bi⋯O inter- and intramolecular close contacts were used as a search criterion and the bond between Bi and O was chosen to be of any type (single, double, triple, etc), 5265 were found in in 852 crystals. Of these, those occurring between 3.15 and 4.50 Å could be considered weak to moderate strength Bi⋯O pnictogen bonds.

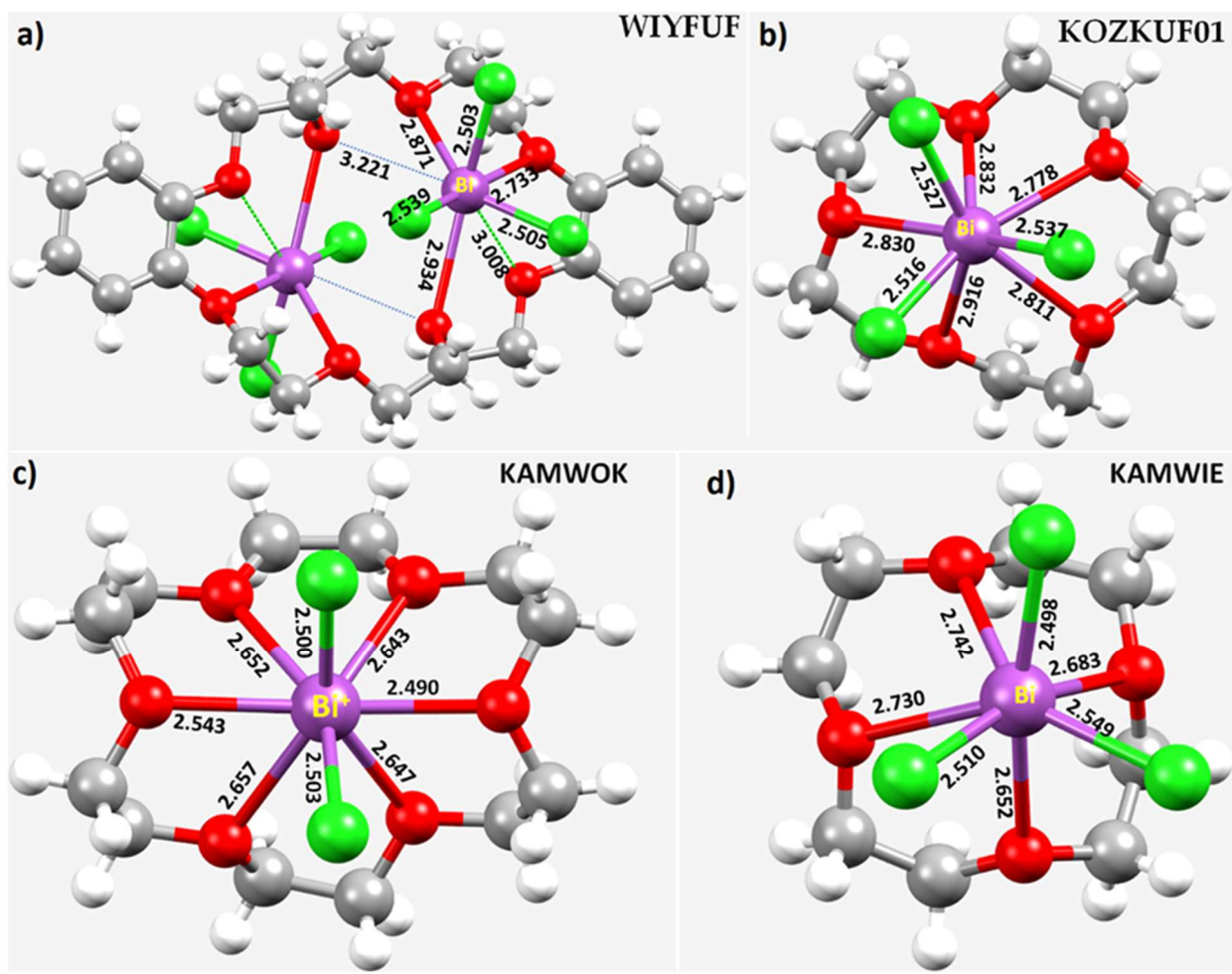

**Figure 14.** Ball-and-stick models of the host-guest interactions between BiCl₃ and crown ethers in some crystals, showing the emergence of strong coordination between Bi and Cl and the formation of potentially strong non-covalent interactions. The $Bi^{3+}$ structures illustrated (CSD ref code given) are of the crown ethers a) $\mu_2$-dibenzo-24-crown-8 [149]; b) 15-crown-5 [153]; c) 18-crown-6 [151]; and d) 12-crown-4 [151]. For clarity, other molecular entities in the crystals in a) and c) are omitted. Bonds are shown as sticks in atom color: Bi – purple; Cl – green; C – gray; H – white-gray; O – red. Long close contacts are shown as dotted lines.

Our IGM-$\delta g$-based results corresponding to the interactions between the bonded Pn site in BiCl₃ and the negative sites of the crown ether in Fig. 14a, 14b and 14d are shown in Fig. 15a, 15b and 15c, respectively. As we inferred from the intermolecular distances between Bi and O, we indeed observed IGM-$\delta g$ based isosurfaces between these atomic basins, indicative of genuine intermolecular bonding interactions. However, we found that there are two additional isosufaces that show up between the Bi center in one BiCl₃ unit and the bonded Cl atom the another same molecule in the 12-crown-4 complex [151]. The Bi···Cl bond distance is 3.403 Å, and ∠Cl–Bi···Cl = 136.1° (see Fig. 14a). From the colors of the isosurfaces, and the bond distances, we conclude that those thick isosurfaces between Bi and ligating O atoms with a bluish color probably comprise some significant percentage of covalency and hence they may include some coordinate bond character. The remaining longer contacts are undoubtedly pnictogen bonds, and are quite evident in the crystal structure of the $\mu_2$-dibenzo-24-crown-8 complex [149] shown in Fig. 14a and 15a.

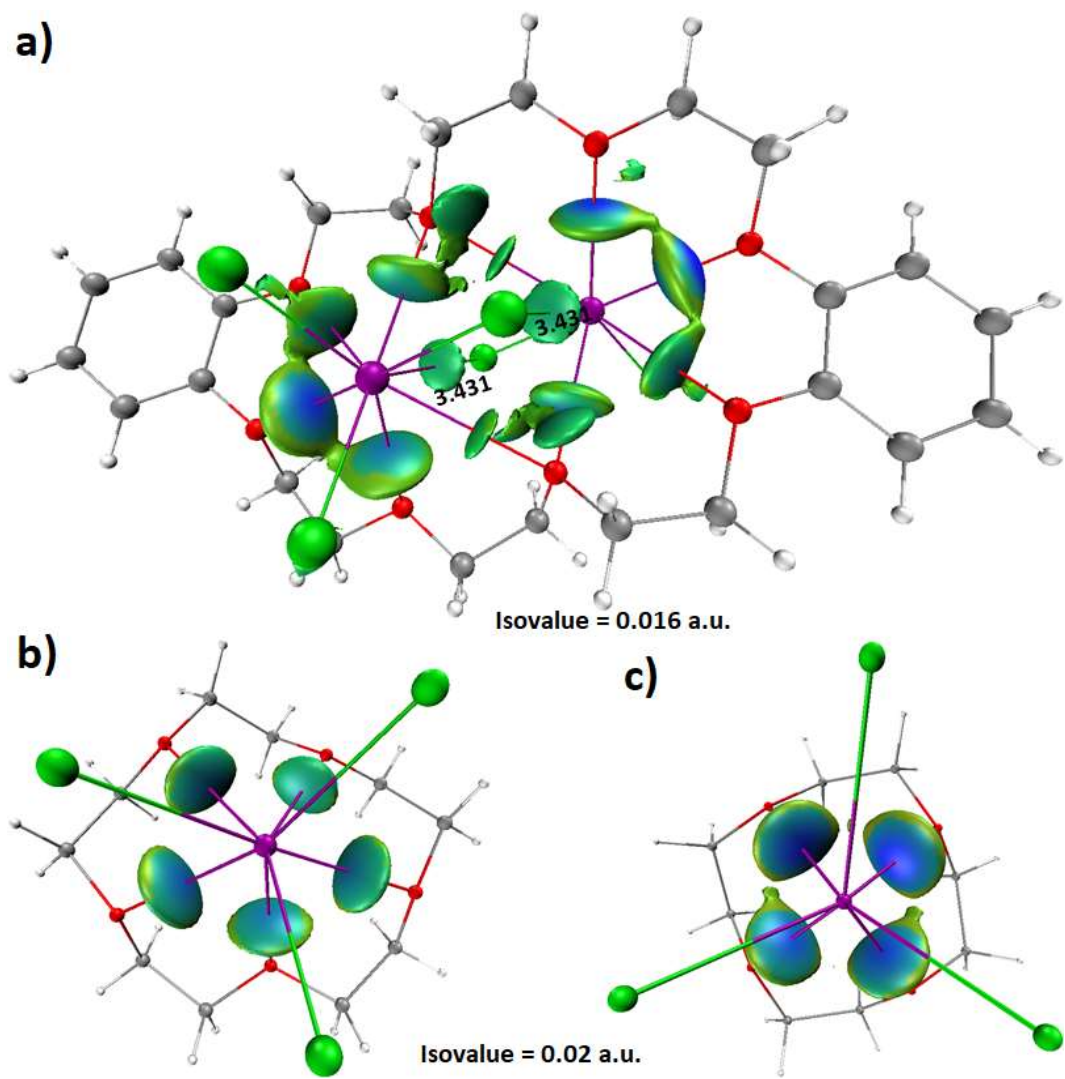

**Figure 15.** IGM-δ$g^{inter}$ based isosurface plots for the Bi$^{3+}$ complexes with a) μ$_2$-dibenzo-24-crown-8 [149]; b) 15-crown-5 [153]; and c) 12-crown-4 [151]. Isosurfaces colored blue and green indicate strong and medium-to-weak interactions, respectively. Isovalues used for the generation of isosurfaces are shown. Atom type and coloring is same as those shown in the ball-and-stick models in Fig. 14a, 14b and 14d, respectively. Selected bond lengths are in Å.

The crystal structure of the complex of BiI$_3$ and 15-crown-5 is shown in Fig. 16a [154]. The bonding modes between Bi and O in the system is similar to that formed by the other members of the pnictogen trihalide family (*vide supra*). However, in the present case the Bi–I bonds are marginally longer than the Bi–O bonds. The former are formal coordinate bonds, expected of the BiI$_3$ molecule. The latter are comparable with that observed in the complexes of BiCl$_3$ with μ$_2$-dibenzo-24-crown-8 [149] (Fig. 14a) and 15-crown-5 [153] (Fig. 14b), and are pnictogen bonds given that they appear along and off the outer extension of the I–Bi bonds in BiI$_3$. One of the special features of this system is that the I sites of the Bi-coordinated crown ether are also engaged in forming long-ranged I⋯I interactions that are directional in nature. The formation of both coordination bonds and I⋯I bonding interactions is in accord with the positive electrostatic potential on Bi and I along the I–Bi and Bi–I bond extensions (Fig. 9d); these are in attractive engagement with the negative lone-pair dominant sites on O and I, respectively. This also explains why the Bi–I⋯I angle deviates from linearity. The I⋯I halogen-halogen bonding interactions are responsible for the formation of the 1D chain-like architecture, Fig. 16a (left).

We used [ωB97XD/def2-TZVPPD] to energy minimize the geometry of [BiI$_3$][15-crown-5] complex. This was done in the gas phase to gain a better understanding of the nature of the bonding interaction between BiI$_3$ and 15-crown-5 in the absence of crystal packing effects. The resulting fully relaxed geometry is shown in Fig. 16b. Two features are evidence of the geometry. First, all three Bi–I coordinate bond distances of BiI$_3$ are reproduced within 0.1 Å (Fig. 16a, right), while all the Bi-to-O bond distances between BiI$_3$ and 15-crown-5 are longer than that found in the crystal (see Fig. 16a vs. Fig. 16b). The latter feature is expected since the effect of packing forces of the crystal lattice causes the Bi⋯O bond distances to shrink. Second, the three angles, ∠I–Bi⋯O, that appear along the extensions of the I–Bi bonds are all quasilinear (Fig. 16b), indicating the presence of σ-hole centered pnictogen bonds. Opposite to the triangular face formed by the three I atoms in BiI$_3$ (Fig. 9d), Bi is completely positive, so it is also linked to the remaining two O-sites of 15-crown-5. The latter two links are relatively long, Bi⋯O bond distances of 3.226 and 3.346 Å, corresponding to bond angles of 138.1° and 134.7°, respectively. We designated the first three Bi⋯O links, which are along the extensions of the I–Bi bonds, as Type-IIa and the latter two as Type-IIb pnictogen bonds; all are σ-hole centered.

The results of our QTAIM and IGM-$\delta g^{inter}$ calculations performed on the wavefunction evaluated using the gas phase geometry of the complex of BiI$_3$-15-crown-5 is shown in Fig. 16b and c, respectively. QTAIM has predicted the expected bond paths and bond critical points of charge density between Bi and O, as well as that between Bi and I. The charge density $\rho_b$ at the Bi–I and Bi⋯O bcps (0.0574 a.u. < $\rho_b$ < 0.0580 a.u. and 0.0091 a.u. < $\rho_b$ < 0.0161 a.u., respectively) are typical of non-covalent interactions, and the Laplacian of charge density $\nabla^2\rho_b$ values at the corresponding bcps were all small and positive (0.0412 a.u. < $\nabla^2\rho_b$ < 0.0438 a.u. at the Bi–I bcps, and 0.0272 a.u. < $\nabla^2\rho_b$ < 0.0498 a.u. at the Bi⋯O bcps), which indicate closed-shell (electrostatic) interactions. However, when the total energy density at the corresponding bcps were analyzed, it was found that $H_b$ is negative for each of the three Bi–I coordinate bonds (–0.0140 < $H_b$ < –0.0148 a.u.), and positive for Bi⋯O close contacts (0.0012 < $H_b$ < 0.0014 a.u.). This provides further evidence that the former are genuine coordinate bonds with appreciable covalent character, and the latter are formal pnictogen bonded interactions. These results are in agreement with IGM-$\delta g^{inter}$ based isosurface charge density topologies (green volumes) between Bi in BiI$_3$ and the five ligating O sites in Benzo-15-crown-5, Fig. 16d. Moreover, the [ωB97XD/def2-TZVPPD] level uncorrected and BSSE corrected binding energies of the complex calculated at the gas-phase geometry were –35.0 and –34.33 kcal mol$^{-1}$, respectively. This signifies that the BSSE corrected stabilization energy of a single Bi⋯O pnictogen bond in the BiI$_3$-15-crown-5 complex is about –6.86 kcal mol$^{-1}$; it is therefore a medium strength interaction [155], and is surely not a typical coordinate bond.

Very similar conclusions can be drawn from the structure of the complex [BiI$_3$][Benzo-15-crown-5], Fig. 16e. A coordination chemist would see this system as a neutral pseudo-octahedral complex, in which the [BiI$_3$][Benzo-15-crown-5] units in the crystal are connected by secondary I⋯I halogen-halogen interactions forming chains in 1D [156]. The pseudo-octahedral nature of the system could be extracted based on the Bi–I and Bi⋯O bond distances (values between 2.8 – 3.0 Å), and that the mean of the former is marginally longer than that of latter bond distances. By contrast, in the gas-phase structure, Fig. 16f, the order of the two types of bonding distances is reversed, i.e., $r$(Bi⋯O) > $r$(Bi–I). This is thought to be due to the absence of packing forces in the gas-phase structure, which allows the interacting molecules to bond freely without any constraints, which causes the appearance of the genuine nature of $r$(Bi⋯O) and $r$(Bi–I). Fiolka and coworkers [156] have claimed that the bonding between [BiI$_3$] and [Benzo-15-crown-5] has to be mainly electrostatic as the interactions of the bismuth 6s lone pair with the 2p orbitals of the oxygen atoms of the crown ether are antibonding.

The bond path and bond critical point topologies of charge density, Fig. 16g, indicate both the Bi–I and Bi⋯O links in [BiI$_3$][Benzo-15-crown-5]. Their strengths are comparable with those found in [BiI$_3$][15-crown-5] (Fig. 16c), and are evidenced by the $\rho_b$ values (0576 a.u. < $\rho_b$ < 0.0586 a.u. at Bi–I bcps and 0.0108 a.u. < $\rho_b$ < 0.0138 a.u. Bi⋯O, respectively).

Because 0.0423 a.u. (–0.0145 a.u.) < $\nabla^2 \rho_b$ ($H_b$) < 0.0432 a.u. (–0.0150 a.u.) at the Bi–I bcps, and 0.0344 a.u. (–0.0013 a.u.) < $\nabla^2 \rho_b$ ($H_b$) < 0.0433 a.u. (–0.0014 a.u.) at the Bi⋯O bcps, it may be concluded that the Bi–I bonds comprise both ionic and covalent character and the Bi⋯O links are largely electrostatically driven. Since the [ωB97XD/def2-TZVPPD] level uncorrected and BSSE corrected binding energies of the gas-phase geometry of the complex were –32.84 and –32.16 kcal mol$^{-1}$, respectively, the corrected energy of a single Bi⋯O pnictogen bond in the [BiI$_3$][Benzo-15-crown-5] complex is about –6.43 kcal mol$^{-1}$, meaning it is a medium strength interaction, and is surely not a typical coordinate bond. Provided the empirical relation, $E_b = – ½V_b$, is used, where $E_b$ represents the energy of a bond with potential energy density $V_b$ at bcp, the energies of all the five Bi⋯O close contacts lies between – 3.7 and –5.2 kcal mol$^{-1}$, which supports the view that the Bi⋯O interactions have the characteristic of medium strength pnictogen bonds.

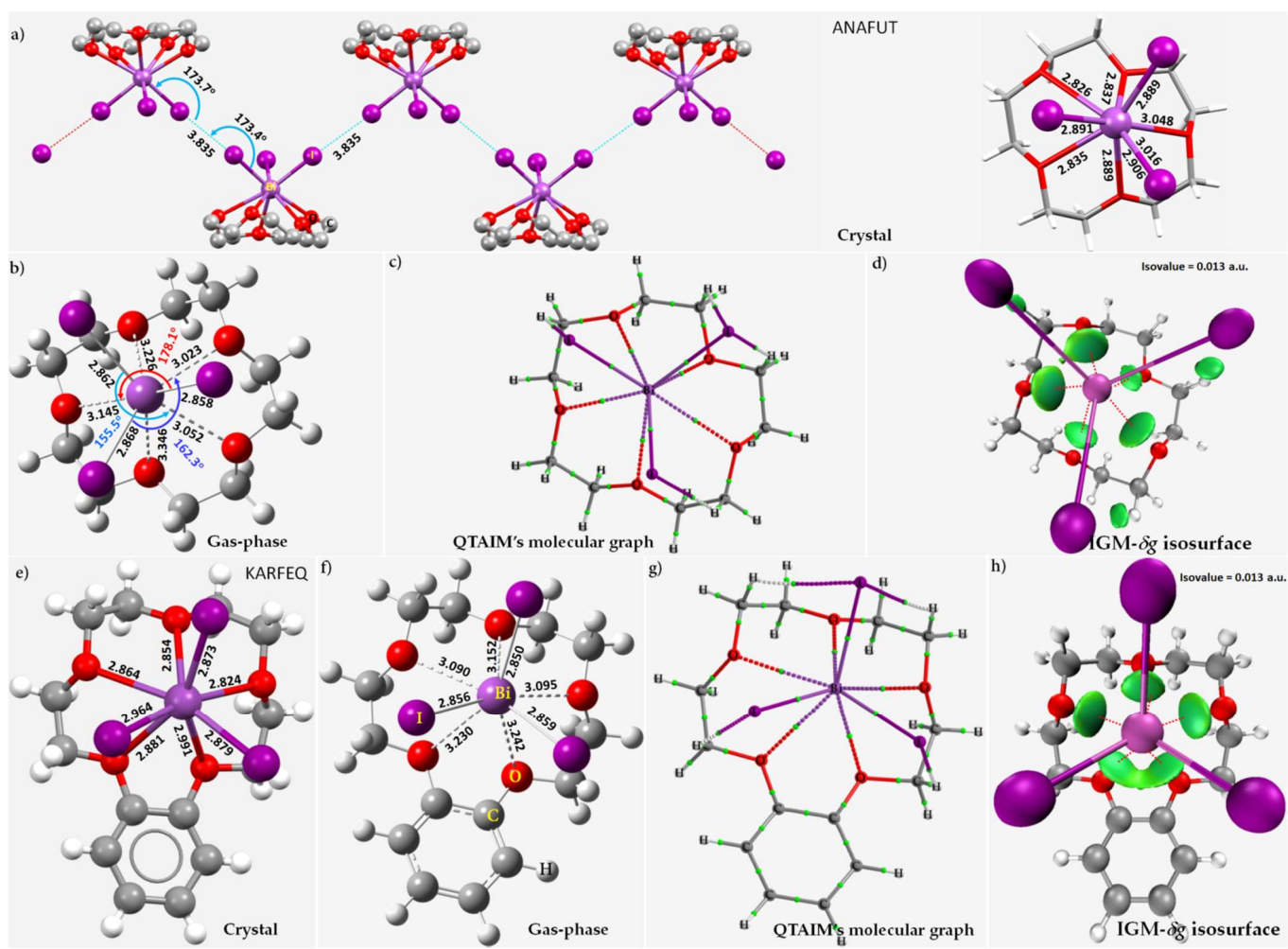

**Figure 16.** Nature of intermolecular bonding in the crystal structure of [BiI$_3$][15-crown-5] (CSD ref code AMAFUT) [154];H atoms are omitted from figure on left; the I⋯I contacts are shown as dotted lines in cyan. b), c) and d) Gas-phase fully-relaxed geometry, QTAIM-based molecular graph and IGM-$\delta g^{inter}$ based isosurface charge density topologies between Bi in BiI$_3$ and O donors in 15-crown-5, respectively, obtained using [ωB97XD/def2-TZVPPD]. e), f), g) and h) The crystal geometry, gas-phase fully-relaxed geometry, QTAIM-based molecular graph, and IGM-$\delta g^{inter}$ based isosurface charge density topologies between Bi in BiI$_3$ and donors in Benzo-15-crown-5, respectively, with the latter three obtained using [ωB97XD/def2-TZVPPD]. Bond paths as sticks and dotted lines in atom color between bonded atomic basins, and bond critical points between bonded atomic basins as tiny spheres in green in c) and g) are shown. Selected bond lengths (Å) and bond angles (degrees) are depicted. The CSD references for the crystals are shown in a) and e) in uppercase letters.

*c. Bi⋯O, Bi⋯P, Bi⋯S, Bi⋯N, Bi⋯C$_\pi$, Bi⋯Se, and Bi⋯X (X = Cl, Br, I) Pnictogen Bonds in Miscellaneous Crystals*

The crystal structure of [[RBi(Cl)Cl]$_2$·Et$_2$O]$_n$ (R = CH(Si(CH$_3$)$_2$) [157], Fig. 17a, consists of infinite chains and is the result of a complex network of non-covalent interactions, driven largely by C–H⋯O hydrogen bonding and other non-covalent interactions. The compound has a large number of sterically demanding Si(CH$_3$)$_2$ groups and covalently bound Cl sites, together with diethyl ether solvate. The [[RBi(Cl)Cl]$_2$·Et$_2$O]$_n$ chains are linked through chloride bridges. The diethyl ether molecules are weakly associated to neighboring bismuth atoms in the chain. As shown in Fig. 17b, the Bi$^{3+}$ center is pseudo-octahedral, comprising of three formal Bi–X (X = Cl, C) bonds, one Bi⋯Cl halogen bridge, two Bi⋯O contacts, and one Bi⋯Cl contact. The first three are ionic bonds with significant covalent character. The last three are substantially longer, and are produced through intermolecular interactions. They are characteristic of Type-IIa and -IIb pnictogen bonding interactions, in which each negative site on the O atom on the diethyl ether couples bifurcately with two nearest Bi sites (Fig. 17b). Because of steric crowding of H atoms that are involved in hydrogen bonding, the Type-IIa/-IIb Cl–Bi⋯O and C–Bi⋯O pnictogen bonded interactions found in this system are not linear (∠Cl–Bi⋯Cl = 169.2°; ∠C–Bi⋯O = 149.9°/144.1°). The intermolecular contact distances (Bi⋯O) are all significantly smaller than the sum of the vdW radii, 4.05 Å, of the Bi and O atoms ($r_{vdW}$(Bi) = 2.54 Å and $r_{vdW}$(Bi) = 1.50 Å).

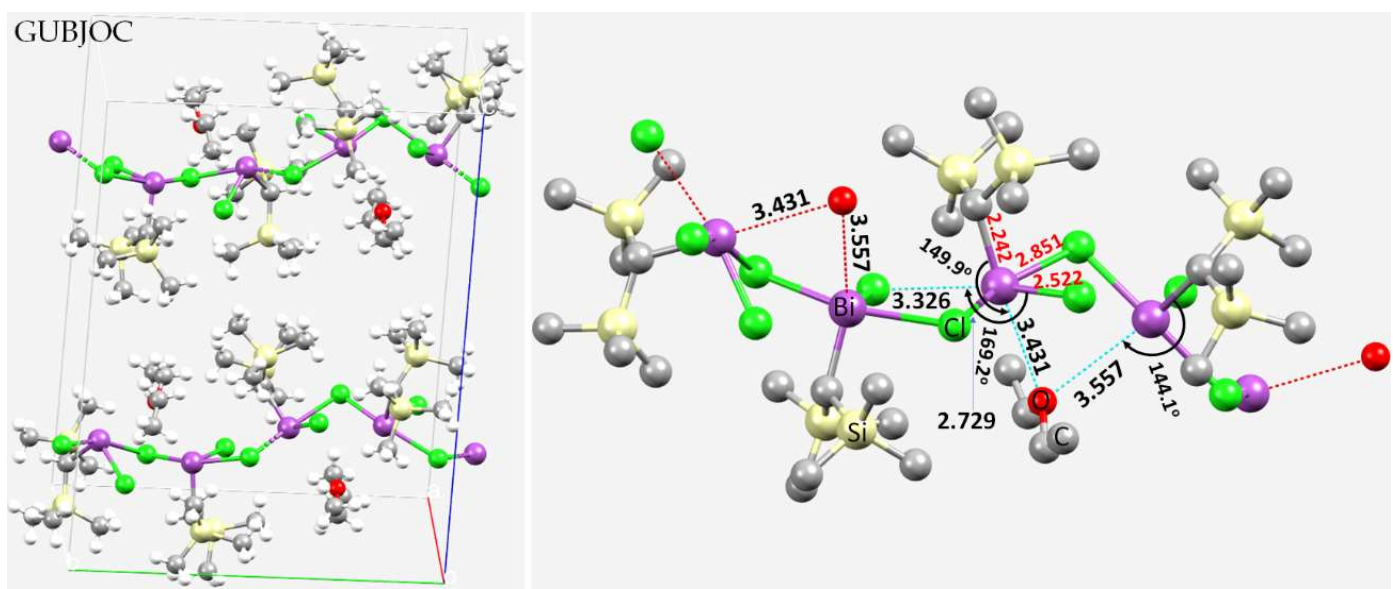

**Figure 17.** a) Ball-and-stick model of the crystal structure of [[RBi(Cl)Cl]$_2$·Et$_2$O]$_n$ (R = CH(Si(CH$_3$)$_2$) [157]; b) Illustration of pnictogen bonding in the [[RBi(Cl)Cl]$_2$·Et$_2$O]$_n$ chain (H atoms are omitted for clarity). Selected bond distances and bond angles are shown in Å and degree, respectively. Specific atom type is labeled.

An example of a chemical system where Bi in a molecule is involved in making both intra- and inter-molecular interactions is shown in Fig. 18a; it was observed in the crystal structure of Bi(C$_6$F$_5$)$_3$ [158]. Each Bi$^{3+}$ center is linked to three C$_6$F$_5^-$ anions, forming three Bi–C covalent coordinate bonds ($r$(Bi–C) in the range 2.245 – 2.290 Å). The arrangement between the three C$_6$F$_5^-$ anions around each Bi center is such that it results in the formation of two additional Bi⋯F interactions between the positive sites on Bi along the C–Bi bond extensions and the negative sites on F. They are non-linear intramolecular pnictogen bonds. Each O site of a 1,4-dioxane makes a Bi⋯O with a Bi center. The Bi⋯O distances are equivalent and quasi-linear. These bonding features constitute the pseudo-octahedral coordination of the Bi ion in the crystal, as seen in the isosurface plot of Fig. 18b. Given the size and thickness of the isosurfaces between the bonded atomic basins, it may be

concluded that the quasi-linear Bi⋯O pnictogen bond is stronger than the non-linear Bi⋯F pnictogen bonds.

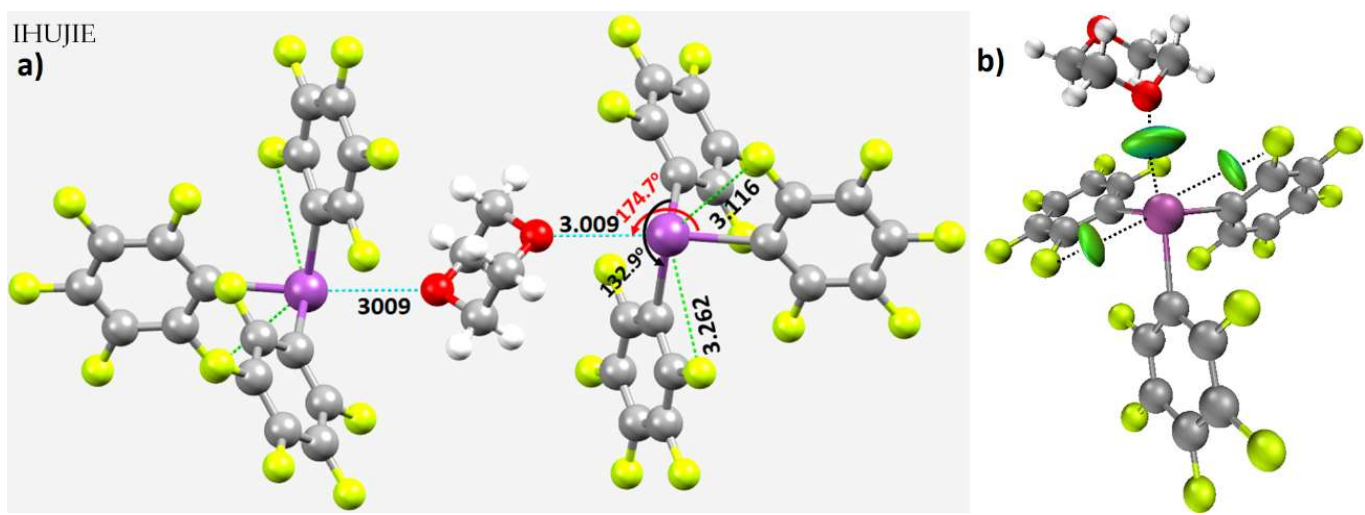

**Figure 18.** a) Nature of Bi⋯O and Bi⋯F pnictogen bonding in the crystal of tris(perfluorophenyl)bismuthine 1,4-dioxane solvate (CSD ref. code IHUJIE) [158]. Bond lengths and bond angles in Å and degrees, respectively. b) The IGM-$\delta g$ based isosurface plot showing the presence of such interactions. Selected bond distance and bond angles in a) are in Å and degree, respectively. CSD ref code is marked in uppercase letters in a).

Breunig and Althaus reported a number of alkylantimony(III) and alkylbismuth(III) halides, $R_2EX$ and $REX_2$, (E = Sb, Bi; X = Cl, Br, I; R = $CH_3$, $(Me_3Si)_2CH$) and demonstrated that they form oligomeric or polymeric structures through halogen bridges between the pnictogen atoms [159]. $Bi^{3+}$ in these compounds is genuinely five-coordinate. In the complex with $CH_3BiBr_2$ and with THF solvate in the crystal structure (Fig. 19a), in addition to binding to the methyl and bromide ligands, $Bi^{3+}$ is coordinated by lone pairs on oxygen from two THF solvent molecules.

The possibility of pnictogen bonding in either of the $R_2EX$ and $REX_2$ systems was not discussed in the original study [159]. However, the authors suggested the existence of halogen bridges between the covalently bound pnictogen atoms and the halide. To verify this, we performed an IGM-$\delta g$ analysis on a dimer $CH_3BiBr_2 \cdot 2THF$. The result, shown in Fig. 19b, suggests that one monomer is linked to the other through a Type-IIa ($H_3$)C–Bi⋯Br(Bi) pnictogen bond, but this has nothing to do with halogen bridges. The presence of the former interaction type is evidenced by the greenish isosurface between the Br and Bi atomic basins. We classify them as pnictogen bridges; there are two such equivalent pnictogen bridges, each with $r$(Bi⋯Br) = 3.979 Å and ∠($CH_3$)C–Bi⋯Br = 167.5°, and are markedly longer than the Bi–O ( = 2.60 Å), Bi–C ( = 2.25 Å) and Bi–Br (= 2.70 Å) coordinate bonds of the complex.

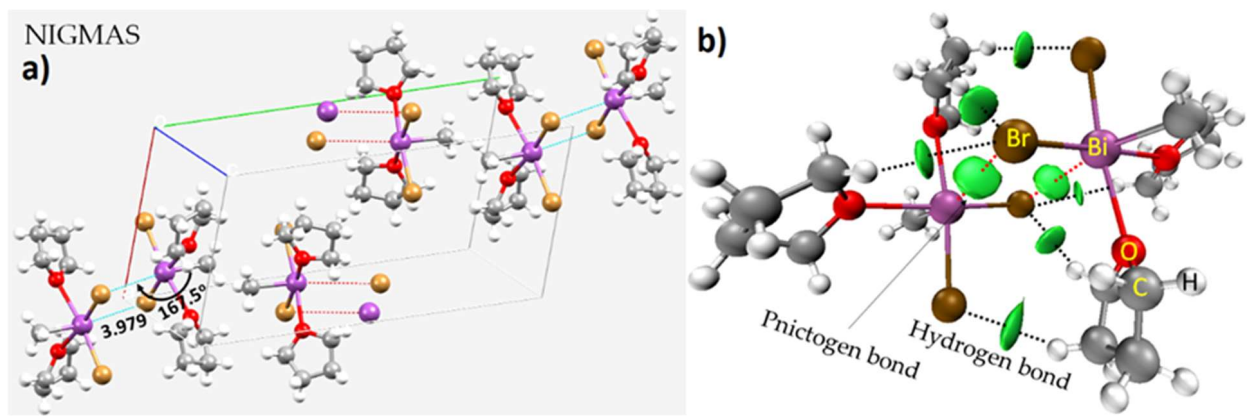

**Figure 19.** a) The crystal structure of CH$_3$BiBr$_2$.THF, CSD ref. code NIGMAS [159]. b) The IGM-δ$g$ based isosurface topologies (isovalue = 0.01 a.u.) of bonding observed between bonded atomic basins in the dimer [CH$_3$BiBr$_2$.2THF]$_2$. Selected bond lengths and bond angles are in Å and degrees, respectively. Non-covalent interactions are shown as dotted lines.

The formation of these pnictogen bridges is not very surprising given that the Bi site along the (H$_3$)C–Bi bond in a molecular entity features a positive σ-hole and is engaged with the lateral negative portion of the covalently bound Br atom in a neighboring molecular entity. The (H$_3$)C–Bi⋯Br(Bi) pnictogen bonds are somewhat deviated from linearity, which is understandable since Br in each molecular entity also participates in Br⋯H(CH) hydrogen bonded with THF (see Fig. 19b). Clearly, the packing between the molecules in the crystal is a consequence of the joint involvement of pnictogen bonds and hydrogen bonds. It is worth nothing that Bi is four-coordinate in the sterically crowded crystal of (Me$_3$Si)$_2$CH) BiCl$_2$.THF (CSD ref: DABWOT) [159]. The Bi center in this system was bonded to the Cl site in a neighboring molecule in the crystal, forming two equivalent Bi⋯Cl long bonds ($r$(Bi⋯Cl) = 3.163 Å; Cl–Bi⋯Cl = 160.4°) that are characteristic of Type-IIa pnictogen bonds.

Shown in Fig. 20a-h are a set of crystals in which covalently or coordinately bound Bi is in attractive engagement with donors such as O, N, Cl, S, and I in interacting partner spices. Except for the crystals 2(C$_{10}$H$_{12}$S$_8^+$)(Bi$_3$Cl$_{11}^{2-}$) [160] and (C$_{12}$H$_{10}$BiI$_2^-$)(C$_8$H$_{20}$N$^+$) [161] shown in Fig. 20d and 20g, respectively, the Bi center in the remaining crystals is positive. Although a Type-IIa bonding featuring Bi is occurs in all eight crystals of Fig. 20, this is not the case in the structures shown in Fig. 20d and 20g because Bi is entirely negative in these two crystals. This means that the appearance and subsequent stability of the Bi⋯Cl and Bi⋯I directional interactions in these two crystals are driven by the cation, and hence may characterized as Type-III interactions. On the other hand, the Bi⋯I/Bi⋯C$_\pi$, Bi⋯O, Bi⋯N/C≡N, Bi⋯Se, Bi⋯π(C$_6$) and Bi⋯Cl in the crystals shown in Fig. 20a, Fig. 20b, Fig. 20c, Fig. 20e, Fig. 20f and Fig. 20h, respectively, have the characteristics of pnictogen bonds since they appear along R–Bi bond extensions, and are less than the sum of the vdW radii of the respective interacting atomic basins.

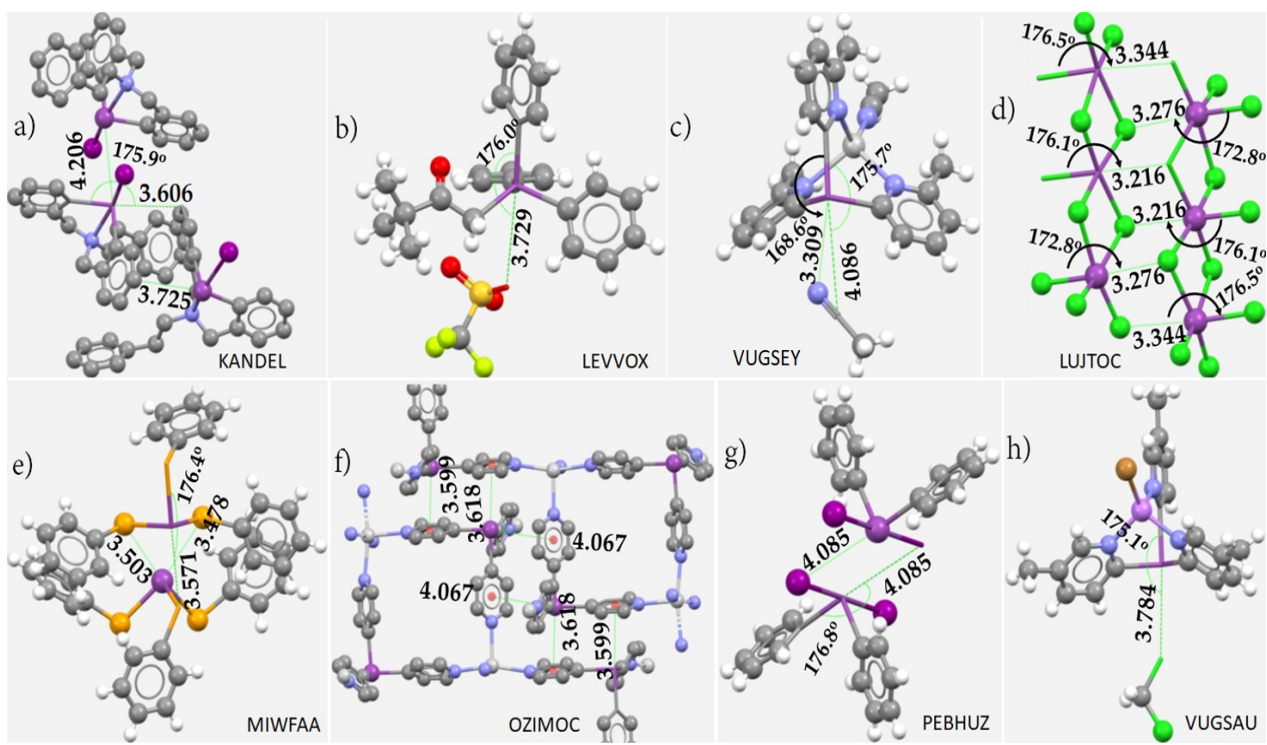

**Figure 20.** Crystal structures showing close contacts of covalently or coordinately bound Bi with O, N, Cl, S, $C_\pi$ and I sites in interacting partner spices: a) (2,2'-[[(2-phenylethyl)azanediyl]bis(methylene)]di(benzen-1-yl))-iodo-bismuth(III) ($C_{22}H_{21}BiIN$) [100]; b) (3,3-dimethyl-2-oxobutyl)-triphenyl-bismuth trifluoromethanesulfonate ($C_{24}H_{26}BiO^+$)($CF_3O_3S^-$) [162]; c) acetonitrile-[2,2',2''-bismuthanetriyltris(6-methylpyridine)]-silver trifluoromethanesulfonate acetonitrile solvate ($C_{20}H_{21}AgBiN_4^+$)($CF_3O_3S^-$)($C_2H_3N$) [134]; d) bis(tetrakis(methylthio)tetrathiafulvalene radical cation) tetrakis(μ-chloro)-heptachloro-tri-bismuth(III) 2($C_{10}H_{12}S_8^+$),($Bi_3Cl_{11}^{2-}$) [160]; e) tris(phenylselenolato)-bismuth ($C_{18}H_{15}BiSe_3$) [163]; f) catena-[nonakis(m-pyridin-4-yl)-tri-bismuth(III)-di-silver(I) bis(hexafluoroantimonate) [($C_{45}H_{36}Ag_2Bi_3N_9^{2+}$)$_n$,2($SbF_6^-$)] [164]; g) tetraethylammonium di-iodo-diphenyl-bismuth ($C_{12}H_{10}BiI_2^-$)($C_8H_{20}N^+$) [161]; h) [2,2',2''-bismuthanetriyltris(5-methylpyridine)]-bromido-lithium dichloromethane solvate ($C_{18}H_{18}BiBrLiN_3$)($CH_2Cl_2$) [134]. Selected bond lengths and bond angles are in Å and degrees, respectively. Non-covalent interactions are shown as dotted lines. H atoms in a) are omitted for clarity. The tiny dot at the center of the aromatic ring in f) represents the centroid.

In the crystals shown in Fig. 21a-e, the covalently or coordinately bound Bi is positive along the R–Bi bond extensions, except for the building block $Bi_4I_{16}^{4-}$ in ($C_{24}H_{20}Bi^+$)$_4$($Bi_4I_{16}^{4-}$) [114]. The ∠R–Bi⋯D (D = N, O, Cl, and $C_\pi$) between the interacting units in these crystals lie between 175° and 180°, regardless of the donors concerned. In all cases, the Bi⋯D pnictogen bond distances are less than the sum of the vdW radii of the respective atomic basins. They undoubtedly follow a Type-IIa topology of bonding (Scheme 1). In case of (2-((dimethylamino)methyl)phenyl)-diphenyl-bismuth ($C_{21}H_{22}BiN$), (Fig. 21e) [165], the N of the dimethylamino group pnictogen bonds with Bi forming a Bi⋯N intra-molecular contact ($r$(Bi⋯N) = 2.851 Å; ∠C–Bi⋯N = 161.9°) (not marked), which is markedly longer than Bi–C coordinate bonds ($r$(Bi–C) between 2.255–2.295 Å) but significantly shorter than the Bi⋯$C_\pi$ pnictogen bond ($r$(Bi⋯$C_\pi$) = 3.859 Å).

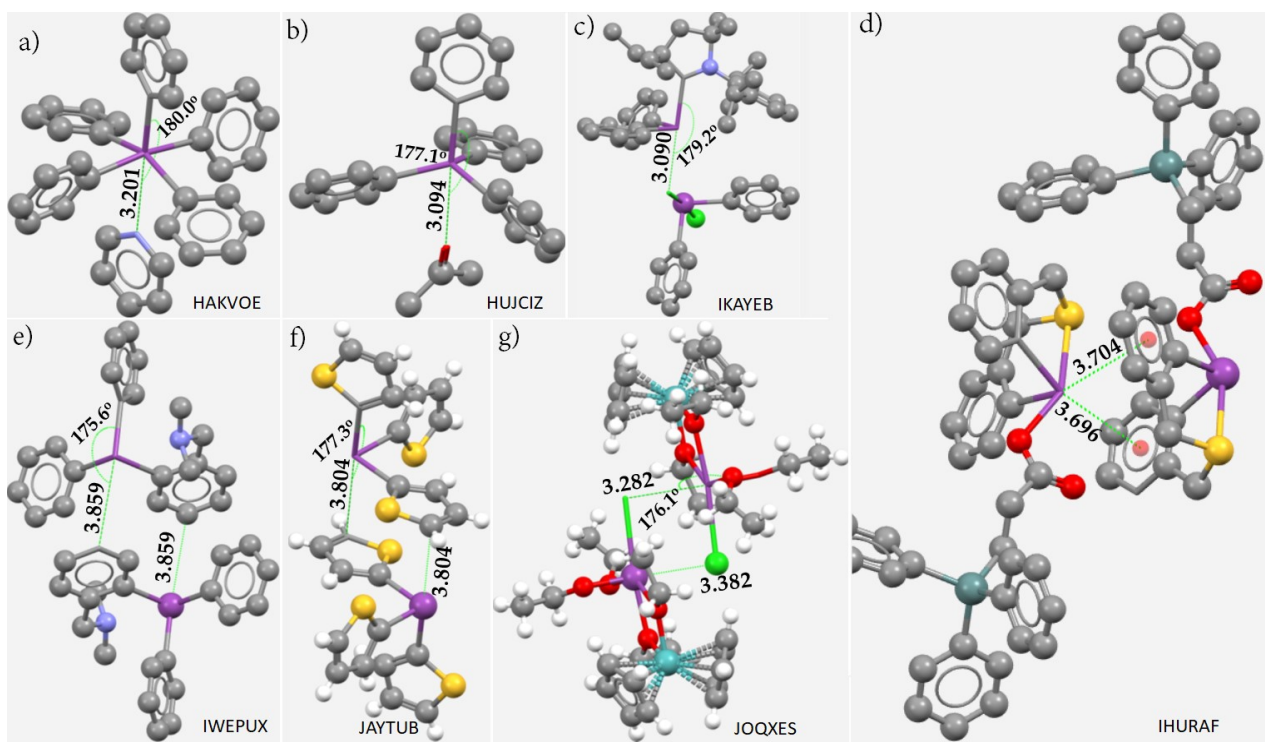

**Figure 21.** Some crystal structures showing close contacts of covalently or coordinately bound Bi with O, N, Cl, and C$_\pi$ sites in interacting partner species. a) pentaphenyl-bismuth pyridine solvate, C$_{30}$H$_{25}$Bi)C$_5$H$_5$N) [166]; b) tetrakis(tetraphenyl-bismuth) bis(μ$_3$-iodo)-tetrakis(μ$_2$-iodo)-decaiodo-tetra-bismuth acetone solvate (C$_{24}$H$_{20}$Bi$^+$)$_4$(Bi$_4$I$_{16}$$^{4-}$)2(C$_3$H$_6$O)] [114]; c) (1-[2,6-bis(propan-2-yl)phenyl]-3,3-diethyl-5,5-dimethylpyrrolidin-2-ylidene)-diphenyl-bismuth(iii) dichloro-diphenyl-bismuth(III) (C$_{34}$H$_{45}$BiN$^+$)(C$_{12}$H$_{10}$BiCl$_2$$^-$) [167]; d) (bis(o-phenylenemethylene)thioxy)-(3-triphenyl-germylpropionato)-bismuth(III) [C$_{35}$H$_{31}$BiGeO$_2$S] [168]; e) (2-((dimethylamino)methyl)phenyl)-diphenyl-bismuth (C$_{21}$H$_{22}$BiN) [165]; f) tris(thiophen-2-yl)bismuthane (C$_{12}$H$_9$BiS$_3$) [169]; g) bis(μ$_2$-ethoxo)-chloro-bis(η$^5$-cyclopentadienyl)-bis(ethanolato)-bismuth-molybdenum ethanol solvate (C$_{18}$H$_{30}$BiClMoO$_4$)0.25(C$_2$H$_6$O) [170]. Selected bond lengths and bond angles are in Å and degrees, respectively. Non-covalent interactions are shown as dotted lines. H atoms in a)-e) and (C$_2$H$_6$O) in g) are omitted for clarity. The tiny dot at the center of the aromatic ring in d) represents the centroid.

As shown in several cases above, pnictogen derivatives in molecules can form neutral or anionic square pyramidal structures. Shown in Fig. 22 is another such instance, in which the Bi center in the 1 × 2 × 1 extended crystal structure of [CH$_3$BiCl$_2$(bipy)] [171] adopts a square-pyramidal geometry. This crystal system provides evidence of pnictogen bonding between the covalently/coordinately bound Bi$^{3+}$ in one molecule and coordinated Cl$^-$ in a neighboring molecule. The C–Bi⋯Cl pnictogen bonding interaction is at 3.638 Å with ∠C–Bi⋯Cl = 158.9°, displaying therefore a Type-IIa bonding topology. The stability of the crystal arises not only from pnictogen bonding, but also from significant π⋯π stacking interactions existing between the aromatic rings of the bipy units and from Cl⋯H(bipy) hydrogen bonds (not shown).

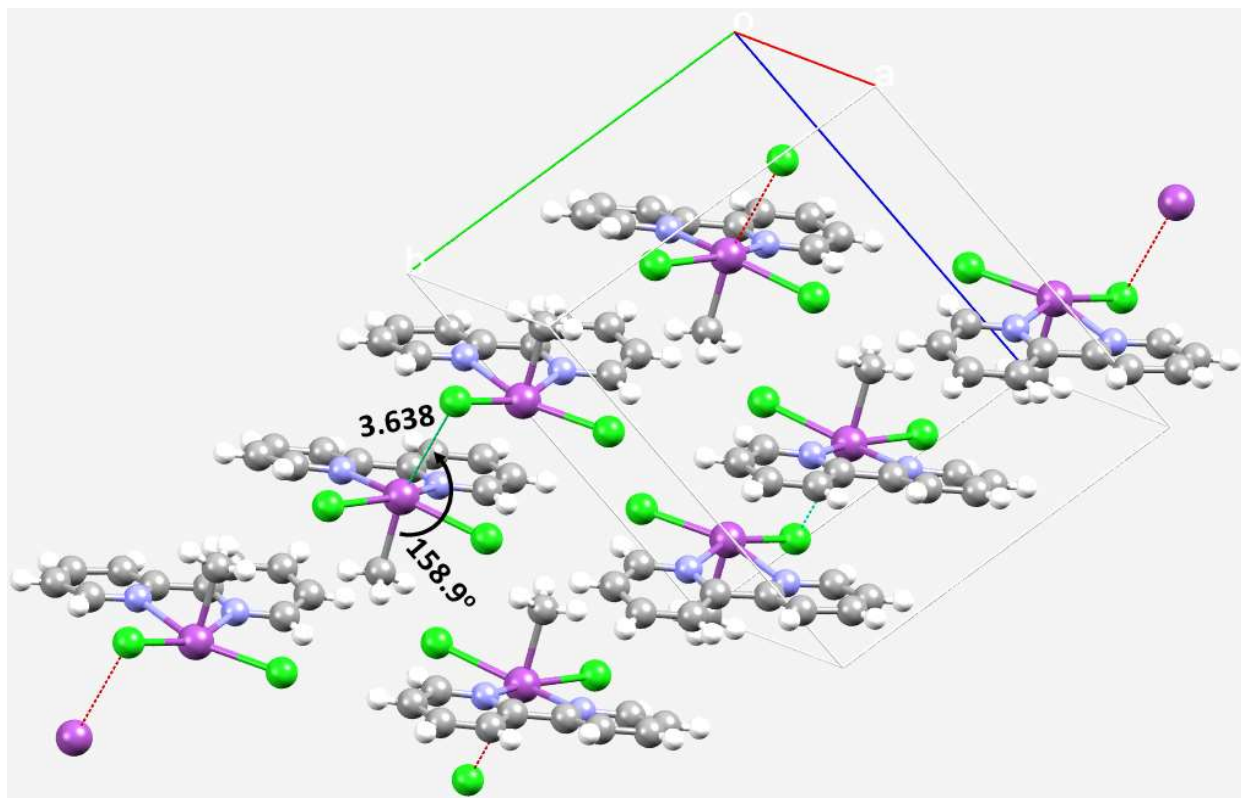

**Figure 22.** The ball-and-stick model of the crystal structure of [CH$_3$BiCl$_2$(bipy)] (bipy = 2,2′-bipyridyl), CSD ref. code EDIGED, space group *P*–1 [171]. Selected bond lengths and bond angles are in Å and degrees, respectively. Atoms: Bi – purple; Cl – green; C – gray; N – blue; H – white.

*d. Bi⋯C$_\pi$(arene/acetylene) Pnictogen/Bismuth Bonds in Crystals*

Our CSD searches revealed many crystals that display Bi⋯C$_\pi$(arene) interactions as the primary driving forces for assembly of building blocks. Shown in Fig. 23 includes a few of them. In each of them the covalently or coordinately bonded Bi is engaged in an attractive interaction with the π-density of six-membered aromatic rings. The intermolecular distances are shown between Bi in one molecular entity and the centroid of the C$_6$ moiety in the moiety with which it is interacting attractively. Except for the case of the structure (NH$_4$)$_2$[Bi$_2$(C$_6$H$_4$O$_2$)$_4$] (Fig. 23c, CSD ref. YEGDUJ [172]), the intermolecular interactions are quasi-linear, and they are no other than bismuth centered Type-IIa pnictogen bonds (π-density as donor). In case of the crystal structure of [(NH$_4^+$)$_2$(C$_{24}$H$_{16}$Bi$_2$O$_8^{2-}$)(C$_6$H$_6$O$_2$)$_2$]2(H$_4$N$^+$)·2(H$_2$O)] (YEGDUJ), the Bi center is entirely negative. The Bi⋯C$_\pi$(arene) interactions between C$_{24}$H$_{16}$Bi$_2$O$_8^{2-}$ units are driven by H$_2$O and NH$_4^+$ that bring the anions in close proximity through hydrogen bonding. Because the Bi⋯C$_\pi$(arene) interaction is formed between negative sites, and considering the directional nature of the interaction, we regard this as Type-III.

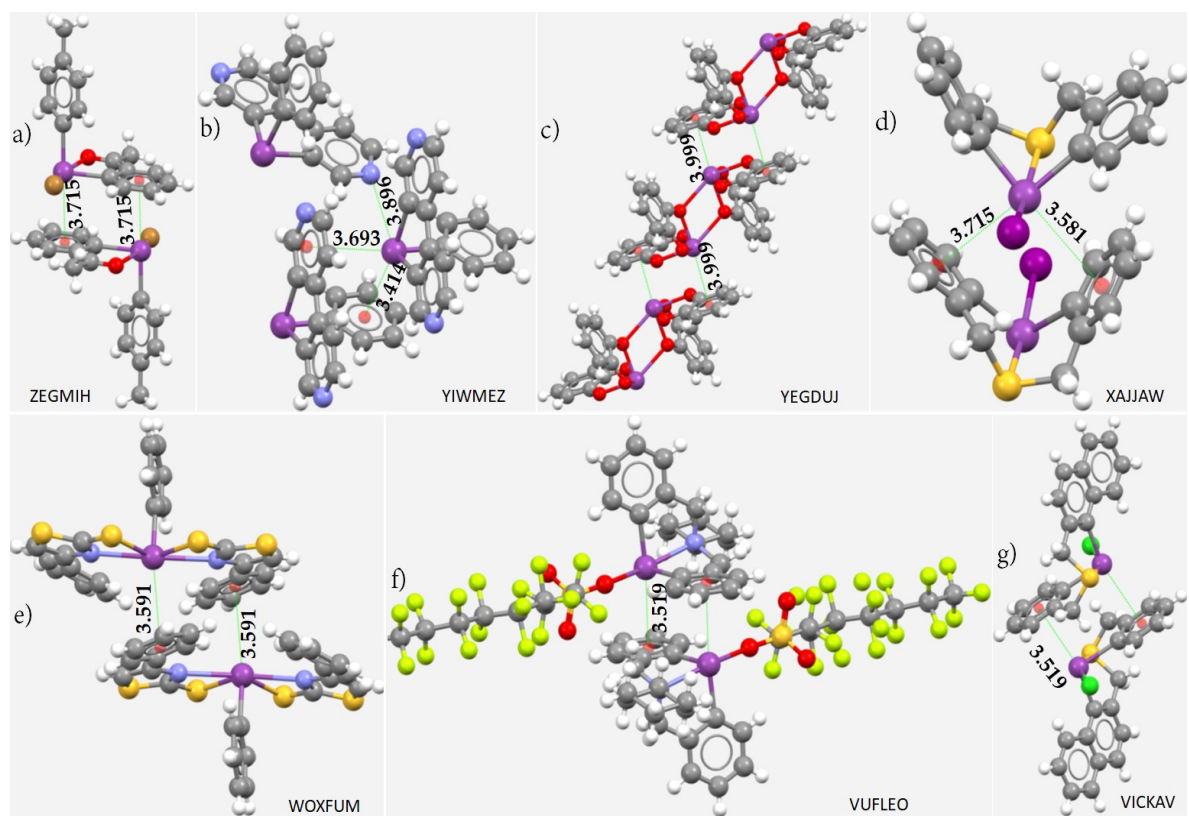

**Figure 23.** Examples of Bi⋯π close contacts in some selected crystal systems. a) (2-acetylphenyl-C,O)-bromo-(4-methylphenyl)-bismuth ($C_{15}H_{14}BiBrO$) [173]; b) 9-phenyl-9H-bismolo[2,3-c:5,4-c']dipyridine [$C_{16}H_{11}BiN_2$] [174]; c) ammonium bis(μ2-catecholato-O,O,O')-bis(catecholato-O,O')-di-bismuth(III) catecholate dihydrate [$(NH_4^+)_2(C_{24}H_{16}Bi_2O_8^{2-})(C_6H_6O_2)_2]2(H_4N^+)·2(H_2O)$] [172]; d) (2,2'-[sulfanediylbis(methylene)]di(phenyl))-iodo-bismuth(III) [$C_{14}H_{12}BiIS$] [175]; e) bis(4-phenyl-1,3-thiazole-2-thiolato)-phenyl-bismuth(III) [$C_{24}H_{17}BiN_2S_4$] [176]; f) 6-Cyclohexyl-6,7-dihydrodibenzo[c,f][1,5]azabismocin-12(5H)-yl-heptadecafluorooctane-1-sulfonate [$C_{28}H_{23}BiF_{17}NO_3S$] [177]; g) (2-([(benzen-2-idylmethyl)sulfanyl]methyl)naphthalen-1-ide)-chloro-bismuth(III) [$C_{18}H_{14}BiClS$] [178]. Bond distances are in Å, and the centroids of some selected Bi-bonded aromatic moieties are represented by tiny spheres in red. The CSD ref. code is shown for each case. Non-covalent interactions are shown as dotted lines in green. Molecular entities $C_6H_6O_2$, $H_4N^+$ and $H_2O$ in c) are omitted for clarity.

In the crystal structure of 10-(4'-chlorophenylethynyl)phenothiabismin-5,5-dioxide ($C_{20}H_{12}BiClO_2S$), CSD ref. JOXTAR [179], the tricoordinate Bi center in $C_{20}H_{12}BiClO_2S$ is non-covalently bonded with the mid-point of the C≡C bond of another similar unit in the crystal. This is a good example of Bi⋯$C_π$(acetylenic) bismuth bonds that hold the building blocks together in the crystal. The interaction is directional (∠C–Bi⋯(C≡C)$_π$ = 167.1°), and $r$(Bi⋯(C≡C)$_π$) is significantly shorter than the sum of the vdW radii of Bi and C, 4.31 Å ($r$(Bi⋯(C≡C)$_π$) = 3.548 Å). Similar intermolecular close contacts were observed in other crystal lattices such as (N-t-butyl-N,N-bis((o-phenylene)methyl)amine)-(phenylethynyl)-bismuth (CSD ref. ABONAH [180]), 2-(phenylethynyl)-1,3-bis(trimethylsilyl)-2,3-dihydro-1H-naphtho[1,8-de][1,3,2]diazabismine (CSD ref. KIKRON [181]), and (2-(1-(t-butylamido)-2-(methoxycarbonyl)prop-2-ene-1,3-diyl)-6-(t-butyliminomethyl)phenyl)-(2-(methoxycarbonyl)ethynyl)-bismuth (CSD ref. SUCJUY) [182]. ∠C/N–Bi⋯(C≡C)$_π$, and $r$(Bi⋯(C≡C)$_π$) in the corresponding systems are 173.4° (4.362 Å), 166.0° (3.819 Å), and 155.1° (3.654 Å), respectively.

*e. Type-III Pnictogen Bonds in Crystals*

A number of cases referred to above contained examples of Type-III interactions (sometimes called counter-intuitive interactions [78,183,184] in crystal lattices). In Fig. 24, another set of examples are given where covalently or coordinately bound Bi acts both as an acceptor and a donor of charge density when in close proximity. In all cases, the Bi···Bi close contact distance is significantly less than twice the vdW radius of Bi, 5.08 Å. The contact angle ∠R–Bi···Bi varies between 171° and 176°. The interacting surface regions on Bi are both negative in the crystal structures shown in Fig. 24a-b, and both positive in the crystals shown in Fig. 24c-i, but have unequal charge densities. A weakly positive (or negative) region on Bi located off the R–Bi bond axis in a molecular entity is in attractive engagement with a more strongly positive (or negative) region on Bi lying along the outer extension of the R–Bi bond axis in the molecule with which it interacts. This is a signature of a Type-III interaction (Scheme 1c). In systems such as [2,6-bis(4,4-dimethyl-4,5-dihydro-1,3-oxazol-2-yl)phenyl]- bismuth (CSD ref: IBEBUQ [185]), the Bi···Bi interaction is significantly non-linear (r(Bi···Bi) = 3.857 Å; ∠R–Bi···Bi = 162.6°), but is still representative of a Type-III interaction.

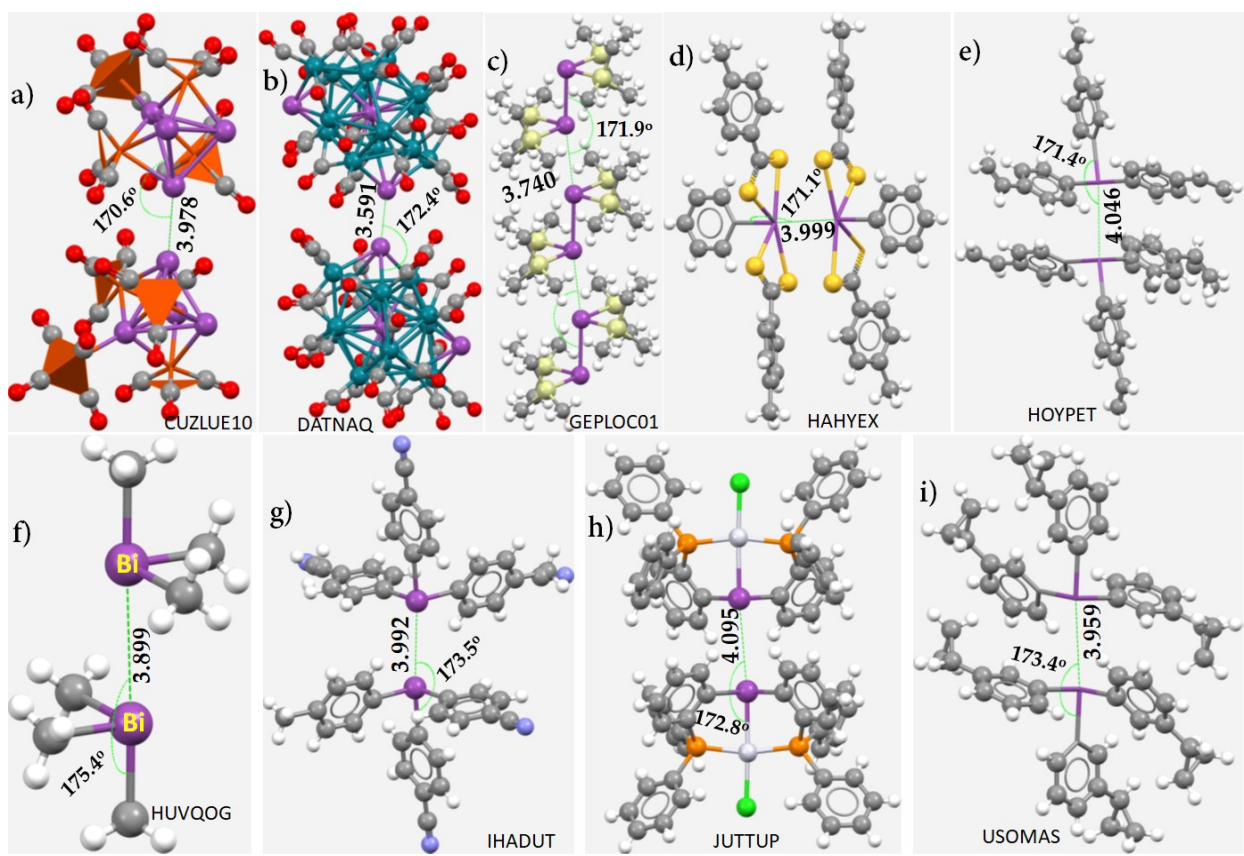

**Figure 24.** a) bis(tetraethylammonium) tris(μ3-tricarbonyl-iron)-(tetracarbonyl-iron)-tetra-bismuth $(C_8H_{20}N^+)_2(C_{13}Bi_4Fe_4O_{13}^{2-})$ [186]; b) tris(tetramethylammonium) pentadecakis(μ-carbonyl)-dodecacarbonyl-tri-bismuth-tetradeca-rhodium acetonitrile solvate $(C_4H_{12}N^+)_3(C_{27}Bi_3O_{27}Rh_{14}^{3-})\cdot3(C_2H_3N)$ [187]; c) 2-[bis(trimethylsilyl)-bismuthanyl]-1,1,1,3,3,3-hexamethyldisilabismane $(C_{12}H_{36}Bi_2Si_4)$ [188]; d) bis(4-methylbenzene-1-carbodithioato)-phenyl-bismuth $(C_{22}H_{19}BiS_4)$ [189]; e) tris(4-ethenylphenyl)-bismuth $(C_{24}H_{21}Bi)$ [190]; f) trimethylbismuthine $(C_3H_9Bi)$ [130]; g) 4,4'-[(4-methylphenyl)-bismuthanediyl]dibenzonitrile $(C_{21}H_{15}BiN_2)$ [191]; h) bis(μ2-2-(diphenylphosphino)phenyl-C,P)-chloro-bismuth-platinum dichloromethane solvate $(C_{36}H_{28}BiClP_2Pt)\cdot CH_2Cl_2$ [192]; i) tris(3-cyclopropylphenyl)-bismuth $(C_{27}H_{27}Bi)$ [193]. Selected bond distances and bond angles are in Å and degree, respectively. The CSD ref. code in uppercase letters is shown for each case. Non-covalent interactions are shown as dotted lines in green. Molecular entities viz. $C_8H_{20}N^+$ in a), $C_4H_{12}N^+$ and $C_2H_3N$ in b) are omitted for clarity.

To confirm that a Bi⋯Bi Type-III bond exists between a pair of molecular entities shown in Fig. 24, we examined the MESP plot of trimethylbismuthine, shown in Fig. 25a. This was obtained on the fully relaxed monomer geometry of the system with ωB97XD/def2-TZVPPD. As can be seen from Fig. 25a-b(left), there are three σ-holes on Bi along the C–Bi bond extensions, each associated with a $V_{S,max}$ = 16.2 kcal mol$^{-1}$. The lateral side of Bi, which is facing the reader, is very negative, with $V_{S,min}$ = −6.3 kcal mol$^{-1}$. The carbons of the methyl groups attached to Bi are all negative and H atoms are all positive along the Bi–C and C–H bond extensions, respectively. The opposite side of Bi (Fig. 25a-b, right), also shown facing the reader, is positive. Clearly, Bi in Bi(CH$_3$)$_3$ has both positive and negative regions, and can react with another identical or different system both as an acid and a base. The quasilinear Bi⋯Bi bonding that exists between the Bi atoms in Fig. 24f is a result of attraction between two positive sites of unequal charge density located along and slightly off the C–Bi bond axes in Bi(CH$_3$)$_3$. This is readily appreciated by looking at the space-filling model of the (Bi(CH$_3$)$_3$)$_2$ dimer arrangement (Fig. 25c) extracted from the crystal of the system shown in Fig. 25d. It is also possible that there are some attractive interactions existing between Bi and C domains that might simultaneously cause the formation of weakly bound Bi⋯C pnictogen bonds.

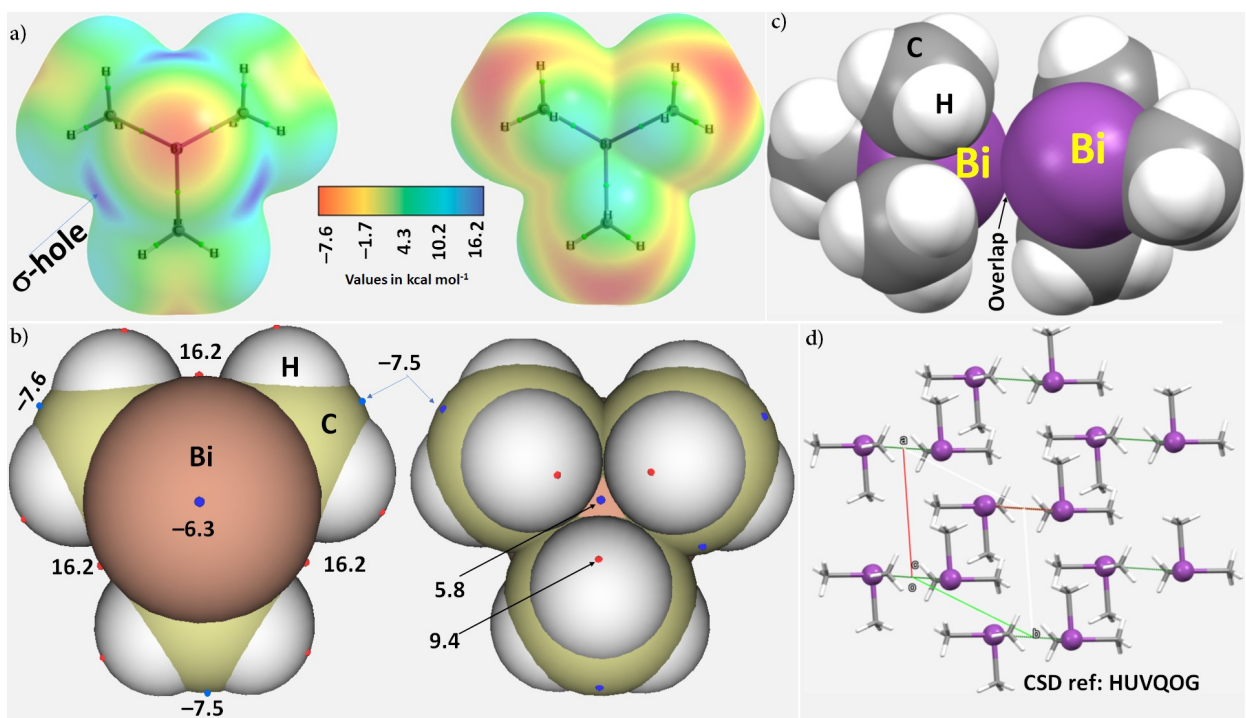

**Figure 25.** a) Two views of the 0.001 a.u. isoelectron density envelope mapped potential on the surface of the Bi(CH$_3$)$_3$ molecule, obtained using its ωB97XD/def2-TZVPPD geometry at the same level: (left) Bi facing the reader; (right) Opposite site of Bi facing the reader. b) Two views of the van der Waals surface of the corresponding same system, with the minimum and maximum of potential as tiny circles in blue and green, respectively. c) The space-filling model representing the extent of overlap between the two interacting Bi atoms of two nearest Bi(CH$_3$)$_3$ molecules in the crystal. d) The 2 × 2 × 2 representation of the Bi(CH$_3$)$_3$ crystal, with the Bi⋯Bi links as dots in cyan. The CSD ref. code is shown in upper case letters in d).

*f. Intramolecular pnictogen bonds*

Shown in Fig. 26 are a set of three flexible accordion-like molecular complexes of the composition [P($C_6H_4$-*o*-$CH_2SCH_3$)$_3$Bi$X_3$], (X = Cl, Br, I) [194]. They feature attraction between the positive electrostatic potential on the Bi center and four negative sites in the phosphine thioether. The $Bi^{3+}$ center is pseudo-octahedral, with three Bi–X and three Bi–S bonds; these are formally polar covalent/ionic bonds. There is an interaction between $Bi^{3+}$ and the P atom in (P($C_6H_4$-*o*-$CH_2SCH_3$)$_3$). The distance between the bridgehead P atoms and Bi in these systems increases from 3.365(1) Å in Fig. 26a through 3.759(7) Å in Fig. 26b and 3.792(9) Å in Fig. 26c, indicating that there is a systematic increase in the Bi⋯P close contact distance as the halogen derivative in Bi$X_3$ changes from Cl to Br to I. Even though there is no strong directional feature in the Bi⋯P pnictogen bonds (∠X–Bi⋯P values range between 121.4° and 132.0°), these are Type-IIb intramolecular contacts (Scheme 1).

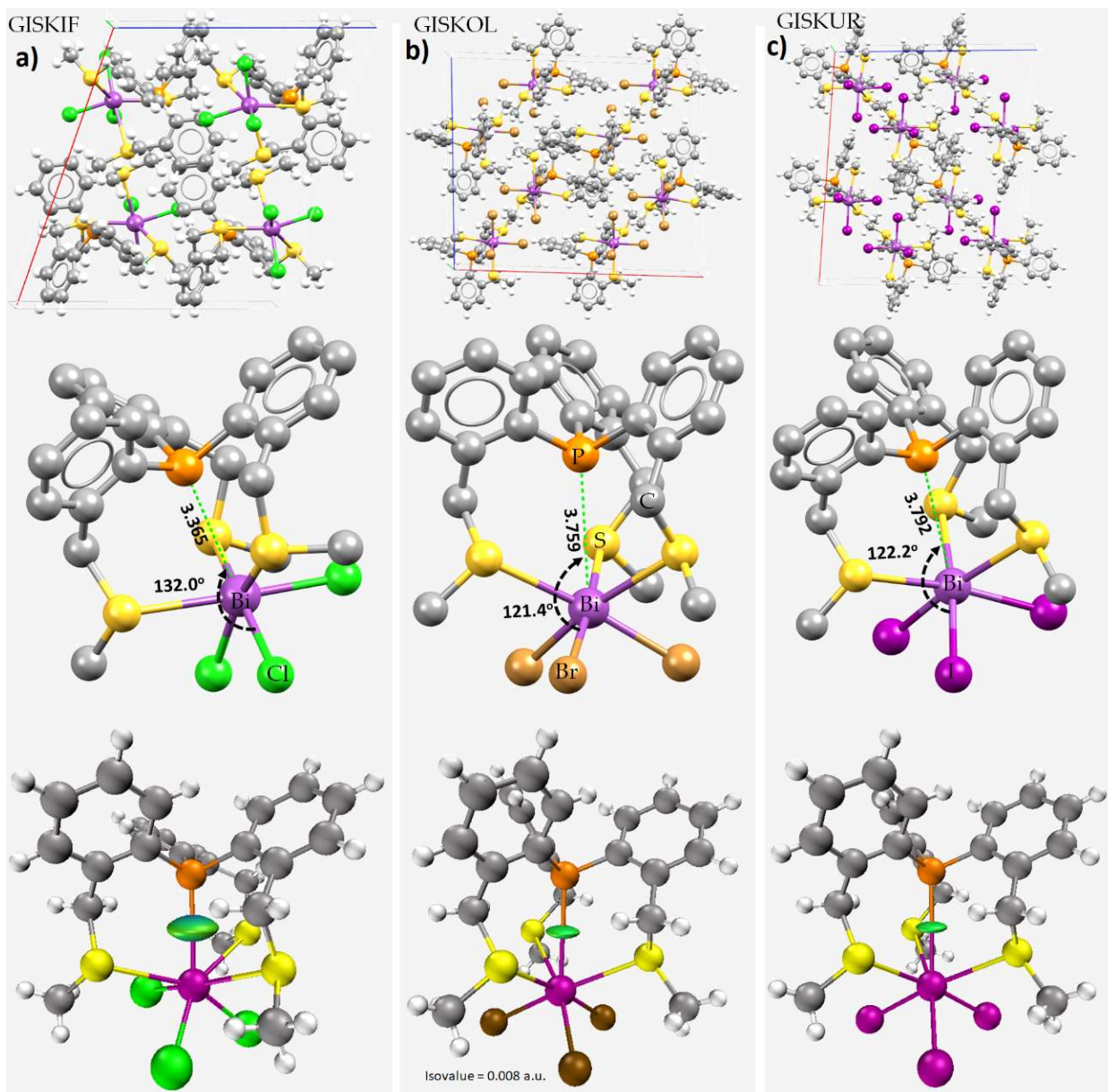

**Figure 26.** Top: Unit-cells of crystals of complexes of Bi$X_3$ with the chelating phosphane, tris-[2-[(methylsulfanyl)methyl]phenyl]phosphane [194]. a) X = Cl; b) X = Br; c) X = I. Middle: Illustration of Bi⋯P pnictogen bonds as dotted lines in green for the corresponding systems, respectively, with bond distances and bond angles in Å and degree, respectively. Bottom: Illustration of the IGM-$\delta g$ based isosurface plots for each of the three systems. CSD ref codes are shown in upper-case letters.

Since the ³¹P NMR resonance in solution and in the solid state lies at a lower frequency than in the case of the free ligand, and there is ²⁰⁹Bi-³¹P coupling, this provides experimental evidence of the pnictogen interaction between the Bi and P centers [194]. An NBO and QTAIM analysis corroborated the experimental evidence for Bi···P pnictogen bonds in the three systems, calculated to be 8.8 kcal mol⁻¹ when X = Cl; 8.0 kcal mol⁻¹ when X = Br; and 7.1 kcal mol⁻¹ when X = I. These results are in good agreement with our IGM-$\delta g$ analysis that show relatively less depletion of charge density (a larger isosurface) in the Bi···P bonding region when X = Cl compared to when X = Br and X = I. This is also in agreement with the trend in the charge density at the Bi···P bond critical points (0.0131 a.u. where X = Cl; 0.0121 a.u. when X = Br; and 0.0103 a.u. when X = I).

The reaction of $(C_6F_5)_2GeH_2$ with $BiEt_3$ produces $[(C_6F_5)_2Ge]_3Bi_2$ [195]. It features a trigonal bipyramid in which the two apical Bi atoms are linked covalently with three bis(pentafluorophenyl)germyl bridges. The molecular framework has $D_{3h}$ symmetry, and is shown in Fig. 27a.

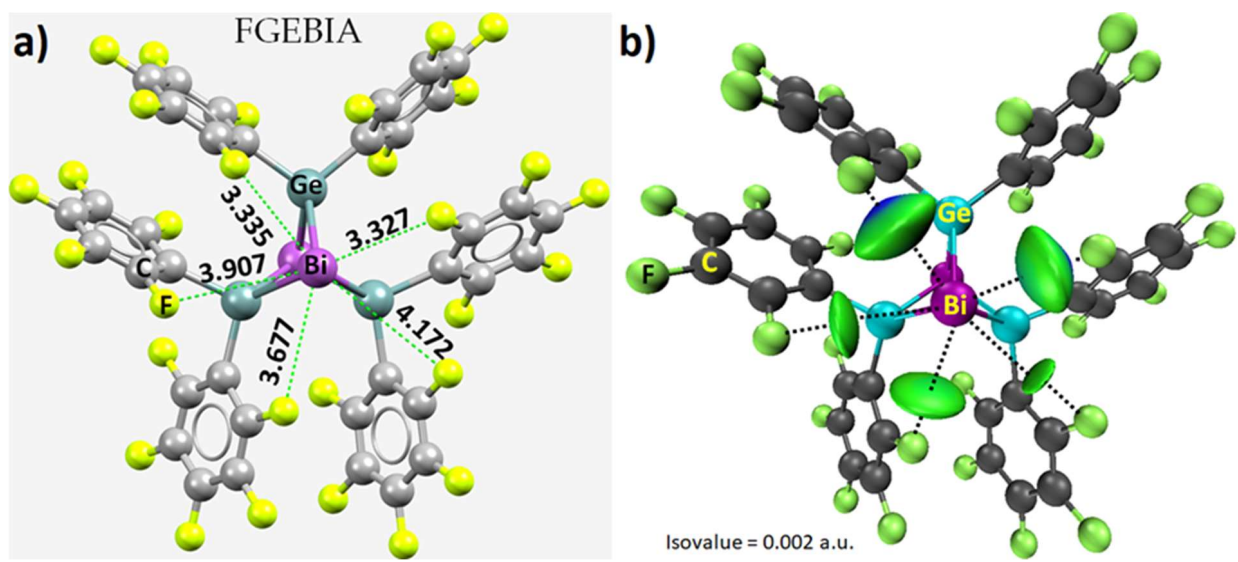

**Figure 27.** a) The tris($\mu_2$-bis(pentafluorophenyl)-germyl)-di-bismuth molecule, $[(C_6F_5)_2Ge]_3Bi_2$, found in the a heteroelemental crystalline network (CSD ref. FGEBIA), showing a number of intramolecular Bi···F interactions. b) The IGM-$\delta g$ based isosurface plot. Selected bond lengths are in Å. Non-covalent interactions are shown as dotted lines between bonded atomic basins.

The average value of the Ge–Bi bond length is 2.739(1) Å and the Bi···Bi distance is 4.005 Å, with the latter substantially smaller than twice the vdW radius of a Bi atom, 5.08 Å. There are additional intramolecular interactions within the framework of $[(C_6F_5)_2Ge]_3Bi_2$. Each apical Bi bonds non-covalently with the five nearest F atoms of the three bis(pentafluorophenyl) moieties. This became evident on exploring the system with an IGM-$\delta g$ based isosurface analysis (Fig. 27b). Although the very long Bi···F interaction ($r$(Bi···F) = 4.172 Å) was evident only when a very small isosurface value of 0.003 a.u. was used in the analysis, the two that are described by the thick greenish-blue volumes between the Bi and F atomic basins were always present regardless of the isovalue used. These two interactions are stronger than the remaining three Bi···F interactions which fall into the weak-to-vdW regime. These results demonstrate that Bi has the capacity to form at least five intramolecular interactions that are a characteristic of non-linear Type-IIb pnictogen bonds in addition to the three formal Bi–Ge ionic bonds with covalent character.

The molecular structures of two hybrid dibismuthines, $(O[(CH_2)_2BiPh_2]_2)$ and $(S(CH_2\text{-}2\text{-}C_6H_4BiPh_2)_2)$ that form crystalline materials are shown in Fig. 28a-b (left) [125]. In these two systems, intramolecular interactions dominate within the molecular frameworks which we assign as C–Bi···O and C–Bi···S intramolecular pnictogen bonds, respectively.

The latter are comparable to the C–Sb⋯S interaction observed in an analogous crystal system, ([S(CH$_2$-2-C$_6$H$_4$SbMe$_3$)$_2$]I$_2$) (Fig. 28c, left). They are the result of attractive engagements between the positive site on the electrostatic surfaces of the Bi/Sb atoms along the C–Bi/C–Sb bond extensions and negative sites localized on the O/S atom. As such, each molecular framework comprises of two intramolecular interactions. They are nearly equivalent in O[(CH$_2$)$_2$BiPh$_2$]$_2$ and S(CH$_2$-2-C$_6$H$_4$BiPh$_2$)$_2$, and significantly non-equivalent in [S(CH$_2$-2-C$_6$H$_4$SbMe$_3$)$_2$]$^{2+}$. This conclusion is arrived at based on the observed differences in the two (C–)Bi⋯O/(C–)Bi⋯S/(C–)Sb⋯S bond distances in O[(CH$_2$)$_2$BiPh$_2$]$_2$ or S(CH$_2$-2-C$_6$H$_4$BiPh$_2$)$_2$, or [S(CH$_2$-2-C$_6$H$_4$SbMe$_3$)$_2$]$^{2+}$. For instance, in the latter system, the (C–)Sb⋯S bond distances are very different, 3.555 and 4.419 Å (Fig. 28c), whereas the (C–)Bi⋯O bond distances in Fig. 28a are 3.203 and 3.126 Å, and (C–)Bi⋯S bond distances in Fig. 28b are 3.301 and 3.325 Å. Moreover, the long and short C–Bi⋯O pnictogen bonds in O[(CH$_2$)$_2$BiPh$_2$]$_2$ have ∠C–B⋯O values of 140.5° and 142.8°, respectively. These are Type-IIb interactions and non-linear pnictogen bonds. For S(CH$_2$-2-C$_6$H$_4$BiPh$_2$)$_2$, the corresponding angles associated with the short and long bonds are 161.4° and 163.1°, respectively, which are clearly Type-IIa interactions. However, in [S(CH$_2$-2-C$_6$H$_4$SbMe$_3$)$_2$]$^{2+}$, the ∠C–Sb⋯S associated with the short and long bonds are 169.8° and 125.9°, respectively, suggesting the former is a Type-IIa and the latter a Type-IIb interaction. Since the development of both these interactions occur with the same dicationic molecule – one quasi-linear and one non-linear – the former and latter may fall into the Type-III and Type-Ib category of pnictogen bonding (as there is no involvement of a negative site on the S in making the Sb⋯S bonds).

The characterization of (C–)Bi⋯O/(C–)Bi⋯S/(C–)Sb⋯S pnictogen bonding interactions in the three systems discussed above and based on the nature of the bond distances and bond angles is in agreement with the IGM-δ$g$ based isosurface plots shown in Fig. 28a-c (right). The isosurface between Sb and S atomic basins appears at a very small isovalue of 0.003 a.u., suggesting that this interaction is of the vdW type.

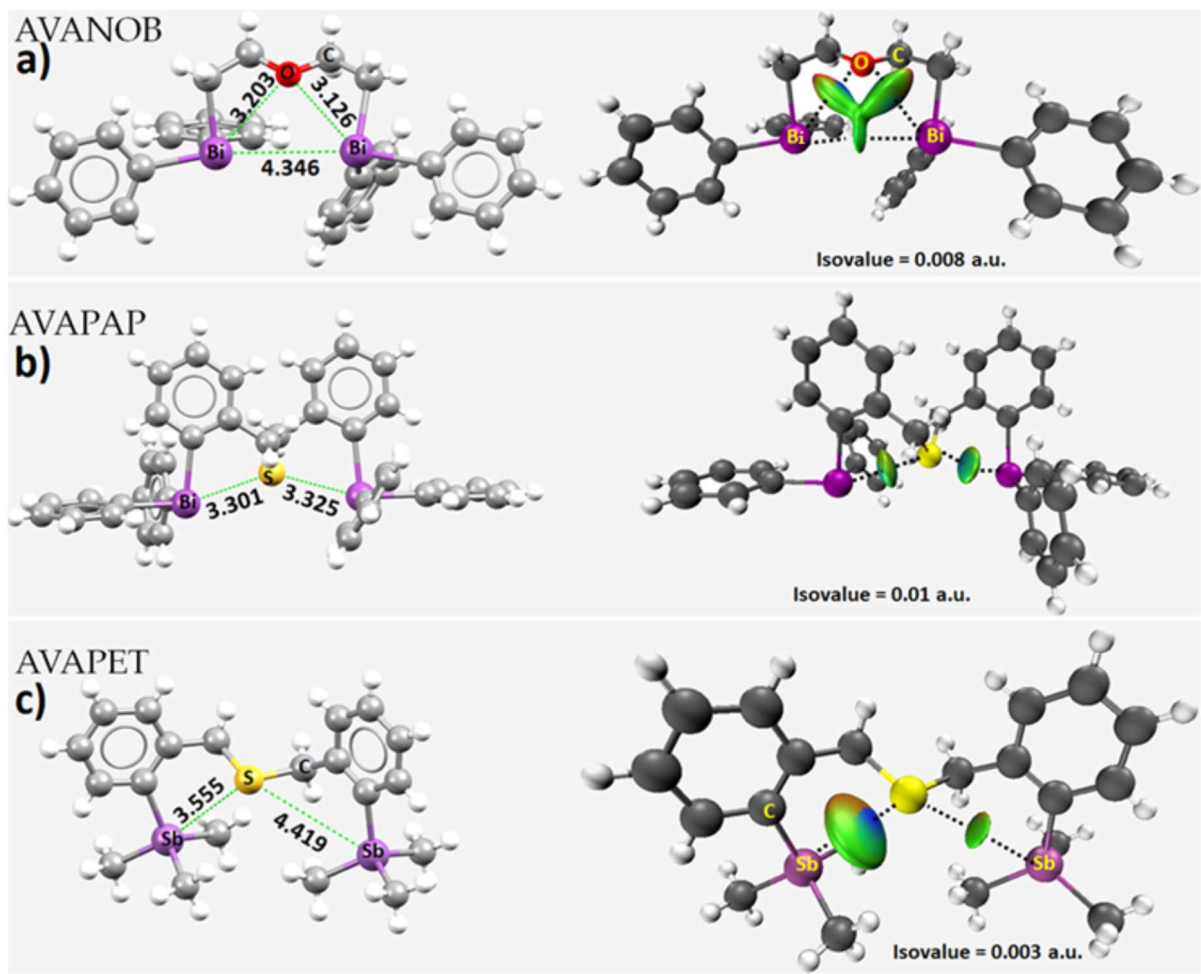

**Figure 28.** a)-c) (Left) Ball-and-stick models of the structure of some molecules displaying the involvement of Bi/Sb in intramolecular interactions [125]; (Right) The corresponding IGM-$\delta g$ based isosurface plots. The crystalline materials are a) ($\mu_2$-oxydiethan-2-yl)-tetraphenyl-di-bismuth; b) ($\mu_2$-2,2'-(2-thiapropan-1,3-diyl)diphenyl)-tetraphenyl-di-bismuth; c) ($\mu_2$-2,2'-(2-thiapropan-1,3-diyl)diphenyl)-hexamethyl-di-antimony bis(iodide). Bond lengths are shown in Å, and the dication in c) is stabilized with the assistance of two iodide anions (not shown). The isosurface colored blue and green signifies the interaction between bonded atomic basins that are strong and weak/vdW attractions, respectively, and that colored red is indicative of repulsion. .

There is evidently C–Bi⋯Bi–C pnictogen bonding in (O[(CH$_2$)$_2$BiPh$_2$]$_2$) (Fig. 28a), but not in the other two systems. Although Bi in this system is positive, the Bi⋯Bi interaction is the result of attraction between two sites of different electron density. This is also evident from the angle of interaction, ∠C–Bi⋯Bi = 166.7°. Moreover, the Bi⋯Bi bond distance, 4.346 Å, is longer than the (C)Bi⋯O contact distances, yet it is much smaller than the sum of the vdW radii of the two Bi atomic basins (5.08 Å). Nevertheless, these are not the key driving forces responsible for the molecular packing in the crystals. There are C–H⋯O/C–H⋯S and C–H⋯π(C$_6$), –CH$_2$⋯C(π) and various π⋯π stacking interactions that drive the packing in the solid state (not shown).

*g. Bi- and Sb-centered Pnictogen Bonds in Functional materials*

In an endeavor to discover new hybrid organic-inorganic materials, Leblanc and coworkers [196] prepared a series of nine new hybrids of Bi$^{3+}$ or Sb$^{3+}$ halides [M(X$_{5-x}$X′$_x$)]$^{2-}$ (M = Bi, Sb; X, X′ = Cl, Br, I) and methyl viologen dications, (N,N′-dimethyl-4,4′-bipyridinium, paraquat, MV$^{2+}$), including (MV)[BiCl$_{3.3}$Br$_{1.7}$], (MV)[BiCl$_{1.3}$Br$_{3.7}$], (MV)[BiBr$_{3.2}$I$_{1.8}$],

(MV)[SbCl$_5$], (MV)[SbBr$_5$], (MV)[SbCl$_{3.8}$Br$_{1.2}$], (MV)[SbCl$_{2.4}$Br$_{2.6}$], (MV)[SbI$_3$Cl$_2$], and (MV)[SbBr$_{3.8}$I$_{1.2}$]. The [M$^{3+}$X$_{5-x}$X'$_x$]$^{2-}$ units in some of these systems are linked with each other by M···X links, thereby forming a 1D linear chain-like architecture in the crystallographic *c*-direction, and are separated by the MV$^{2+}$ units that act as spacers between the inorganic entities. Two examples are shown in Fig. 29. In both cases, the M···X intermolecular links responsible for the 1D chain are significantly longer in length than the ordinary M–X bonds (viz. 3.46/3.56 Å vs. 2.78 Å in Fig. 29a; 3.603 Å vs. 2.87 Å in Fig. 29b). A very similar trend was found in the structures of (MV)[SbBr$_5$] and (MV)[SbCl$_5$] (CSD ref. codes: CEHSUF and CEHSOZ, respectively), in which the bond distances associated of Sb–Br (Sb–Cl) and Sb···Br (Sb···Br) are 2.541 (2.363), 3.669 (3.317) and 2.541 (2.263) Å, respectively, with ∠Br–Sb···Br (∠Cl–Sb···Cl) = 158.9° (175.8°)).

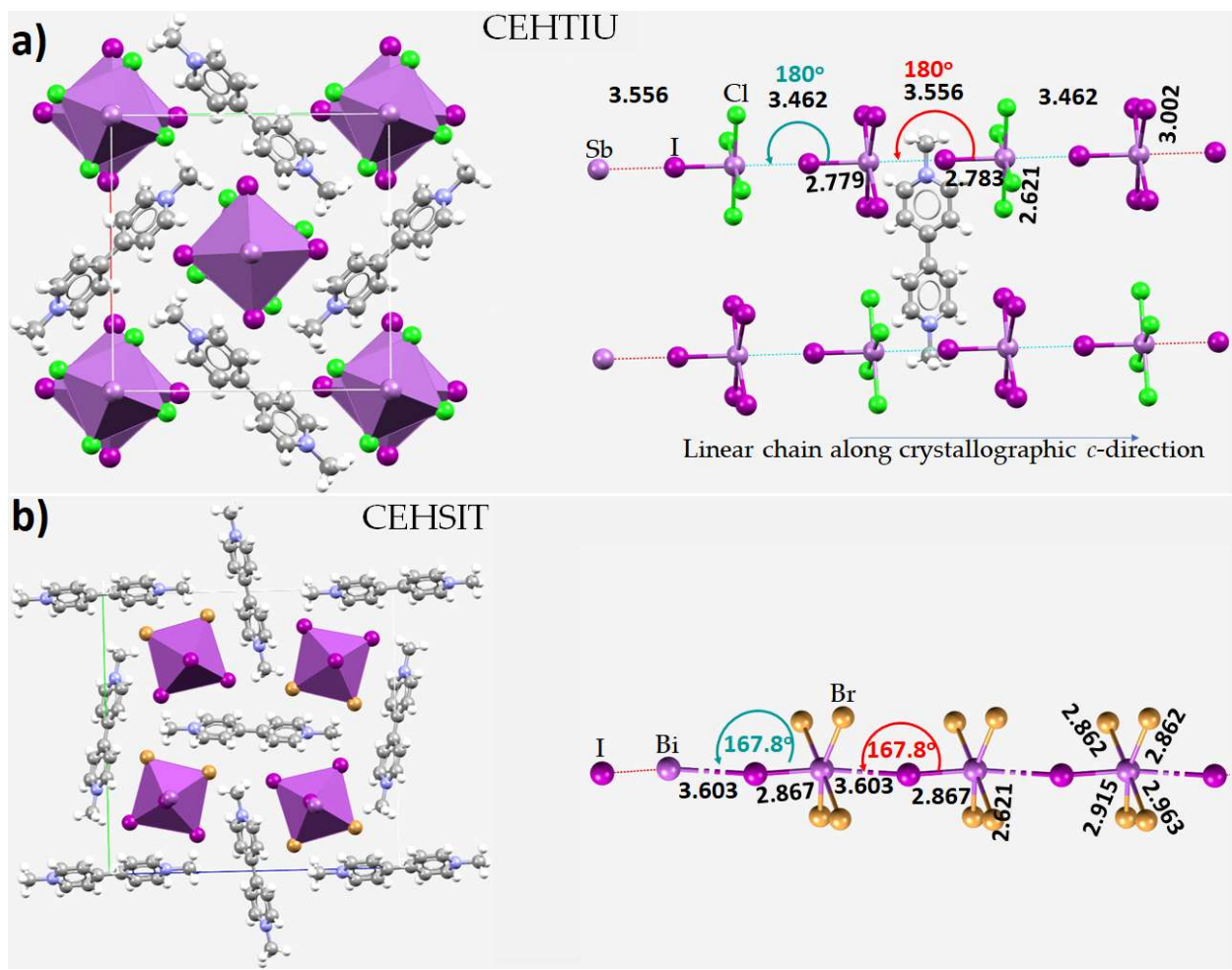

**Figure 29.** Polyhedral models of the unit cells of two hybrid organic-inorganic ferroelectric systems that feature Sb···I/Bi···I long-range interactions leading to the formation of linear or quasi-linear chains along the crystallographic *c*-axes. a) [(MV)$_2$(SbI$_5$)(SbCl$_4$I$_5$)]; b) [(MV)(BiBr$_4$I$_2$)]$_n$. Bond lengths and bond angles are in Å and degree, respectively. The CSD ref code is shown in each case.

The M···X bonds in the chains are characteristic of non-covalent interactions, yet they are formed between entirely negative [MX$_{5-x}$X'$_x$]$^{2-}$ units; they could be categorized as pnictogen-centered counterintuitive interactions that are a result of attraction between like charges [183,184,197]. Of the nine crystals systems listed above, it was observed that some feature a *syn* coupling of polar chains; this results in high experimentally-determined polarization values – similar to that observed for the well-known room temperature ferroelectric material, (MV)[BiI$_3$Cl$_2$] [198].

Another set of examples that involve linear or zigzag 1D chain-like architectures formed of pentagonal pyramidal MX$_3{}^{2-}$ dianions in an environment of organic dications

is shown in Fig 30. The same topology of M···X and M–X bonding modes noted above is seen in these systems.

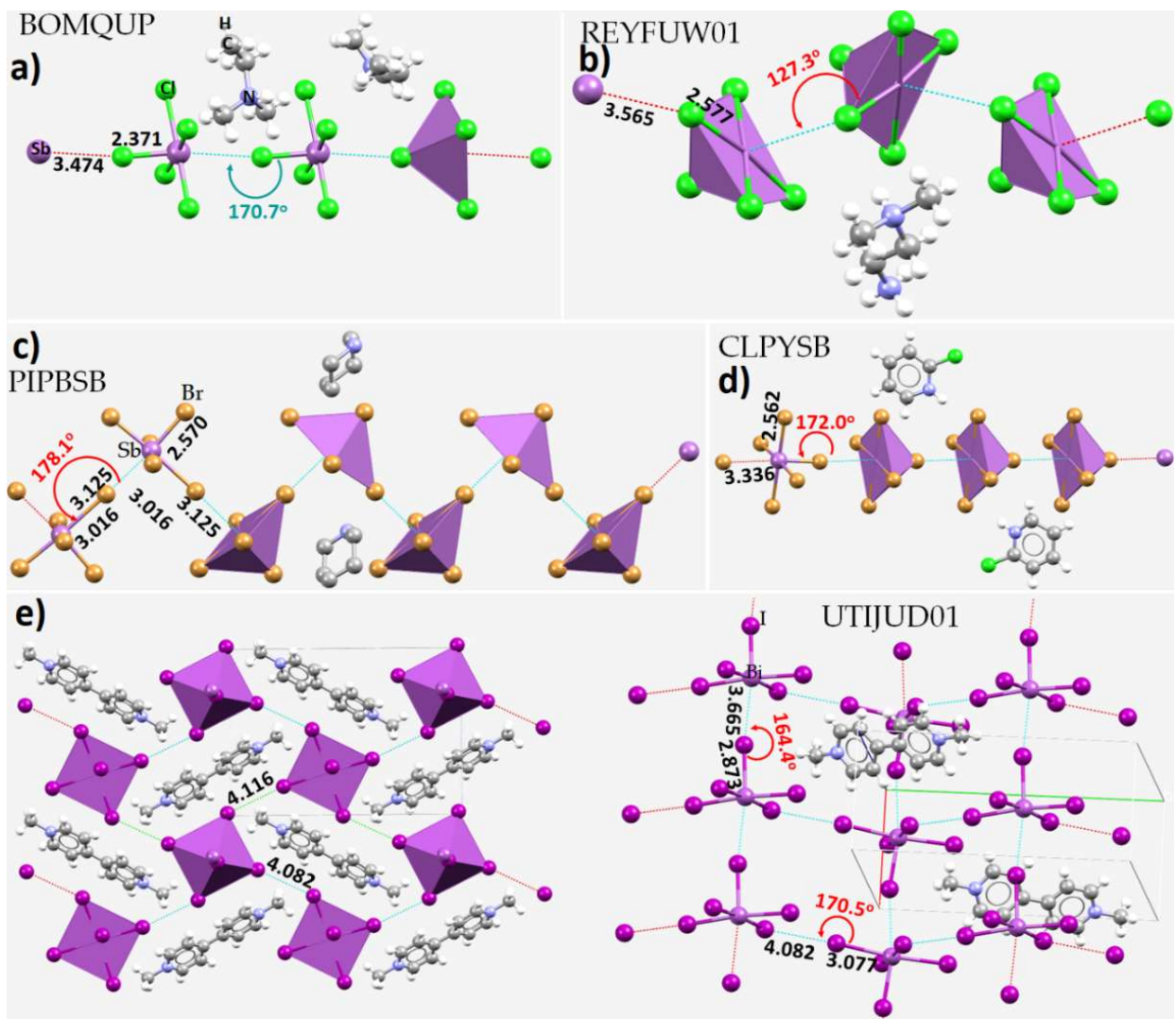

**Figure 30.** Further examples of hybrid organic anion-inorganic cation systems that feature 1D and 3D [PnX$_5$]$^{2n-}$ architectures though pnictogen-centered non-covalent links between the [PnX$_5$]$^{2-}$ entities. a) [En][SbCl$_5$] (En = [C$_2$H$_4$(NH$_3$)$_2$]$^{2+}$ [199]; b) [(CH$_3$)$_2$En][SbCl$_5$] (CH$_3$)$_2$En = [(CH$_3$)$_2$NHCH$_2$CH$_2$NH$_3$]$^{2+}$ [200]; c) (Pip)$_2$[SbBr$_5$] (Pip = piperidinium);[201] [2Cl-py][SbBr$_5$] [202]; and e) (MV)[BiI$_3$] [203].

Probably one of the most important hybrid organic-inorganic materials for photovoltaics is MAPbI$_3$ (MA = methylammonium). While MAPbI$_3$ has been an important photovoltaic material, it is unstable and toxic, and hence environmentally unfriendly. Many attempts have been made to discover lead-free halide-based perovskites for photovoltaics and other optoelectronic applications. Several antimony- and bismuth-based halide perovskites have been synthesized and reported which may have future potential as perovskite-based solar-cell absorbers because of their lower toxicity. For instance, Wang and co-workers recently reported a lead-free, pseudo-3D perovskite optoelectronic material, (MV)BiI$_5$ [203]. Although the authors have assigned the pentaiodobismuth cation to have a charge of +2, and noted that there are I···I contacts in the crystal, this may be misleading. Our analysis suggests that the methylviologen units act as spacers between the [BiI$_5$]$^{2-}$ quasi-linear chains along the crystallographic *a*-axes, and [BiI$_5$] carries a charge of –2. The Bi···I links between the [BiI$_5$]$^{2-}$ units causing the 1D pseudo-linear chains are longer than the Bi–I bonds (*r*(Bi···I) values 3.665 Å *vs.* 2.873 Å). As explained above, the former links

are typical of non-covalent interactions and the latter are bonds with mixed ionic and covalent character. Since the [BiI$_5$]$^{2-}$ units that cause the development of the long-range non-covalent interactions are entirely negative and there are no positive sites involved in making these interactions, it would be misleading to name the Bi⋯I links as pnictogen bonds even though they feature a directionality synonymous with Type-IIa pnictogen bonds. Because of this, we characterize them as Type-III pnictogen bonds. There are I⋯I links between linear chains. They are very long ($r$(I…I) = 4.082 and 4.112 Å), slightly longer than twice the vdW radius of I, and more directional than the Bi⋯I long bonds. These, together with π⋯I interactions between MV$^{2+}$ and [BiI$_5$]$^{2-}$, probably engineer the overall structure of the crystal system as pseudo three dimensional.

The (MV)BiI$_5$ perovskite has a narrow band gap of 1.48 eV, electrical conductivity of 0.73×10$^{-10}$ S cm$^{-1}$ and a better photoresponse than (MV)BiCl$_5$, with its 1 D structure formed by the non-covalent links between the [BiCl$_5$]$^{2-}$ units, and which has a bandgap of 2.59 eV. Comparable 1D systems include (TMP)[BiCl$_5$], (TMP)[BiBr$_5$] (TMP = tetramethylpiperazine) [204] and (DMEDA)BiI$_5$ (DMEDA$^{2+}$ = CH$_3$NH$_3$CH$_2$CH$_2$NH$_3$CH$_3$$^{2+}$) [205]; they have bandgaps of 3.21, 2.67 and 1.82 eV, respectively. It was suggested in those studies that (MV)BiI$_5$ is the first Bi-based perovskite compound with a band gap energy comparable with (CH$_3$NH$_3$)PbI$_3$, which is encouraging for optoelectronic applications. This perovskite may open a pathway to the design of pseudo-3D Bi-based perovskites with performance comparable with the widely examined APbX$_3$ absorbers.

## 8. The crystals of Bismuth

Crystals of Bi are known in several phases both in 2D and 3D, with space groups *R-3m*, *P2$_1$/n*, *P2$_1$/m*, *I$_4$/mcm*, *I$_4$/mmm*, *Pm-3m*, *Im-3m*, *Cmca*, and *C$_2$/m* found in the ICSD. Stable, free-standing, 2D single-layer phases of Bi, called bismuthene, have been reported, including, for example, the buckled honeycomb or hexagonal (h-Bi), symmetric washboard (w-Bi), asymmetric washboard (aw-Bi), and square-octagon (so-Bi) structures; aw-Bi is less stable than w-Bi [206].

The rhombohedral A7 structure of Bi is stabilized by Jones–Peielrs distortion [207]. It has two atoms in a primitive cell. The bulk bismuth with *R-3m* space group [207] has a layered structure. Each atom bonds covalently to its three nearest neighbors forming buckled bilayer with a σ bond, a characteristic of pnictogen bonding in a semimetal [208]. The interaction between the adjacent bilayers is much weaker than the intra-bilayer bonding; hence bismuth cleaves along the (111) plane.

Shown in Fig. 31a-c is the structure of Bi in the *R-3m* space group. The coordinately bound Bi in a given layer links with the equivalent atom in the neighboring layers by means of long-range contacts. Depending on the number of layers and packing of atoms in the unit-cell, the inter-layer distance between the monolayers can be determined. In the structure shown in Fig. 31a-b, the Bi⋯Bi inter-layer distance is 3.529 Å; each covalently bonded Bi site in a monolayer is linked with three nearest neighbors forming three Bi⋯Bi equivalent contacts that are directional (∠Bi–Bi⋯Bi) = 169.3°). These are non-covalent interactions. This view is justified since the coordinate bond formed by each Bi atom in each monolayer is 3.071 Å, which is markedly shorter than the long bonds just noted, and Bi is locally trigonal within a monolayer (if one ignores the presence of its stereo-active lone-pairs). Although the directional feature of the Bi⋯Bi links in the crystal is consistent with a Type-IIa interaction, the long-range interactions could be characterized as Type-III (Scheme 1) given the electrostatic potential on the Bi site is entirely positive. This conclusion is supported by the MESP of a Bi$_2$ molecule, computed with MP2/Aug-cc-pVTZ (Fig. 31d). The outer cap on Bi along the Bi – Bi bond extensions is positive with a local minimum of potential $V_{S,min}$ = 4.9 kcal mol$^{-1}$. The lateral portions of the same atom is described by a belt of positive potential ($V_{S,max}$ = 7.1 kcal mol$^{-1}$). The bonding region is described by a belt of negative potential ($V_{S,min}$ = –3.2 kcal mol$^{-1}$). These were 5.0, 7.1 and –3.4 kcal mol$^{-1}$ with MP2/def2-TZVPPD, suggesting that the magnitude, but not sign, of potential is marginally affected by changing the size of the pseudopotential-based basis set. Clearly, the quasi-linearity of Bi⋯Bi close-contacts originates from the attraction between regions on

Bi of unequal charge density, i.e., the portions with $V_{S,min}$ and $V_{S,max}$ along and around one Bi in a monolayer are attracting the opposite portions described by $V_{S,max}$ and $V_{S,min}$ on the same atom in the interacting monolayer, respectively. While this conclusion is drawn using the potentials computed on the fully relaxed geometry of the Bi$_2$ molecule that has an $r$(Bi–Bi) of 2.675 Å (and 2.663 Å) with MP2/Aug-cc-pVTZ (MP2/def2-TZVPPD), each monolayer in the crystal has an $r$(Bi–Bi) of is 3.071 Å. Our single point calculation with MP2/Aug-cc-pVTZ using the crystal geometry of Bi$_2$ has altered $V_{S,min}$ to $V_{S,max}$ on the surface of the Bi atom along the bond extensions with $V_{S,max}$ = 7.1 kcal mol$^{-1}$, but the character of $V_{S,min}$ ($V_{S,min}$ = –3.2 kcal mol$^{-1}$) remains unchanged at the bonding region. Further elongation of $r$(Bi–Bi) close to $r$(Bi⋯Bi) = 3.529 Å resulted in an MESP shown in Fig. 31e. While the surface region of the molecule is dissected into lateral and axial regions of positive potentials, the charge density at the bonding region is largely depleted, giving rise to a belt a positive potential. This is not unexpected since *elongated* molecules have *mobile electron density*, *increasing* their polarizability, and thus strengthening the dispersion forces between the bonded atomic basins.

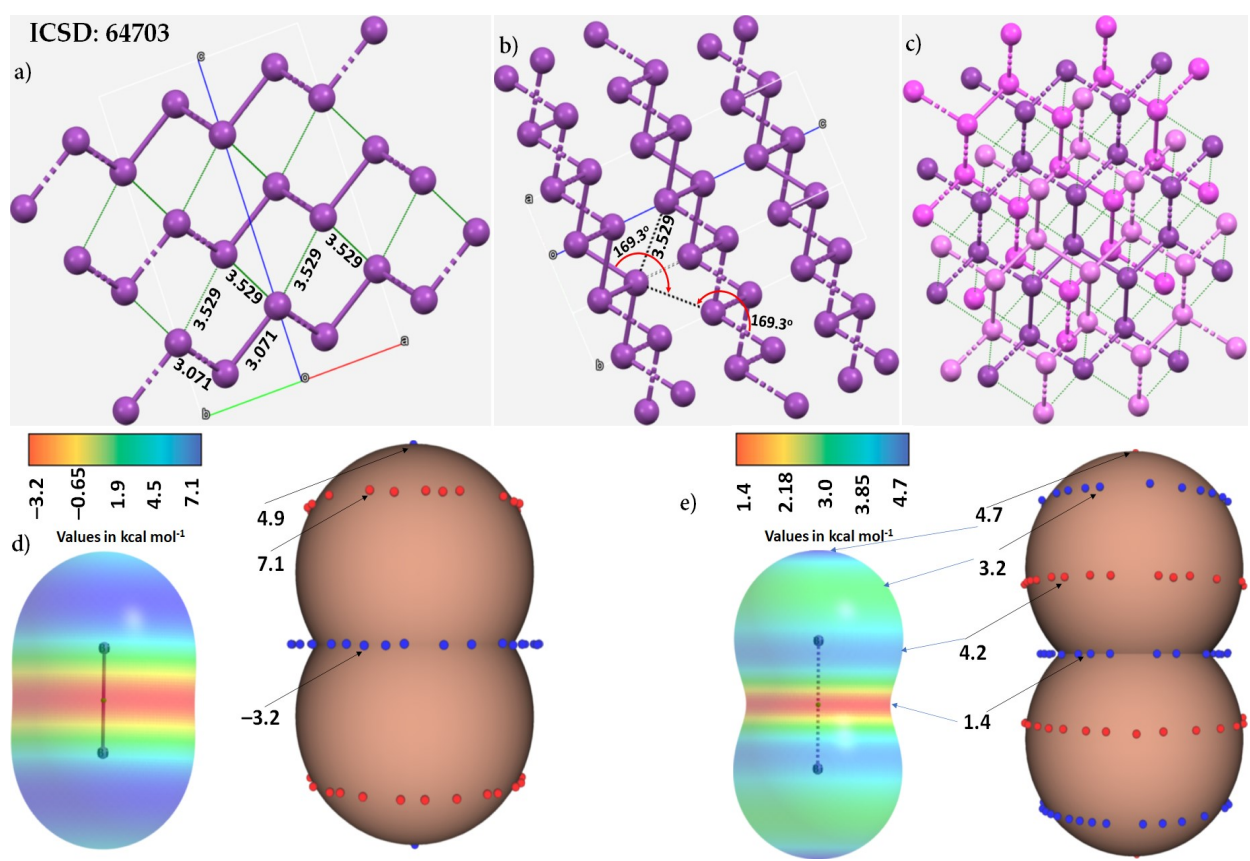

**Figure 31.** a)-c) Ball-and-stick model of \three different views of the crystal of Bi (space group: *R-3m*), showing the inter-layer interactions between the bonded Bi sites that hold the monolayers together. b) Illustration of the typical nature of link formed by a given Bi site in a monolayer with three nearest neighbor Bi sites in a neighboring monolayer. c) Illustration of Bi⋯Bi contacts formed by each Bi site in each of the three layers (the three layers colored in purple, pink and faint-pink, respectively). The thin dotted lines between Bi sites represent Bi⋯Bi close contacts. Selected bond distances and bond angles are in Å and degree, respectively. d) and e) The MP2/Aug-cc-pVTZ level 0.001 a.u. isoelectron density envelope mapped potential on the surface of a Bi$_2$ molecule calculated on the MP2/Aug-cc-pVTZ optimized and (fixed) crystal geometries, respectively. The tiny blue and red dots on van der Waals surfaces of Bi$_2$ (right) represent the local most minimum and local most maximum of potential, respectively. The ICSD ref. for the crystal is shown in a).

## 9. Discussion and Conclusion

From our search of the CSD and ICSD we note that the coordination number of Bi in its most common oxidation state of 3+, varies between 7 and 10, although in many of its compounds this can range from 2 to 9 or 10 [104]. This, of course, is dependent on its ligands. Not all coordination modes are equivalent, and we note that the bonds with bond lengths >2.6 Å have predominantly non-covalent character but may have some covalent character as well. For instance, Bi in the catena-(($\mu_5$-ethylenediamine-N,N,N',N'-tetra-acetato)-triaqua-(nitrato-O)-(nitrato-O,O')-bismuth-praseodymium nitrate dehydrate, [($C_{10}H_{18}BiN_4O_{17}Pr^+$)$_n$]·2n($H_2O$)·n($NO_3^-$), (CSD ref: HUMQUC [209]), is 10-coordinate; there are four Bi···O($NO_2^-$) bonds (one pair with $r$(Bi···O) = 2.774/2.889 Å and ∠N–Bi···O = 141.6/151.9°, and the other pair with $r$(Bi···O) = 2.664/2.791 Å and ∠N–Bi···O/∠O–Bi···O = 154.5/146.2°) that are clearly predominantly ionic interactions, whereas the bonds to the N donors of the ethylenediamine ligands (2.48 and 2.51 Å) are coordinate covalent bonds. Similarly, some of the Bi sites in [($H_4Bi_6O_8^{6+}$)($NO_3^-$)$_6$]·$H_2O$ (CSD ref: YEGDIZ [210]) is nine-coordinate, with five Bi–O covalent coordinate bonds (in the range 2.3 – 2.67 Å) and five Bi···O bonds ($r$(Bi···O) in the range 2.9 – 3.10 Å) between Bi in the cage-like cation $H_4Bi_6O_8^{6+}$ and O in the surrounding $NO_3^-$ and $H_2O$ species.

Many crystals show Bi···D directional interactions between entirely negative sites in two close-lying building blocks in presence of a third building block, such as the Bi···O close contacts in the crystal structures of 4($Ph_4Bi^+$)($Bi_4I_{16}^{4-}$), CSD ref: HUJCIZ [114]; $Ba^{+2}$($C_{20}H_{26}Bi_2N_4O_{17}^{2-}$), LOHLUP [211]; 2($C_{10}H_{16}N^+$)($C_4Bi_2Cl_6FeO_4^{2-}$), TINMEI [212]; and ($C_5H_6N^+$)($C_{20}H_{12}BiO_8S_4^-$)·$C_5H_4O_2S$, MARTOR [213]; and the Type-III Bi···S close contact in (18-crown-6)-potassium trichloro-isothiocyanato-bismuth(III) ($C_{12}H_{24}KO_6^+$)($BiCl_3NCS^-$), CSD ref. GIHGEJ, ($r$(Bi···S) = 3.046 Å and ∠Cl–Bi···S = 173.5° [214]).

Searching the CSD for a specific geometry does not always give all possible close contacts in crystals. For instance, a single Bi···O interaction was identified by a CSD search in ($^t$BuO)$_3$Bi (HURSUJ [215]), with $r$(Bi···O) = 3.327 Å and ∠O–Bi···O = 140.8°; yet our inspection of the coordination environment around Bi showed up two more such interactions with $r$(Bi···O) = 3.327 Å and ∠O–Bi···O = 140.8°. Similarly, there should be two Bi···S interactions in bis(*N*,*N*-dimethyldithiocarbamato)-(2-(dimethylaminomethyl)phenyl)-bismuth [$C_{15}H_{24}BiN_3S_4$] (EBUXUY [216]), but CSD search gave a single contact ($r$(Bi···S) = 4.060 Å and ∠C–Bi···S = 150.3°). This means that the number of close contacts shown in the histogram plots in Section 6 must be regarded as indicative only, since many would undoubtedly have been missed.

As mentioned, while Bi has a variable coordination number, this it is not the case for a covalently or coordinately bound Bi in molecular entities in the majority of chemical systems examined in this study when pnictogen bonds are involved. In this case, the coordination number of Bi could be as large as nine if the pnictogen bonds (as in [($C_6F_5$)$_2$Ge]$_3$Bi$_2$, Fig. 27a), are include. Clearly the coordination number is limited by steric crowding of the metal ion.

Our analysis reveals that arene π-donors interact with covalently bound Bi in many crystals, forming Bi···π interactions. Of course, the number of such instances is larger than that with donors such as I, Br, Se, and Te in partner molecules. Among the lone-pair donors investigated, O in molecules is found to be more active in donating electron density to Bi in forming Bi···O pnictogen bonds when compared to the number of pnictogen bonds found when N, Cl and F sites serve as pnictogen bond acceptors. Formation of bismuth bonds with regions of π-density in $C_π$ (arene), (C≡C)$_π$ and ($C_6$)$_π$ moieties is not very rare.

Type-III Bi···Bi interactions were shown to exist in many crystals containing Bi. They appear either between two negative sites, or between two positive sites. One such readily available system where Type-III Bi···Bi bonds play an important role to provide stability is crystalline Bi($CH_3$)$_3$ (cf. Fig. 25). Whether this type of interaction is seen only in the solid state because of packing, or whether it can be occurred in chemical systems when packing and counterion effects are removed, could be the one of the key focus areas of future computational studies.

Many crystal structures have been published in which covalently or coordinately bound Bi is electrically neutral. However, the anisotropy in the charge density profile of Bi in these systems enables it to form Bi⋯Bi interactions that follow a Type-III bonding topology. We expect that crystals with analogous bonding features (such as Bi⋯Sb and Bi⋯As) may be present in the other crystallographic databases should be of interest to the material science community since chemical systems with such interactions have implications in the design of functional materials such as bismuthene [206,217-220].

A statistical analysis of various Bi-centered interactions suggested that, depending on the size of Lewis base, the intermolecular distance range can vary. There was no close contact observed in the range 2.6 – 3.2 Å for Bi⋯I and Bi⋯Br pnictogen bonds, although this was not the case for Bi⋯Cl and Bi⋯F, in which, the directionality associated with these latter two types of interactions were largely affected by secondary interactions. While a variety of inter- and intra-molecular interactions pnictogen bonds were explored, the Bi⋯O pnictogen bonds are found to be widely scattered in various crystals, and a large number of them were non-linear because of the high electronegativity of O. The Bi⋯$C_6$(centroid) close contact distance was relatively less dense in the range 2.8 – 3.4 Å than in the range 3.5 – 4.5 Å, even though the peak of the bell curve appears at 3.86 Å, yet these were abundant compared to Bi⋯N, Bi⋯Se, Bi⋯Te, Bi⋯S, and Bi⋯Bi pnictogen bonds.

Although there is no general consensus on how to distinguish between coordinate bonds and potentially strong pnictogen bonds, and it could be argued that they form a continuum without sharp boundaries. An interaction can be regarded as a pnictogen-centered coordinate bond if the bonding region presents with a moderate concentration of electron density that causes the Bi⋯D distance to shrink appreciably. We have observed in some systems, such as host-guest complexes of crown ethers, similar to those observed in $BiCl_3$ crystals, that the Bi⋯D bond distances could be used as an indicator to discriminate between coordinate and pnictogen bonds. In most cases, when D is N, O, F, or Cl, the Bi–D coordinate bonding distance is close to or less than 2.6 Å (a lower limit of intermolecular bond distances used for most of the contacts used in the statistical analysis in this study). When S, Se, Te, Br, or I atoms are the donors, the Bi–D coordinate bonding distances are much longer since the vdW radii of these atoms are larger. The identification and subsequent characterization of interactions between Bi and D as pnictogen bonds in the illustrative crystal systems was made possible by comparing them to bonds that were unquestionably coordinate bonds, and when they were found to be reasonably or notably longer than the coordinate bonds (see, for example, the discussion on the $BiCl_3$ crystal). The use of the concept of the size of the vdW radii of the atoms involved in the formation of the coordinate and pnictogen bonds was useful in this regard; of course, this is in addition to the exploration of the directional feature and the nature of the MESP-based extrema that appear along the R–Bi bond extensions in molecular entities. As demonstrated in a recent study by Santos et al. [5], the strength of pnictogen bonds can be very significant (viz. –89.2, –52.0 and –45.8 kcal mol$^{-1}$ for $Br_3Sb$⋯$F^-$, $Br_3Sb$⋯$Cl^-$ and $Br_3Sb$⋯$Br^-$, respectively, with ZORA-M06/QZ4P), which not only crosses the covalent limit (–40.0 kcal mol$^{-1}$) of hydrogen bonds (viz. FH⋯$F^-$) proposed by Desiraju [221], but also comparable with, or even larger than, the strength of coordinate bonds. Future computational studies on a variety of chemical systems such as those illustrated in this study are expected to clarify the boundaries that can discriminate between coordinate and pnictogen bonds.

This overview summarized the occurrence and versatility of Bi-centered non-covalent interactions in crystals deposited in the CSD and ICSD databases, but did not provide any detailed analysis of these using advanced computational approaches. Therefore, theoretical and computational efforts to characterize the chemistry of Bi-centered non-covalent interactions in molecules, molecular complexes, crystals, and nanomaterials have not yet been fully explored. Such studies are surely necessary since they assist us in developing our basic knowledge on the non-covalent chemistry of Bi; the presence of Bi in molecular entities has already been played a critical role in many areas of science and technology, including catalysis, photovoltaics, and other optoelectronic engineering.

**Author Contributions:** Conceptualization, project design, and project administration, P.R.V.; formal analysis and investigation, P.R.V. and A.V.; supervision, P.R.V.; writing—original draft, P.R.V. and A.V.; writing—review and editing, P.R.V., H.M.M., A.V., and K.Y. All authors have read and agreed to the published version of the manuscript.

**Funding:** This research received no external funding.

**Institutional Review Board Statement:** Not applicable.

**Informed Consent Statement:** Not applicable.

**Data Availability Statement:** This research did not report any data.

**Acknowledgments:** This work was entirely conducted using the various computation and laboratory facilities provided by the University of Tokyo and the Research Center for Computational Science of the Institute of Molecular Science (Okazaki, Japan). P.R.V. is currently affiliated with the University of the Witwatersrand (SA) and Nagoya University, Aichi 464-0814, Japan. A.V. is currently affiliated with Tokyo University of Science, Tokyo, Japan 162-8601. K.Y. is currently affiliated with Kyoto University, ESICB, Kyoto, 615-8245, Japan. H.M.M. thanks the National Research Foundation, Pretoria, South Africa and the University of the Witwatersrand for funding.

**Conflicts of Interest:** The authors declare no conflict of interest. The funders had absolutely no role in the design of the study; in the collection, analyses, or interpretation of data; in the writing of the manuscript; or in the decision to publish the results.